\documentclass[aps,prb,reprint,showpacs]{revtex4-2}
\usepackage{graphicx}
\usepackage{latexsym}
\usepackage{amsmath}
\usepackage{amssymb}
\usepackage{amsthm}
\usepackage{amsfonts}
\usepackage{color}
\usepackage{xcolor}
\usepackage{verbatim}
\usepackage{url}
\usepackage{array}
\usepackage{multirow}

\def\bx{\mathbf{x}}
\def\br{\mathbf{r}}
\def\bR{\mathbf{R}}

\def\bk{\mathbf{k}}
\def\bG{\mathbf{G}}

\def\h2o{\mathrm{H}_2\mathrm{O}}

\def\001{$\langle 001 \rangle$}
\def\111{$\langle 111 \rangle$}
\def\icomp{\mathrm{i}}

\newcommand\refeq[1]{Eq.~(\ref{#1})}

\newcommand\fabien[1]{\textcolor{black}{#1}}

\newcolumntype{C}[1]{>{\centering\arraybackslash }b{#1}}
\newcolumntype{L}[1]{>{\flushleft\arraybackslash }b{#1}}

\graphicspath{{./Figures/}{./}}

\begin{document}

\title{\fabien{$GW$ density matrix to estimate self-consistent $GW$ total energy in solids}}

\author{Adam Hassan Denawi}
\affiliation{Universit\'e Paris-Saclay, CEA, Service de recherche en Corrosion et Comportement des Matériaux, SRMP, 91191 Gif-sur-Yvette, France}

\author{Fabien Bruneval}
\affiliation{Universit\'e Paris-Saclay, CEA, Service de recherche en Corrosion et Comportement des Matériaux, SRMP, 91191 Gif-sur-Yvette, France}

\author{Marc Torrent}
\affiliation{CEA, DAM, DIF, F-91297 Arpajon, France}
\affiliation{Universit\'e Paris-Saclay, CEA, Laboratoire Mati\`ere en Conditions Extr\^emes, 91680 Bruy\`eres-le-Ch\^atel, France}

\author{Mauricio Rodr{\'i}guez-Mayorga}
\affiliation{Universit\'e Paris-Saclay, CEA, Service de recherche en Corrosion et Comportement des Matériaux, SRMP, 91191 Gif-sur-Yvette, France}
\affiliation{Department of Physical Chemistry, University of Alicante, E-03080 Alicante, Spain}

\begin{abstract}
The $GW$ approximation is a well-established method for calculating ionization potentials and electron affinities in solids and molecules.
For numerous years, obtaining self-consistent $GW$ total energies in solids has been a challenging objective that is not accomplished yet.
However, it was shown recently that the linearized $GW$ density matrix permits
a reliable prediction of
the self-consistent $GW$ total energy for molecules
[F. Bruneval \textit{et. al.} J. Chem. Theory Comput. \textbf{17}, 2126 (2021)]
for which self-consistent $GW$ energies are available.
Here we implement, test, and benchmark the linearized $GW$ density matrix for several solids.
We focus on the total energy, lattice constant, and bulk modulus obtained
from the $GW$ density matrix and compare our findings to more traditional results obtained within the random phase approximation (RPA).
We conclude on the improved stability of the total energy obtained from the linearized $GW$ density matrix with respect to the mean-field starting point.
We bring compelling clues that the RPA and the $GW$ density matrix total energies
are certainly close to the self-consistent $GW$ total energy in solids if we use 
hybrid functionals with enriched exchange as a starting point.
\end{abstract}
\maketitle

\section{Introduction}

\fabien{
While a few self-consistent $GW$ calculations (sc$GW$) for the band gaps of real solids
are available 
\cite{ku_prl2002,cao_prb2027,grumet_prb2018},
sc$GW$ total energies are still not available today, certainly because of their high computational cost.}
However, there exist hints that sc$GW$ could be accurate:
first, the results on the homogeneous electron gas are extremely good
\cite{holm_prl1999,garcia_prb2001,kutepov_prb2009};
second, the random-phase approximation (RPA) which derives from the same family
has been shown to yield total energies capable of describing the tenuous van der Waals interactions
\cite{dobson_prl1999,garcia_prl2002,harl_prb2008,lu_prl2009,schimka_nm2010,lebegue_prl2010,bruneval_prl2012}.
Unfortunately, sc$GW$ calculations are very involved in real solids.
That is why it would be highly desirable to obtain sc$GW$ quality energies without actually
performing the cumbersome self-consistency.

Pursuing the quest for a non-self-consistent approximation to sc$GW$,
a series of studies have been published in the early 2000s
\cite{almbladh_ijmpb1999,dahlen_jcp2004,dahlen_prb2004,dahlen_pra2006,stan_epl2006,hellgren_prb2015}.
More recently, some of us proposed an alternative non-self-consistent total energy expression based
on the $GW$ density matrix, labeled $\gamma^{GW}$~\cite{bruneval_jctc2019,bruneval_jctc2021}.
Benchmarks on small molecules for which sc$GW$ are possible \cite{caruso_prb2012,caruso_prl2013}
confirmed the remarkable properties of the $\gamma^{GW}$ total energies:
although it is evaluated non-self-consistently using a generalized Kohn-Sham (gKS) input,
the resulting total energy remains quite insensitive to the gKS choice and
approximates very well the reference sc$GW$ total energies.

In this context, this study focuses on the evaluation of total energies in solids.
We port to the solid systems the $\gamma^{GW}$ total energy with the sensible prospect that
it will remain a good approximation to the sc$GW$ total energy.
In doing so, we obtain the correlated density matrix $\gamma^{GW}$
as a physically meaningful intermediate object with unique properties due
to correlation.

When considering solid systems several technical questions have to be addressed.
Firstly, the closed formulas obtained for finite systems \cite{bruneval_prb2019,bruneval_jctc2019}
have to be adapted for numerical efficiency.
Secondly, the pseudopotentials \cite{martin_book} that are customary in the plane-wave basis codes
are typically designed to be used in conjunction with standard semi-local approximations
to density-functional theory.
Therefore, it is necessary to investigate which type of pseudopotential is suitable for obtaining consistently $GW$-type total energies.

With this, we will study the performance of the $\gamma^{GW}$ total energy for solids.
We will compare it to the popular RPA total energy which may be derived
as the $GW$ approximation of the Klein functional \cite{pines_book,klein_pr1961,almbladh_ijmpb1999}.
As all these calculations are performed as a one-shot procedure,
memory about the mean-field starting point is present.
We will particularly investigate this issue by varying the content of exact exchange $\alpha$
in the hybrid functional PBEh($\alpha$),
zero being the standard PBE \cite{perdew_prl1996}
and 0.25 yielding regular PBE0 \cite{adamo_jcp1999}.
It is therefore always necessary to specify the exact procedure
used to obtain a one-shot energy.
We do so here using the @ notation (e.g. RPA@PBE stands for the RPA total energy
evaluated with self-consistent PBE inputs).

The article is organized as follows:
In Sec.~\ref{sec:theory}, we recapitulate the theoretical foundation
for the $GW$ density matrix and derive the working equations;
In Sec.~\ref{sec:computation}, we detail the technical aspects of the 
implementation in a plane-wave code and assess
the pseudopotential choice;
Sec.~\ref{sec:sigwdm} shows some of the unique properties of the $GW$ density matrix,
exemplified with bulk silicon;
In Sec.~\ref{sec:results}, we benchmark the total energies
obtained with the different approximations with a test set of
7 standard covalent semiconductors
and one layered material.
Finally, Sec.~\ref{sec:conclusion} concludes our work.

\section{Theory and working formulas}
\label{sec:theory}

\subsection{Green's function-derived density matrix}

In the many-body perturbation theory, the central quantity, namely the one-particle Green's function
$G(\br,\br',\omega)$, contains a great deal of information.
In particular, by virtue of the Galistkii-Midgal formula \cite{galitskii_jetp1958},
it is sufficient to calculate the total energy of an electronic system.

Also it straightforwardly yields the density matrix $\gamma(\br,\br')$
\begin{equation}
 \label{eq:gamma}
 \gamma(\br,\br') = -\frac{\icomp}{2\pi}  \int_{-\infty}^{+\infty} d\omega 
             e^{i\eta \omega}  G(\br,\br',\omega) ,
\end{equation}
where $\eta$ is a vanishing positive real number that enforces the sensitivity to the occupied
manifold of the time-ordered Green's function $G$.
Hence, the electronic density can be obtained as the \fabien{diagonal}: $\rho(\br) = \gamma(\br,\br)$.

Therefore, the Green's function methods, such as the $GW$ approximation, 
can give access to an approximate density matrix.

\subsection{Linearized Dyson equation}

In the many-body perturbation theory, the overall strategy is to connect
the exact Green's function $G$ to a known Green's function $G_0$.
The connection between the two is ensured by the complicated self-energy $\Sigma_{xc}$
that is in charge of all the correlation effects.

The expression of $G_0$ that is derived from a mean-field approach (Kohn-Sham, Hartree-Fock, etc.),
is simple:
\begin{equation}
 \label{eq:G0}
    G_0(\br,\br',\omega) =
      2  \sum_{\bk i}
          \frac{\varphi_{\bk i}(\br)
                \varphi_{\bk i}^*(\br')}{\omega - \epsilon_{\bk i} \pm \icomp \eta} ,
\end{equation}
where $\varphi_{\bk i}(\br)$ and $\epsilon_{\bk i}$ are the mean-field
wavefunctions and eigenvalue
for state $i$ at k-point $\bk$
and
the small positive $\eta$ ensures the correct location of the poles for
a time-ordered function (above the real axis for occupied states $\epsilon_{\bk i} < \mu$
and below the real axis for empty states $\epsilon_{\bk i} > \mu$,
$\mu$ being the Fermi level).
Spin-restricted calculations are assumed here and the factor 2 accounts for it.

Then the connection from $G_0$ to $G$ is made with the so-called Dyson equation:
\begin{equation}
\label{eq:dyson}
  G = G_0 + G_0 (\Sigma_{xc} - V_{xc} ) G  ,
\end{equation}
where $V_{xc}$ is the exchange-correlation operator (possibly including non-local exchange)
the space and frequency indices have been omitted for conciseness.

The self-energy $\Sigma_{xc}$ is itself a functional of the exact $G$.
When approximating $\Sigma_{xc}$ and $G$, only a self-consistent solution
ensures the conservation of the electron count \cite{baym_pr1961}.
In particular, for the $GW$ approximation of $\Sigma_{xc}$ that is most often evaluated
with a one-shot procedure,
the violation of electron conservation is well documented
\cite{schindlmayr_prb2001,stan_epl2006,caruso_phd,bruneval_jctc2021}.

In a previous study of ours \cite{bruneval_jctc2021}, it was demonstrated analytically and verified numerically that linearizing the Dyson equation completely cures the problem 
of electron count conservation in the $GW$ approximation.
The linearized Dyson equation (LDE) reads
\begin{equation}
\label{eq:lde}
  G = G_0 + G_0 (\Sigma_{xc} - V_{xc} ) G_0  ,
\end{equation}
where the last $G$ in Eq.~(\ref{eq:dyson}) had been simply replaced by $G_0$.
The LDE is customary in the context of the Sham-Sch{\" u}ter equation \cite{sham_prl1983}.

This electron-conserving equation is then applied with the $GW$ approximation to $\Sigma_{xc}$.

\subsection{$GW$ self-energy based density matrix}

The $GW$ approximation \cite{hedin_pr1965} is simply sketched here,
since it has been the subject of numerous detailed reviews
\cite{hedin_chapter1970,aulbur_review1999,reining_wires2018}.

The screened Coulomb interaction $W$ is defined with the Dyson-like equation:
\begin{equation}
    W = v + v \chi_0 W ,
\end{equation}
where $\chi_0 = -2 \icomp G G$ is the non-interacting polarizability and $v$ is the usual bare
Coulomb interaction.

The $GW$ self-energy then reads 
\begin{equation}
    \Sigma_{xc} = \icomp G W .
\end{equation}

It is convenient to decompose the self-energy into pure exchange and correlation.
The exchange part $\Sigma_x$ is static, whereas the correlation part carries the 
frequency dependence $\Sigma_c(\omega)$.
These quantities are routinely obtained with a one-shot procedure
in standard periodic codes
\cite{kresse_prb1996,gonze_cpc2020}

Some general properties of the density matrix are detailed in App.~\ref{appexA}.
For instance, it is demonstrated that the density matrix can be fully characterized with a single k-point index within the first Brillouin zone,
even though it is a function of two spatial indices.

Now let us focus on the $GW$ density matrix.
It is handy to project into the mean-field orbitals $| \bk i \rangle$,
which form a valid orthogonal basis:
\begin{eqnarray}
\label{eq:gammaij}
  \gamma_{\bk ij} &=& \langle \bk i | \gamma | \bk j \rangle  \nonumber \\
                  &=& \int d\br d\br' \varphi_{\bk i}^*(\br) \gamma(\br,\br') 
                            \varphi_{\bk j}(\br') .
\end{eqnarray}
The first term on the right-hand side of Eq.~(\ref{eq:lde}) is $G_0$.
Let us insert it in Eq.~(\ref{eq:gamma}) and project on the orbital basis to 
obtain the spin-summed density matrix elements
\begin{equation}
  \gamma^\mathrm{gKS}_{\bk ij} = 2 \delta_{i j} \theta(\mu - \epsilon_{\bk i}) .
\end{equation}
This expression has been obtained by closing the contour in the upper part of the complex plane
so that only the poles located above the real axis have survived.

A similar approach technique can be used for the static terms in the right-hand part of 
Eq.~(\ref{eq:lde}), $G_0 (\Sigma_x - V_{xc} ) G_0$: 
\begin{equation}
 \label{eq:dgammahf}
  \Delta \gamma^\mathrm{HF}_{\bk ij}
      = 2 \theta(\mu - \epsilon_{\bk i})  \theta(\epsilon_{\bk j} - \mu)
                 \frac{   \langle \bk i | \Sigma_x - V_{xc} | \bk j\rangle  }
                      { \epsilon_{\bk i} - \epsilon_{\bk j} }  .
\end{equation} 
We denote it with a ``HF'' superscript because this contribution to the (spin-summed) linearized density matrix is obtained from 
a pure exchange self-energy; thus, it vanishes when the HF approximation is employed to obtain the mean-field orbitals
(i.e. $\Delta \gamma^\mathrm{HF}=0$ for $\gamma^\mathrm{gKS}=\gamma^\mathrm{HF} $).

For the last term in Eq.~(\ref{eq:lde}), $G_0 \Sigma_c G_0$, the self-energy
$\Sigma_c$ has a frequency dependence and therefore the calculations cannot be
performed analytically in contrast with the two previous terms, it is 
convenient to perform the integration along the imaginary axis, so as to keep some distance
with the poles of $G_0$ and of $\Sigma_c$.
Closing the contour, we transform the real-axis integration of Eq.~(\ref{eq:gamma})
into
\begin{equation}
 \label{eq:dgammagw}
   \Delta \gamma^{GW}_{\bk ij}
        = \frac{1}{\pi} \int_{-\infty}^{+\infty} d\omega
        \frac{\langle \bk i | \Sigma_c(\mu + \icomp \omega) | \bk j\rangle}
        {(\mu + \icomp \omega-\epsilon_{\bk i}) ( \mu + \icomp \omega-\epsilon_{\bk j})} .
\end{equation}

The complete spin-summed linearized $GW$ density matrix finally reads
\begin{equation}
    \gamma^{GW} = \gamma^\mathrm{gKS} + \Delta \gamma^\mathrm{HF}  + \Delta \gamma^{GW}. 
\end{equation}
Lastly, the corresponding electronic density is $\rho^{GW}(\br) = \gamma^{GW}(\br,\br)$.

\subsection{Total energies from $GW$ density matrix}

In previous studies \cite{bruneval_jctc2019,bruneval_jctc2021},
we introduced a new total energy functional:
\begin{multline}
  \label{eq:etotal_lgw}
  E_\mathrm{total}^{\gamma^{GW}} =
      T[\gamma^{GW}] + V_{ne}[\rho^{GW}] + E_H[\rho^{GW}] \\
         + E_x[\gamma^{GW}] + E_c[G_0] + V_{nn} ,
\end{multline}
where $T$, the kinetic energy, $V_{ne}$, the electron-nucleus interaction, $E_H$,
the Hartree energy, $E_x$ the exchange energy are evaluated with $\gamma^{GW}$
density matrix.
\fabien{Klimes \textit{et. al.} \cite{klimes_jchemphys2015} also used the $GW$ (or RPA) density matrix to improve sub-parts of the energy.}
Just the correlation energy $E_c$ cannot be calculated with $\gamma^{GW}$ and
is pragmatically obtained from the Galitskii-Migdal equation \cite{bruneval_jctc2021}:
\begin{equation}
\label{eq:ecgm}
 E_c[G_0] = \frac{1}{4\pi} \int_{-\infty}^{+\infty} d\omega
   \mathrm{Tr} \{
    v\chi_0(i \omega)
    - v \chi(i \omega) 
     \} ,
\end{equation}
where $\chi = \chi_0 + \chi_0 v \chi$ is the RPA polarizability.

This one-shot energy expression has the desirable property that all the input quantities
conserve the number of electrons.
Of course, being a one-shot total energy, it keeps a dependence with respect to the starting point.
This will be studied in detail in Sec.~\ref{sec:results}.

\subsection{RPA total energy}

For completeness, we report here without derivation the RPA expression for the total energy
as we will extensively compare 
$E_\mathrm{total}^\mathrm{RPA}$
and 
$E_\mathrm{total}^{\gamma^{GW}}$ in the following.

The RPA correlation is defined as \cite{fuchs_prb2002}
\begin{equation}
    \Phi_c[G_0] = \frac{1}{4 \pi}
         \int_{-\infty}^{+\infty} d \omega \,
           \mathrm{Tr} \left\{
                  v\chi_0(i\omega)
                 + \mathrm{ln} \left[ 1 - v \chi_0(i\omega) \right] 
                   \right\}   .
\label{eq:rpa_phi}
\end{equation}

By construction, $\Phi_c[G_0]$ contains the correlation part of the kinetic energy.
The total one-shot energy expression reads
\begin{multline}
  \label{eq:etotal_rpa}
  E_\mathrm{total}^\mathrm{RPA} =
      T[\gamma^\mathrm{gKS}] + V_{ne}[\rho^\mathrm{gKS}] + E_H[\rho^\mathrm{gKS}] \\
         + E_x[\gamma^\mathrm{gKS}] + \Phi_c[G_0] + V_{nn} .
\end{multline}

The one-shot RPA total energy is known to have a noticeable starting point dependence
\cite{dahlen_jcp2004,nguyen_jchemphys2010,angyan_jctc2011,bruneval_jctc2021}.
We will quantify this in Sec.~\ref{sec:results} in comparison with
$E_\mathrm{total}^{\gamma^{GW}}$.

\section{Implementation and computational details}
\label{sec:computation}

The linearized $GW$ density matrix in periodic systems has not been studied
before to the best of our knowledge.
It should be noted though that
the linearized $GW$ density matrix appears as an intermediate quantity
in the RPA forces derived by Ramberger \textit{et al.} \cite{ramberger_prl2018}.

In this section, we provide a detailed description of our implementation of $\gamma^{GW}$ in the ABINIT code \cite{gonze_cpc2020}. We also highlight the key technical aspects that are crucial for producing accurate results.

\subsection{Implementation in a periodic plane-wave approach}

ABINIT is a standard plane-wave-based DFT code.
The core electrons are frozen and hidden in a pseudopotential.
While the Kohn-Sham part of ABINIT is able to use the more accurate and smoother projector augmented-wave atomic datasets \cite{bloechl_prb1994,kresse_prb1999,torrent_cms2008},
the extension to $GW$ is very delicate \cite{arnaud_prb2000,klimes_prb2014}.  As of today,
the $GW$ part of ABINIT is fully validated only for regular norm-conserving pseudopotentials
\cite{kleinman_prl1982,hamann_prb2013}.

The existing implementation in ABINIT provides us with
$\langle \bk i | \Sigma_c(\mu + \icomp \omega) | \bk j\rangle$
for any value of $\omega$.
From this starting point, we have then implemented a Gauss-Legendre quadrature to perform the integral
in Eq.~(\ref{eq:dgammagw}). The symmetry relation
\begin{equation} 
\langle \bk i | \Sigma_c(\mu - \icomp \omega) | \bk j\rangle
    = \langle \bk j | \Sigma_c(\mu + \icomp \omega) | \bk i \rangle^*
\end{equation}
is employed to limit the integration from 0 to $+\infty$.
A grid with typically 50 to 120 grid points is sufficient to ensure a very accurate convergence: we monitor the electron count deviation, which is always kept below
$10^{-3}$.
In the future, grid design could be optimized to minimize the computational burden
\cite{caruso_phd,kaltak_prb2020}.

The static term from Eq.~(\ref{eq:dgammahf}) has been implemented as well, for any
type of exchange-correlation potential $V_{xc}$, including those based on hybrid functionals.
Note that for a Hartree-Fock mean-field starting point, the static term $\Sigma_x - V_{xc}$
vanishes.
Furthermore, it is clear from Eq.~(\ref{eq:dgammahf}) that the linearized density matrix is limited to
systems with a finite band gap, or else diverging denominators would occur.

The matrix representation of $\gamma^{GW}$ is obtained on the gKS states
$| \bk i \rangle$ for $i \leq N_b$.
We then diagonalize it to obtain the natural orbitals in the gKS basis:
\begin{equation}
  \sum_{j=1}^{N_b} \gamma_{\bk i j} U_{\bk j \lambda} = n_{\bk \lambda}  U_{\bk i \lambda} ,
\end{equation}
where $n_{\bk \lambda}$, the eigenvalues, are the so-called natural occupations
and where $U_{\bk j \lambda}$, the eigenvector coefficients, form the natural orbitals.

In other words, the natural orbitals $\phi_{\bk \lambda}(\br)
= \langle \br | \bk \lambda \rangle$ 
can be obtained
from the unitary matrix $U_{\bk i \lambda}$:
\begin{equation}
  \phi_{\bk \lambda}(\br) = \sum_{i=1}^{N_b} U_{\bk i \lambda} \varphi_{\bk i} (\br)  .
\end{equation}

Then all the one-body operators expectation values are readily obtained.
For instance, the kinetic energy $T$ is calculated as
\begin{align}
 T &= \frac{1}{N_k} \sum_{\bk \lambda} n_{\bk \lambda}
        \langle \bk \lambda | \hat{T} |   \bk \lambda \rangle \\
   &= \frac{1}{N_k} \sum_{\bk \lambda} n_{\bk \lambda} \sum_{ij} U_{\bk i \lambda}^* U_{\bk j \lambda}
              \langle \bk i | \hat{T} |   \bk j \rangle .
\end{align}

Finally, we would like to emphasize that the formal proof of the conservation of electron count
\cite{bruneval_jctc2021}
requires that the state range in the internal sum of $G_0$
in Eq.~(\ref{eq:G0}) is the same as the one used in the basis expansion in Eq.~(\ref{eq:gammaij}).
This restriction is enforced in all our calculations.

Table~\ref{tab:parameters} summarizes the numerical parameters used for the 8
crystals considered in this study.
All the calculations for face-centered cubic crystals
reported in this work use four shifted k-point grids, as commonly used in ABINIT.
The grid discretization is $4 \times 4 \times 4$, which yields 256 k-points
in the full Brillouin zone and 10 in the irreducible wedge.
The calculation on hexagonal boron nitride (h-BN) uses a $\Gamma$-centered  $12 \times 12 \times 6$ k-point grid for exact exchange and a $4 \times 4 \times 2$ grid for the rest.
\fabien{We evaluate the computational effort to scale as $O(N^4)$, similar to a conventional 
$GW$ calculation. However, the prefactor is much larger, due to the fine frequency grids for both $W$ and $\Sigma$.}

\begin{table}[b]
  \caption{
    \label{tab:parameters}
    Energy cutoff for the wavefunctions ($E_{cut}^\varphi$, in Hartree),
    energy cutoff for the plane-waves expansion of the screened Coulomb interaction $W$ ($E_{cut}^W$, in Hartree),
    number of bands ($N_{b}$) in $G_0$ and in Eqs.~(\ref{eq:dgammahf},\ref{eq:dgammagw}),
    number of pure imaginary frequencies ($N_{\omega_W}$) used for $W(\icomp \omega)$,
    and number of frequency points ($N_{\omega_\Sigma}$), used for the 
    quadrature along the imaginary axis in Eq.~(\ref{eq:dgammagw}).
  }
  \begin{ruledtabular}
  \begin{tabular}{lccccccccc}
        & Si   & C   & SiC  & zb-BN  & AlP  & AlAs  & Ge & h-BN      \\
    \hline
$E_{cut}^\varphi$ &  12  & 25 & 50  & 40 & 35  & 35 & 30 &  55    \\ 
$E_{cut}^W$ &   8 & 15   & 15 & 15  & 15 & 10  & 10 & 25      \\ 
$N_{b}$&  150 & 175  & 175 & 175 & 175  & 175  & 175 & 1400  \\ 
$N_{\omega_W}$& 120  & 120  & 120 & 120 & 120  & 120  & 120 & 40  \\ 
$N_{\omega_\Sigma}$& 60  & 60  & 60 & 60 & 60  & 60  & 60 & ---  \\ 
  \end{tabular}
  \end{ruledtabular}
\end{table}

\subsection{Adequate norm-conserving pseudopotentials}

\begin{figure}[t]
  \includegraphics[width=0.99\columnwidth]{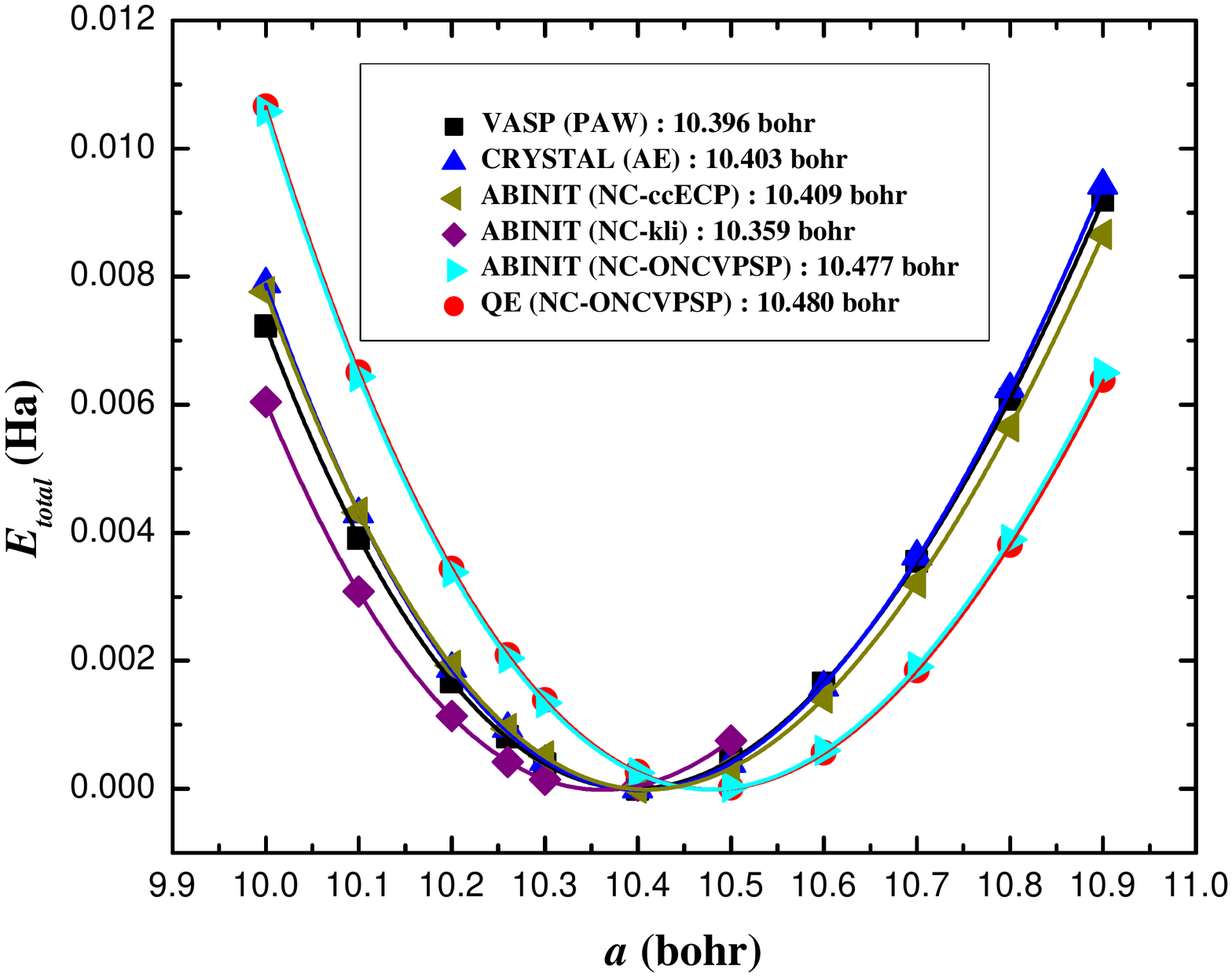}
      \caption{HF total energies as a function of lattice constant with different codes and pseudopotentials for bulk silicon.
      The equilibrium lattice constant for each calculation is given in the legend.
    \label{fig:sihf}
  }
\end{figure}

As mentioned in the previous paragraph,
our implementation uses norm-conserving pseudopotentials.
In the preliminary stages of our study, we concluded that 
while the details of the pseudopotential are not critical when studying band structures, they become of the utmost importance when investigating structural properties.

Norm-conserving pseudopotentials are designed to reproduce the electronic
and energetic properties of a given mean-field approximation.
For instance, using a PBE pseudopotential for a hybrid functional is not advised in principle.
As no ``$GW$-suitable'' pseudopotentials exist, we have enforced the minimal requirement that the selected pseudopotential be able to reproduce HF structural
properties.

In Fig.~\ref{fig:sihf}, we show a wide comparison among codes and techniques for bulk
silicon at the HF level of theory. Silicon is chosen
as a typical example.
The results for two other crystals are reported in the supporting information with
identical conclusions.
The all-electron (AE) of CRYSTAL\cite{dovesi_jchemphys2020}
with the accurate basis set designed by Heyd and coworkers~\cite{heyd_jchemphys2005}
and the projector augmented-wave (PAW) results of VASP~\cite{kresse_prb1996}
agree very well.
We consider them as the reference,
since by construction, the Gaussian basis set used in CRYSTAL describes all the electrons at once
and since in the PAW framework, though frozen-core, the 
core-valence interactions are completely recalculated for each approximation.

Then we turn to the regular PBE pseudopotential obtained from
the pseudodojo suite \cite{vansetten_cpc2018}.
This pseudopotential is highly tested and should be rather transferable as it relies on Hamann's ONCVPSP scheme \cite{hamann_prb2013} that introduce
several projectors per angular momentum.
However, based on Fig.~\ref{fig:sihf}, the HF energy-volume curve 
departs significantly from the reference.
This error is intrinsic to the pseudopotential because using it in Quantum-Espresso \cite{giannozzi_jpcm2009} gives the exact same result.
\fabien{In our opinion, the inability of the ONCVPSP pseudopotentials to reproduce HF
energy-volume curves is not due to the Hamann's scheme itself, but rather
due to the practical choice of large cutoff radii selected in the pseudodojo initiative.
Generating our own dedicated ONCVPSP pseudopotentials would be possible of course, however, would require a significant effort. Fortunately, alternatives already exist.}

In a previous work \cite{bruneval_prl2012}, one of us mitigated this problem
by using pseudopotentials generated for KLI\cite{krieger_pla1990}
which devises a local potential that simulates the non-local exact-exchange operator. This improves over the ONCVPSP pseudopotentials but is not quantitative enough.

Recently, in the quantum Monte Carlo community, there has been an effort to 
support the design of ``correlation consistent'' effective core potentials (ccECP)
\cite{bennett_jchemphys2017}.
These norm-conserving pseudopotentials are meant to be used in combination with correlated methods beyond the usual mean-field ones.
Figure~\ref{fig:sihf} shows that this type of pseudopotential produces results in close agreement with AE and PAW for HF: the lattice constants match within 0.1~\%.

We conclude that the ccECP pseudopotentials are our preferred norm-conserving pseudopotentials
to obtain quantitative results:
i) they have been designed specifically for explicit-correlation methods and $GW$ belongs to this family;
ii) they are best to reproduce HF lattice constants.
The main drawback of these pseudopotentials is the high cutoff energy that is necessary
to converge the total energies in plane waves.
In the following, all the reported results employ ccECP pseudopotentials.

\section{$GW$-density matrix in crystalline solids}
\label{sec:sigwdm}

\begin{figure}[t]
  \includegraphics[width=0.99\columnwidth]{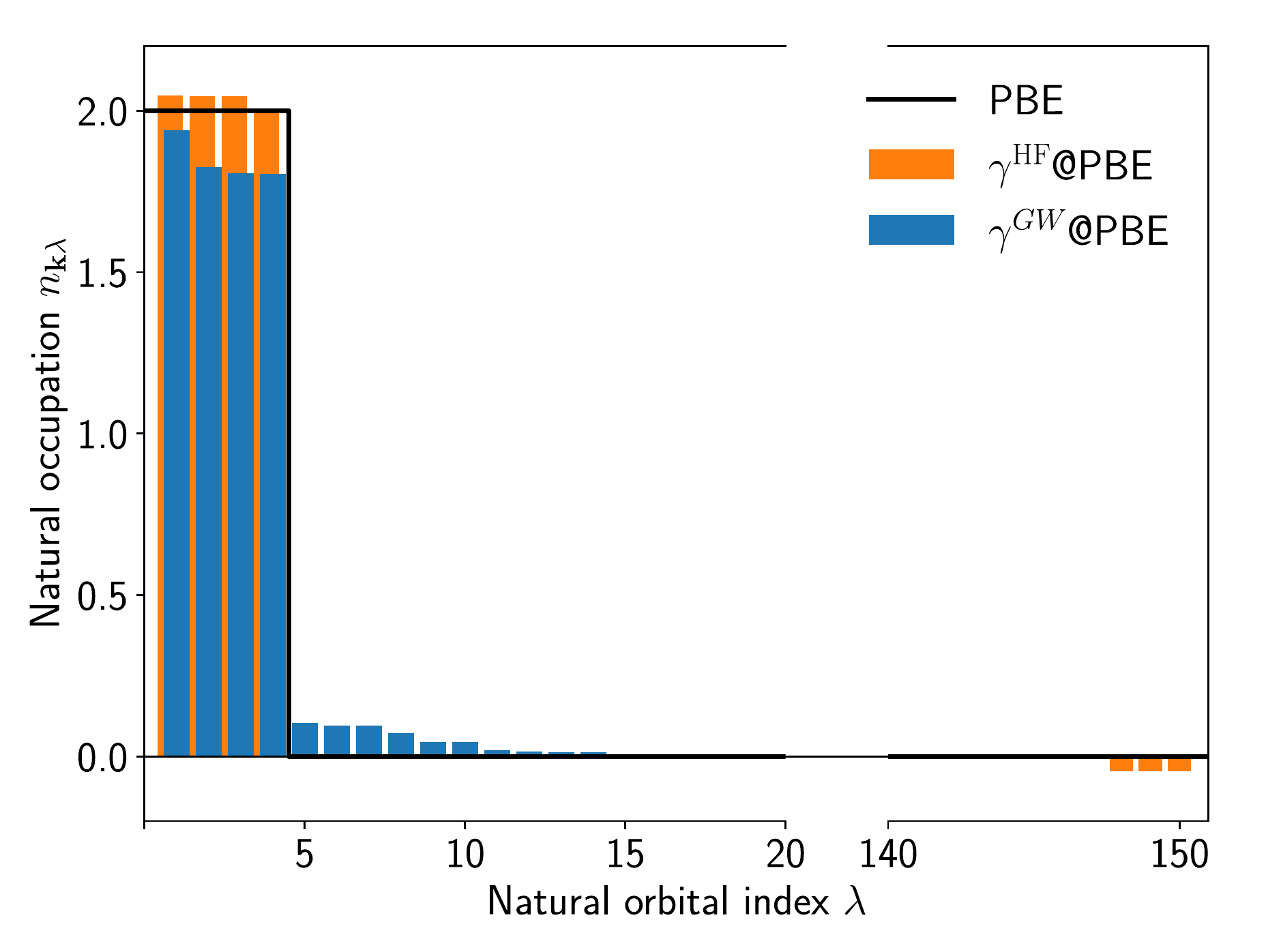}
    \caption{Spin-summed natural occupations obtained from PBE, $\gamma^\mathrm{HF}$@PBE, and
    $\gamma^{GW}$@PBE density-matrices
    in bulk silicon for k-point (1/8, 0, 0) in reciprocal lattice vectors.
    The natural occupations are ordered by descending values, from 1 to 150,
    which is $N_b$, the dimension of the matrix.
    The x-axis has been cut to show the first and the last values.
    \label{fig:si_occ}
    }
\end{figure}

As summarized in Appendix~\ref{appexA}, the spin-summed natural occupations $n_{\bk \lambda}$
should continuously span the range from 0 to 2 at variance with regular
Fermi-Dirac ground-state occupations $f_{\bk i}$ that are only 0 or 2.

These natural occupations for realistic crystalline solids can be 
compared to the 
momentum distribution function $n_{\bk}$
for the homogeneous electron gas in Refs.~\cite{hedin_chapter1970,olevano_prb2012}.
But for the  homogeneous electron gas, the momentum $\bk$ is enough
to uniquely characterize the quantum state.
For solids, we need an additional quantum number $\lambda$ (similar to a band index).

In Fig.~\ref{fig:si_occ}, we represent the natural occupations $n_{\bk \lambda}$ for
a fixed k-point (1/8, 0, 0). This particular k-point was selected as an example:
the other k-points produce very similar results.
The PBE occupations $f_{\bk i}$ are shown as a reference.
Then the natural occupations for $\gamma^\mathrm{HF}$@PBE, the static part of the
density matrix, are plotted.
While their sum precisely equalizes the number of electrons $N_e$, the values
can exceed 2 and be below 0.
These occupation values violate the constraints of the exact density matrix.
However, after adding the dynamic correlation, the $\gamma^{GW}$@PBE has all its spin-summed natural occupations between 0 and 2.
Four natural orbitals have an occupation close to 1.8-1.9 and then
many more (15 or so) have a non-vanishing occupation.
\fabien{A PBE mean-field starting point was chosen to magnify the effect.
Starting from HF would yield perfectly sane natural occupations for
$\gamma^\mathrm{HF}$}
The overall shape of the occupation is similar to the homogeneous electron
gas result \cite{hedin_chapter1970,olevano_prb2012}.

However one notices that $N(\bk) = \sum_{\lambda} n_{\bk \lambda}$ for a given
$\bk$ slightly deviates from the number of electrons $N_e$.
This intriguing observation does not violate an exact constraint.
We only proved mathematically \cite{bruneval_jctc2021}
that the sum of the natural occupation over the whole Brillouin zone
\begin{equation}
  \sum_\bk \sum_\lambda n_{\bk \lambda} = N_e
\end{equation}
is valid.
Appendix~\ref{appexB} demonstrates that this variation with $\bk$ is possible.

\begin{figure}[t]
  \includegraphics[width=0.99\columnwidth]{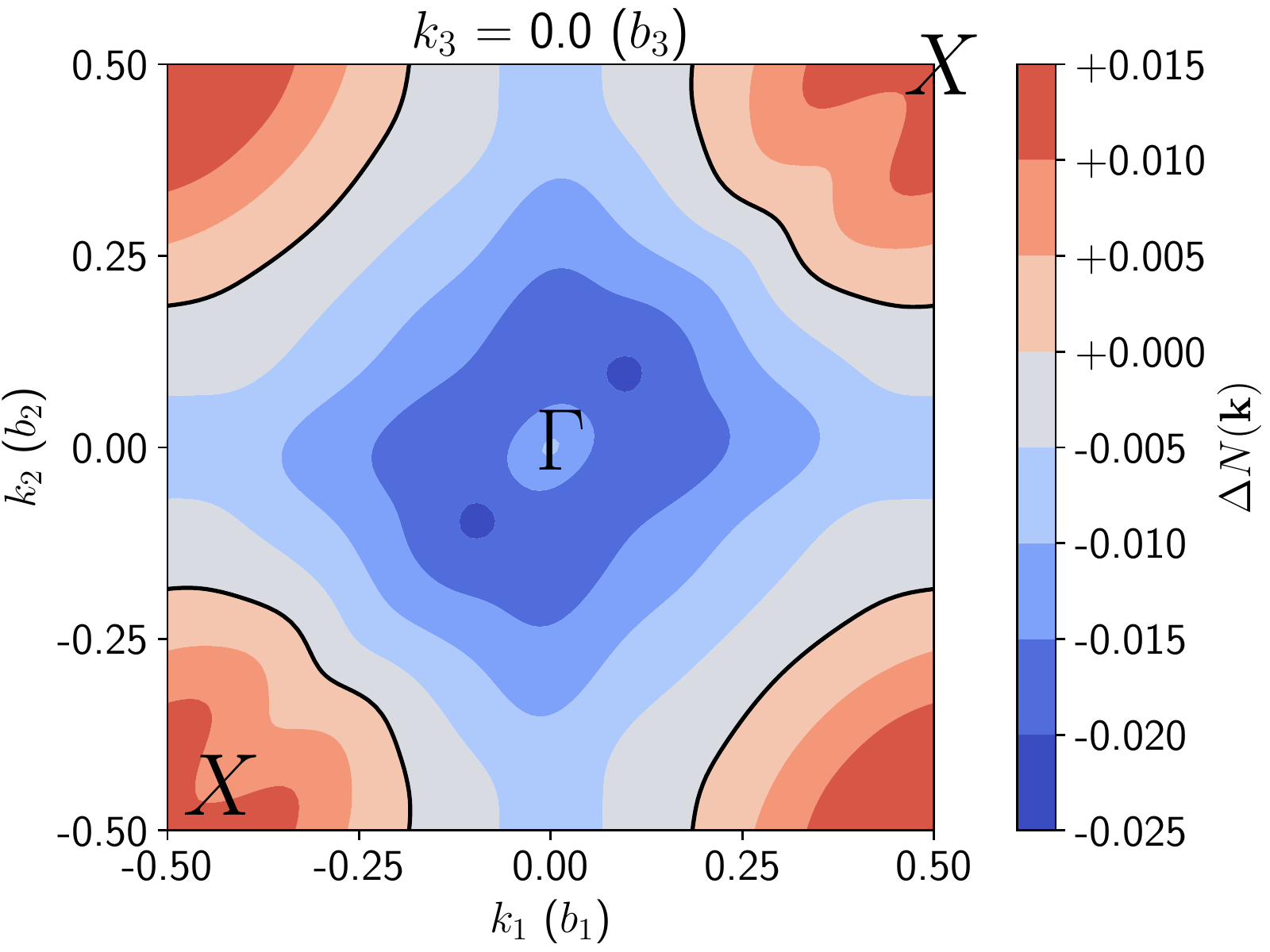}
    \caption{Electron count deviation $\Delta N(\bk)$ as a function of the 
    wavevector $\bk$ in the first Brillouin zone for bulk silicon.
    $\bk = k_1 \mathbf{b}_1 + k_2 \mathbf{b}_2 + k_3 \mathbf{b}_3$ is reported in
    reduced coordinates ($\mathbf{b}_i$ are the reciprocal lattice vectors).
    The plane $k_3=0$ is represented.
    The isoline $\Delta N(\bk)=0$ is drawn with a bold black line.
    Special points $\Gamma$ and $X$ are marked.
    \label{fig:si_nk}
    }
\end{figure}

As this observation can be considered surprising when compared to the usual
mean-field occupations $f_{\bk i}$,
it is insightful to monitor the sum $N(\bk)$ across the Brillouin zone.
In Fig.~\ref{fig:si_nk}, we report the deviation $\Delta N(\bk) = N(\bk) - N_e$
in a cut plane in the Brillouin zone.
The $\Delta N(\bk)$ function is interpolated from a refined $\Gamma$-centered
$6\times 6 \times 6$ with 4 shifts k-point grid (864 points in the full Brillouin zone).
We use the Shankland-Koelling-Wood interpolation technique
\cite{koelling_ijqc1986} as implemented in \texttt{abipy} \cite{gonze_cpc2020}.
The numerical integration of $\Delta N(\bk)$ over the whole Brillouin
yields $4 \times 10^{-4}$, which is very close to the expected zero.

From Fig.~\ref{fig:si_nk},
we observe an electron transfer from the $\Gamma$~point region
to the Brillouin zone edge.
The weight transfer is not large (at most $\sim 0.01-0.02$), but still sizable.
The electron count $N(\bk)$ is an observable and could be possibly measured in angle-resolved photo-emission spectroscopy \cite{damascelli_rmp2003}.
Note that this electron count transfer is a pure electronic correlation effect.
Any static approximation of the self-energy $\Sigma$ would nullify it.

\section{Structural properties of crystalline solids within $GW$ density matrix and RPA}
\label{sec:results}

\subsection{Covalent-bonded crystals}

\begin{figure*}
   \includegraphics[width=0.99\columnwidth]{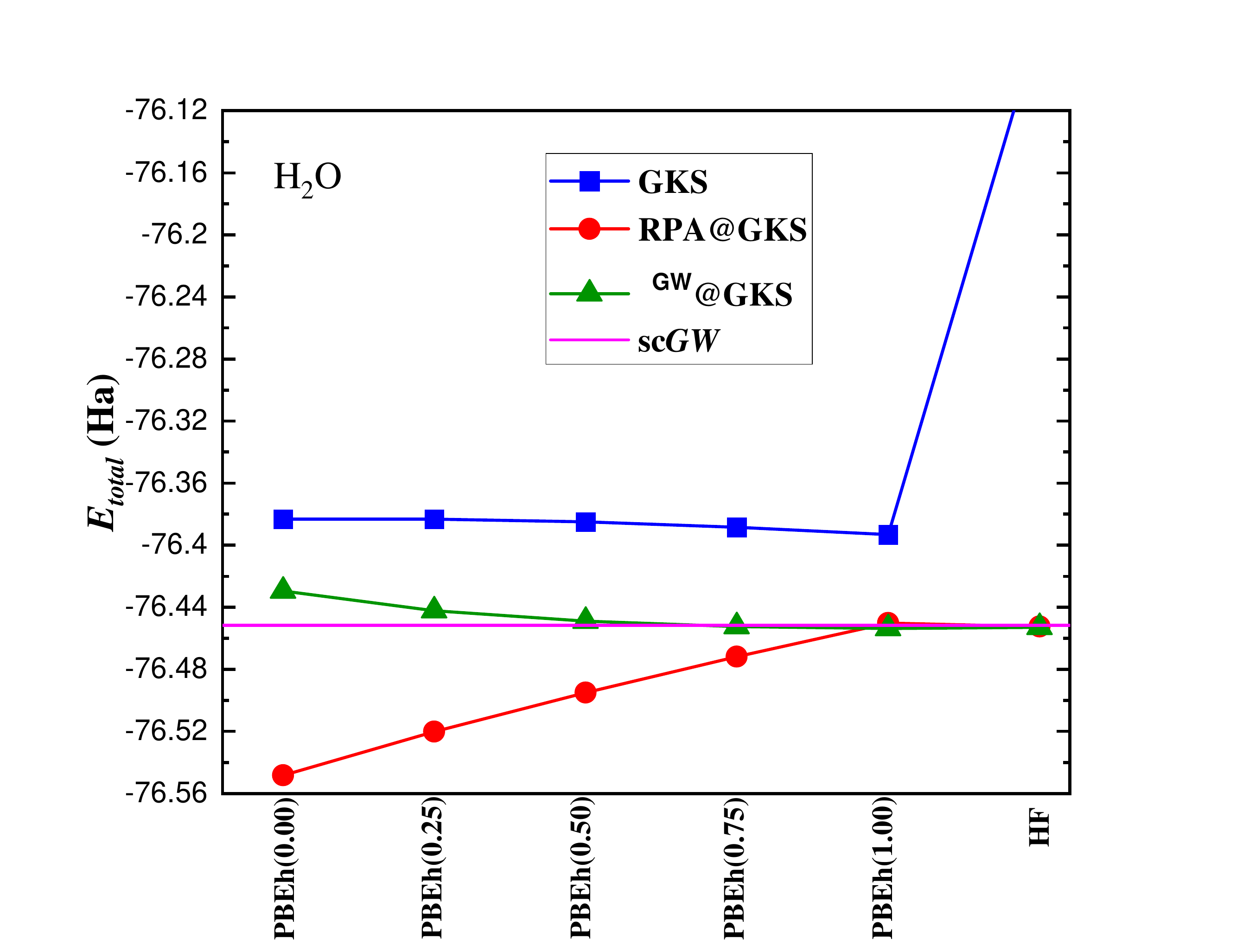}
   \includegraphics[width=0.99\columnwidth]{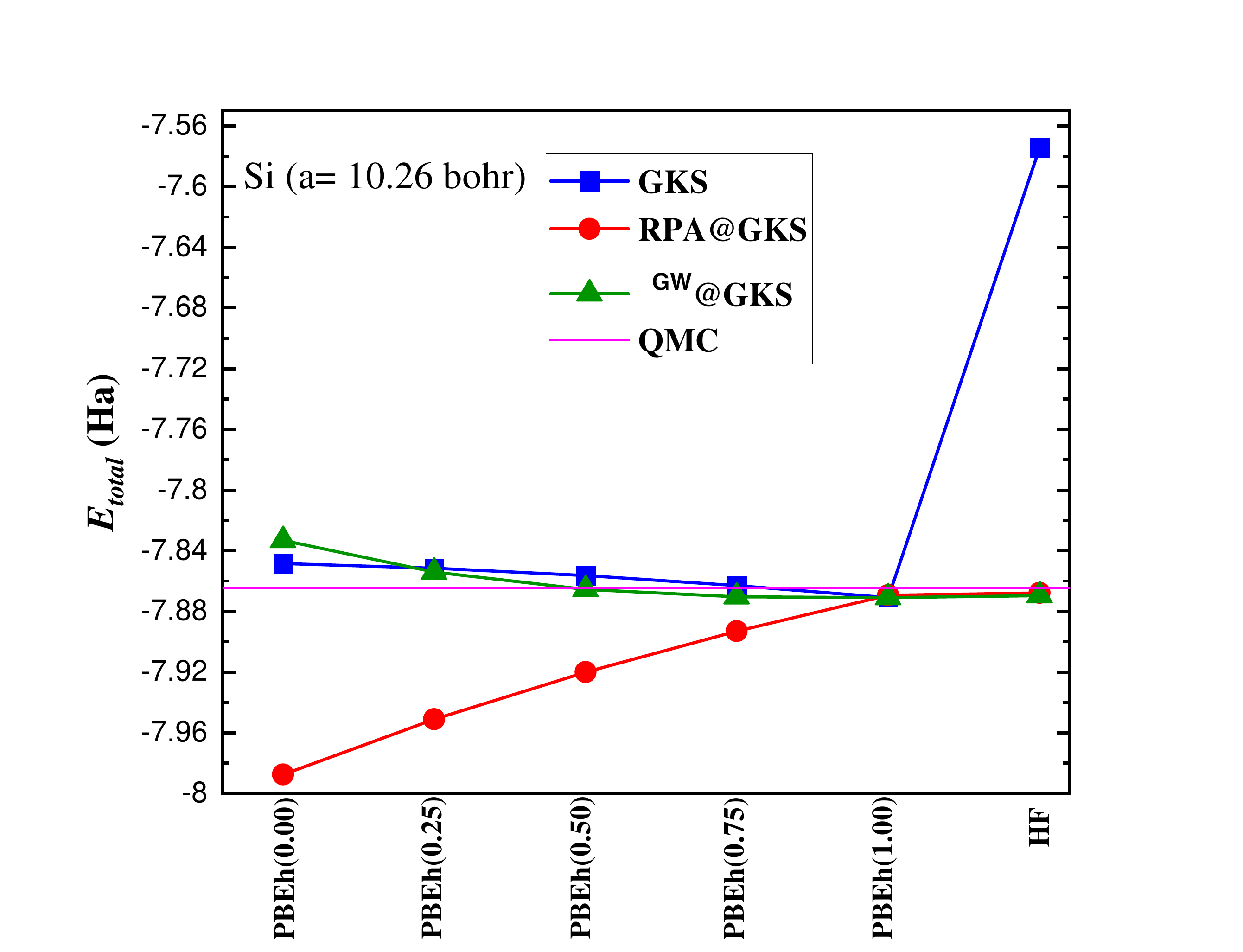}
    \caption{
      Total energies for the water molecule in the gas phase (left-hand panel)
      and crystalline silicon (right-hand panel) as a function of the gKS starting point.
      The water results were extracted from a previous study \cite{bruneval_jctc2021}.
      The silicon Quantum Monte Carlo (QMC) value comes from Ref.~\onlinecite{annaberdiyev_prb2021}.
      \label{fig:h2o_si}
    }
\end{figure*}

In this section, we analyze the calculation of structural parameters for crystalline solids
using the total energy expressions introduced 
in Eqs.~(\ref{eq:etotal_lgw}) and (\ref{eq:etotal_rpa}).
\fabien{
Our main question is which expression works ``best''
in the context of one-shot calculations, i.e. which expression best approximates a hypothetical reference sc$GW$ that is not currently available for crystalline systems.}
In molecules, where reference sc$GW$ were produced \cite{bruneval_jctc2021},
the accuracy of the $\gamma^{GW}$ total energy was demonstrated.

A way to measure the robustness of a one-shot total energy expression is to
explore its sensitivity to the starting point.
Here we use the PBEh($\alpha$) hybrid functional family:
\begin{equation}
  V_{xc} = \alpha \Sigma_x + (1-\alpha) v_x^\mathrm{PBE} + v_c^\mathrm{PBE} ,
\end{equation}
where the parameter $\alpha$ controls the amount of exact-exchange $\Sigma_x$.

Calculations were carried out for seven covalent
crystals (Si, C, SiC, zb-BN, AlP, AlAs, and Ge). In the main text, we
will mostly report silicon results. However, the complete set of results is made available as Supplemental Material \cite{supplmat}.

In Fig.~\ref{fig:h2o_si}, we compare the total energy behavior for 2 different
systems: water, a small molecular system, and crystalline silicon.
The results for the water molecule were extracted from Ref.~\onlinecite{bruneval_jctc2021}
that was using a different implementation based on Gaussian basis \cite{bruneval_cpc2016}.
The figure reports the total energies for PBEh($\alpha$), for
$E_\mathrm{total}^{\gamma^{GW}}$, for $E_\mathrm{total}^\mathrm{RPA}$,
and when available for sc$GW$.
The overall similarity between the two panels is striking:
RPA is rather sensitive to the starting gKS, whereas $\gamma^{GW}$ is much less so.
RPA increases with $\alpha$, whereas $\gamma^{GW}$ decreases.
RPA and $\gamma^{GW}$ rejoin for large values of $\alpha$.

For water, where the sc$GW$ reference exists, the RPA and $\gamma^{GW}$ total energies give the best approximation of the full sc$GW$ total energy when they are equal.
Owing to the similarity between the two panels of Fig.~\ref{fig:h2o_si},
we can reasonably anticipate that in bulk silicon,
the sc$GW$ total energy will be best approximated by
$\gamma^{GW}$ and by RPA with PBEh(0.75), PBEh(1.00) or HF,
however with no formal proof.

If supplied with the sc$GW$ Green's function, all total energy formulas should
match \cite{almbladh_ijmpb1999,dahlen_jcp2004,hellgren_prb2015}.
The above results tend to make us think that the non-interacting Green's function $G_0$
for PBEh(0.75), PBEh(1.00), or HF are close to the sc$GW$ Green's function $G$ for
both the molecular and the solid-state systems.

The striking difference between the molecule and the solid in
Fig.~\ref{fig:h2o_si}
is the agreement or the disagreement with respect to the gKS total energies.
While for the molecule, the RPA and the $\gamma^{GW}$ were undershooting much
the total energy with respect to PBEh($\alpha$) (too negative correlation energies),
the match is very good for crystalline silicon.
The accurate quantum Monte Carlo approach reports -7.8644~Ha for silicon
\cite{annaberdiyev_prb2021},
whereas $\gamma^{GW}$ and RPA respectively give -7.8708 and -7.8692~Ha
with the same ccECP pseudopotentials.
This excellent agreement of the absolute total energies reminds us
about the amazingly accurate sc$GW$ energies in the
homogeneous electron gas \cite{holm_prl1999,garcia_prb2001,kutepov_prb2009} that
precisely match the quantum Monte Carlo values \cite{ceperley_prl1980}.

We conclude with some reasonable confidence that the sc$GW$ total energy in bulk silicon
is certainly close to the gKS, close to $E_\mathrm{total}^{\gamma^{GW}}$, and
close to RPA@PBEh(1.00).
We also stress that RPA@PBE, which is the most commonly accepted implementation
of RPA functional, underestimates noticeably the total energy.
Our confidence in these results is further strengthened when
considering the results for the other 6 crystalline systems reported 
in the Supplemental Material \cite{supplmat}.
They all support the same conclusion.

\fabien{However if sc$GW$ is very accurate for solids and less accurate for finite systems,
the atomization energies that measure the energy gain
when forming a bulk crystal as compared to the isolated atoms are likely to
have a low accuracy in sc$GW$. This should be explored in the future.}

\begin{figure*}[t]
  \includegraphics[width=0.68\columnwidth]{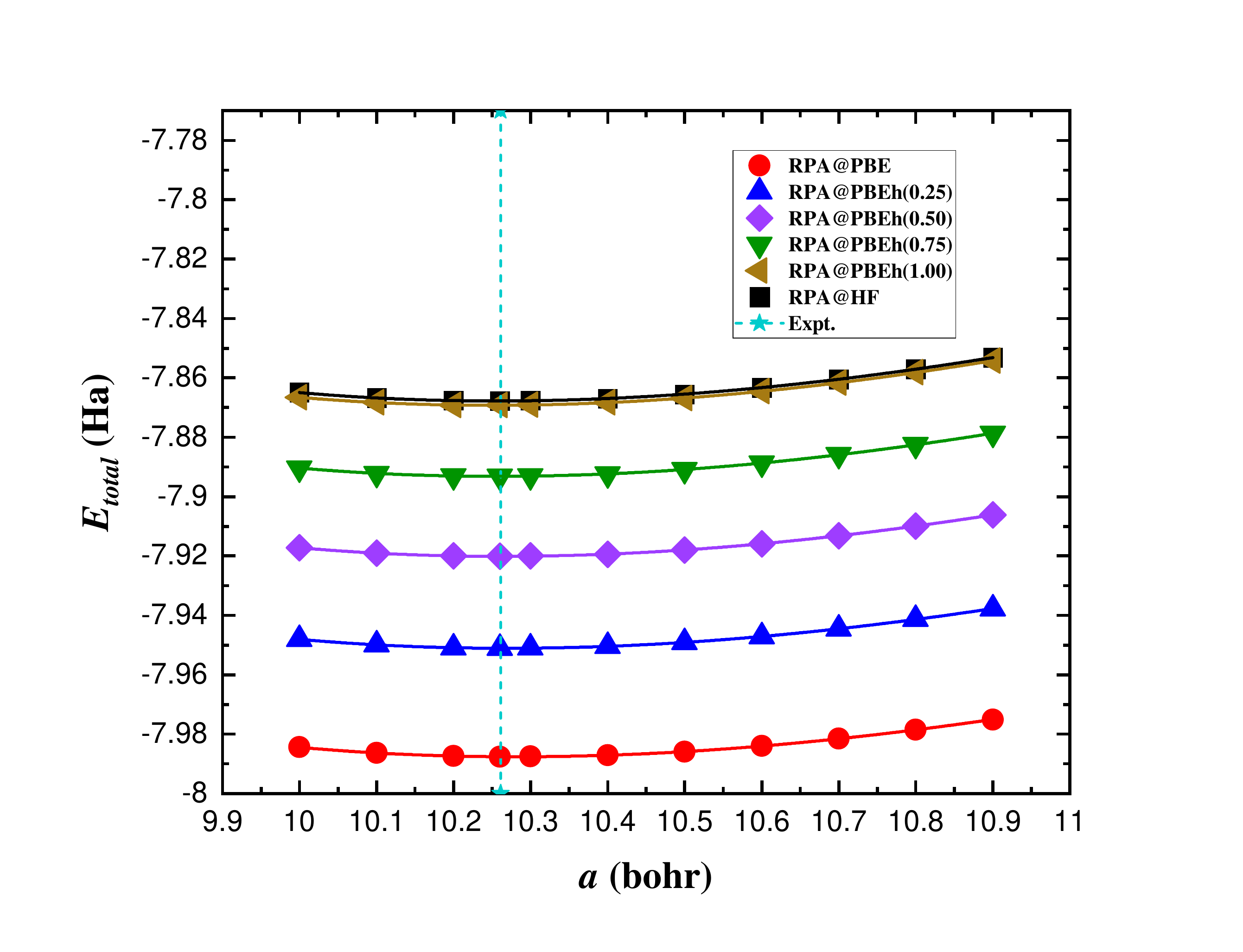}
  \includegraphics[width=0.68\columnwidth]{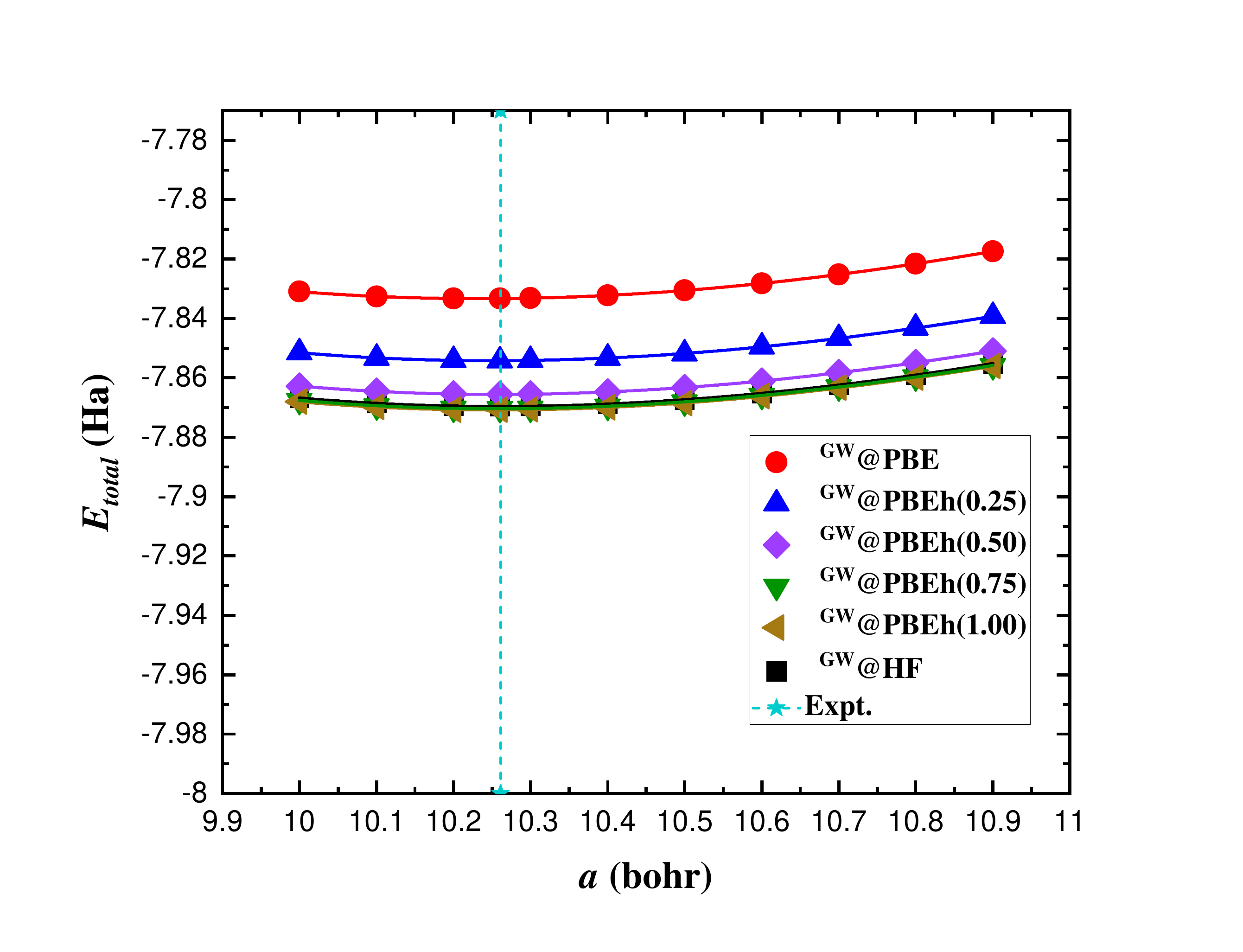}
  \includegraphics[width=0.68\columnwidth]{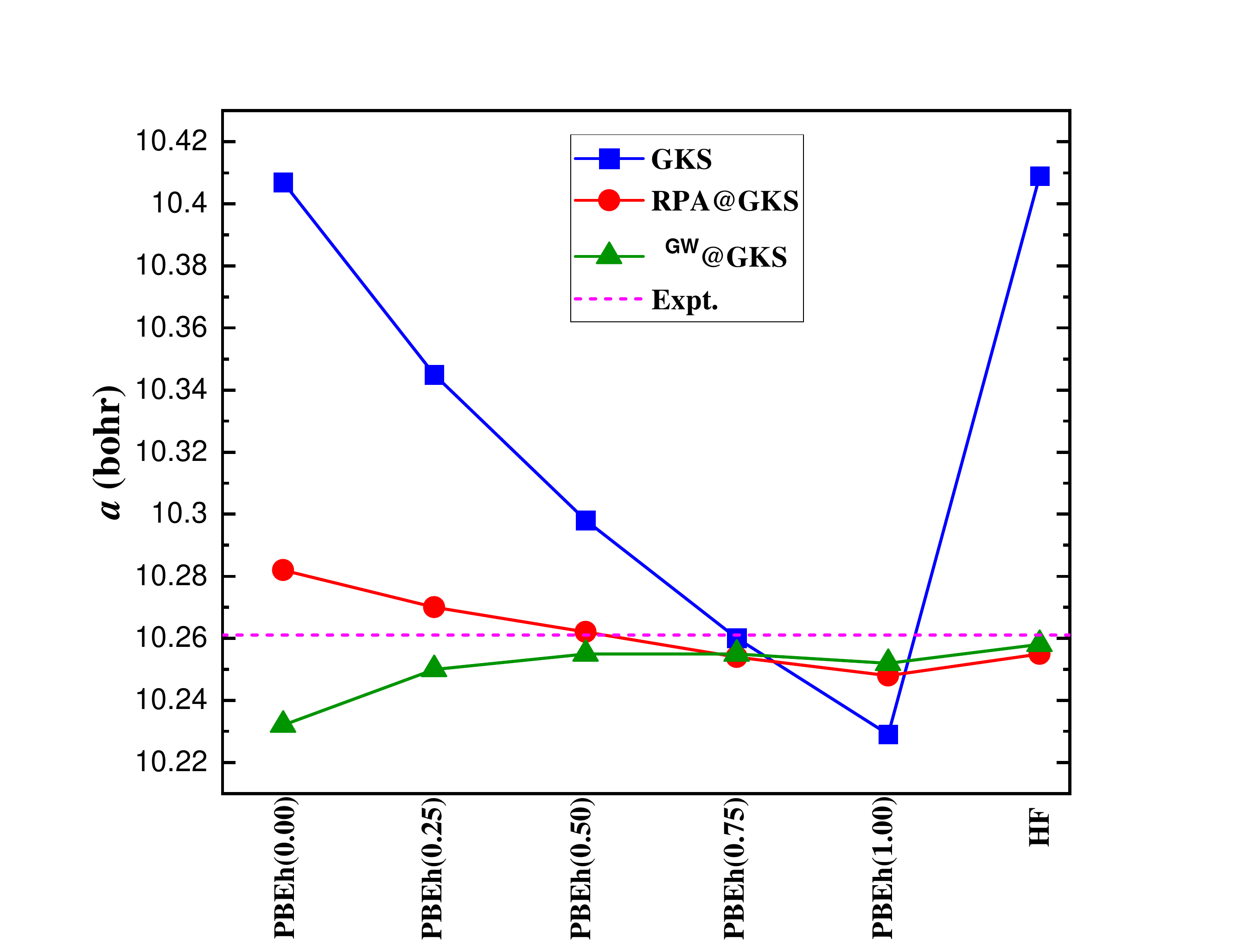}
  \caption{Energy-lattice constant curves for crystalline silicon for
    RPA functional (left-hand panel), $\gamma^{GW}$ energy functional (central panel).
    Equilibrium lattices are summarized in the right-hand panel.
    The experimental lattice constant is given as a reference.
    \label{fig:sigw}
  }
\end{figure*}

Let us now focus on the complete energy versus lattice constant curves and check the sensitivity to the starting point not only for the total energy but also for the equilibrium lattice constant $a$
and for the bulk modulus $B$.

Figure~\ref{fig:sigw} reports on the same scale the RPA and the $\gamma^{GW}$ 
total energies for different gKS starting points and in the right-hand panel
the equilibrium lattice constant as a function of the gKS starting point.
We can see again that the RPA total energy in the left-hand panel of Fig.~\ref{fig:sigw}
is much more sensitive to the starting point compared to $\gamma^{GW}$ total energy in the central panel.
However, when focusing on the equilibrium lattice constant itself,
the sensitivity to the starting point is much weaker: at most 0.03~bohr.

All panels in Fig.~\ref{fig:sigw} support again the same conclusion:
the RPA and $\gamma^{GW}$ total energies agree best when using PBEh($\alpha$) with large $\alpha$ or even 
when using HF.
The statements drawn for silicon perfectly hold for the other 6 crystalline
systems presented in the Supplemental Material \cite{supplmat}.

\begin{table*}[b]
  \caption{
    \label{tab:structure}
    Lattice constants $a$ (bohr), bulk moduli $B$ (GPa), mean absolute error (MAE), and mean absolute percentage error (MAPE) of 7 covalent crystals obtained with
    several total energy methods.
  }
  \begin{ruledtabular}  
  \begin{tabular}{lcccccccccc}
    &    & Si   & C   & SiC  & zb-BN  & AlP  & AlAs  & Ge & MAE & MAPE     \\ \hline

\multirow{2}{2.7cm}{PBEh(0.00)}    &  $a$  & 10.407   & 6.774   &  8.310 & 6.861  &  10.471 &  10.847 & 10.947&0.101 & 1.00      \\ 
                         &  $B$ &  88.44  & 441.76   &  216.68 & 380.18  & 82.72  & 69.98  &   62.24 & 21.961 & 10.5   \\ \hline
 \multirow{2}{2.7cm}{RPA@PBEh(0.00)}   &  $a$  &  10.282  &  6.815  & 8.257  & 6.867  & 10.308  & 10.725  &   10.769 & 0.011 & 0.12  \\
                             &  $B$  & 97.08   &  477.44  &  235.25 & 401.35  & 96.78  & 83.57  &  86.10 & 12.123& 8.7   \\ \hline
 \multirow{2}{2.7cm}{$\gamma^{GW}$@PBEh(0.00)}   &  $a$ &  10.232  &  6.713  &  8.180 &  6.800 & 10.287  & 10.668   &   10.708 & 0.011 & 0.12   \\
                               &   $B$ &  104.58  &  420.75  & 262.57  &  417.06 & 101.56  & 87.73  &   85.70 &11.451 & 8.3   \\
   \hline

\multirow{2}{2.7cm}{PBEh(0.75)}    &  $a$  &  10.260  &  6.675  & 8.170  & 6.728  &  10.322 & 10.680  & 10.682 &0.036 &  0.49   \\ 
                         &  $B$ &  111.93  & 502.28   & 270.10  & 461.11  & 102.44  & 88.02  &  87.30 &  31.026 &  15.7   \\ \hline
 \multirow{2}{2.7cm}{RPA@PBEh(0.75)}   &  $a$  & 10.254   &  6.759  & 8.210  & 6.831  & 10.301  & 10.683  &   10.738 & 0.012 & 0.13   \\
                             &  $B$  & 103.37   &  438.89  & 252.60  &  408.86 & 99.96  & 85.94  &    81.01 & 10.407 & 7.8  \\ \hline
 \multirow{2}{2.7cm}{$\gamma^{GW}$@PBEh(0.75)}   &  $a$ &   10.255 &  6.748  & 8.207  & 6.825  & 10.301  & 10.681  &  10.732 & 0.010 & 0.11    \\
                               &   $B$ &  103.91  &  436.08  & 254.39  & 410.23  & 100.22  & 86.59  &  82.07 & 11.619 & 8.4     \\\hline

   \multirow{2}{2.7cm}{Expt.}    &  $a$  &  10.261  & 6.741   &  8.213 & 6.833  &  10.301 &  10.675 & 10.692 & --- & ---    \\ 
                         &  $B$ &  99  &  443  & 225  &  369-400 & 86  & 77  & 76 & --- & ---      \\ 
   
  \end{tabular}
  \end{ruledtabular}
\end{table*}

Finally, we summarize the lattice constants and bulk moduli of the 7 covalent crystals
that we have studied in Table~\ref{tab:structure}.
We focus on two gKS starting points PBEh(0.00) (i.e. standard PBE)
and PBEh(0.75).
While for PBE starting point the RPA@PBE and $\gamma^{GW}$@PBE lattice
constants differ by about 0.06~bohr,
 the RPA@PBEh(0.75) and $\gamma^{GW}$@PBEh(0.75) lattice
constants always agree within 0.01~bohr.
The same type of conclusion holds for the bulk modulus $B$.
This is another proof that PBEh(0.75) is a good starting point
to evaluate the different $GW$-based total energy expressions.
Most probably, the properties $a$ and $B$ evaluated with RPA@PBEh(0.75)
or $\gamma^{GW}$@PBEh(0.75) are reliable estimates to the sc$GW$ result.

In the end, we also compare to experiment.
Table~\ref{tab:structure} shows that all the $GW$-based energy expressions
yield structural properties in excellent agreement with respect to the experiment:
a 0.1~\% deviation for lattice constants and 8~\% for bulk moduli.
The different expressions and starting points have a minor influence on this.
This conclusion is valid for the covalent crystals.
However, it is worth considering whether this conclusion still holds for weak van der Waals interactions, which are one of the attractive features of RPA.

\subsection{van der Waals bonded layered material}

\begin{figure}[t]
  \includegraphics[width=0.99\columnwidth]{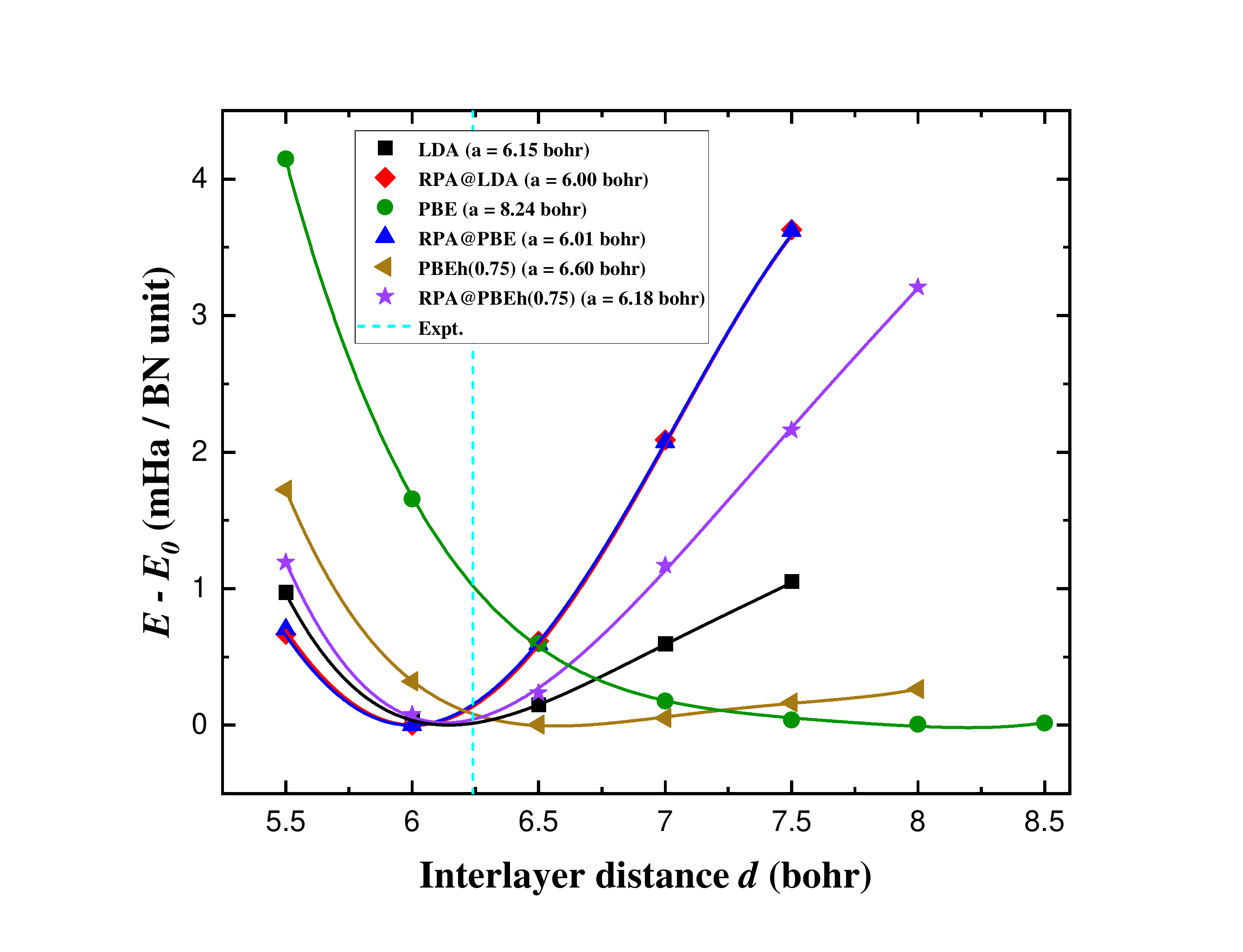}
    \caption{Energy as a function of the spacing between the layers for hexagonal BN
    with different gKS approximations and different starting points for RPA.
    \label{fig:hbn}
  }
\end{figure}

In order to test if the weak van der Waals interactions would correctly be described
by sc$GW$, we analyzed a layered material, namely the hexagonal BN (h-BN).
RPA@LDA or RPA@PBE were proven to be able to describe properly the spacing
between the layers \cite{marini_prl2006,bruneval_prl2012}.

The task of evaluating the $\gamma^{GW}$ for h-BN is beyond our current
computational capabilities.
Indeed, for the weak van der Waals interactions, the energy scales
are so low that extremely converged calculations are required.
Fortunately, based on the previous discussion, we assume that
the sc$GW$ total energy is also well approximated by RPA@PBEh(0.75).

In Fig.~\ref{fig:hbn}, we report the energy versus spacing between the layers.
LDA is known to give the correct spacing thanks to a lucky compensation
of errors \cite{marini_prl2006}, whereas PBE that improves the exchange over LDA does 
not benefit from this and yields a much too large spacing.
Our RPA@LDA reproduces within 0.25~bohr the earlier estimates
from Refs.~\onlinecite{marini_prl2006,bruneval_prl2012}.
This agreement is quite good considering the computational power difference and considering
the fact that we use newly developed pseudopotentials.

Next, let us comment that the RPA@PBEh(0.75) result (see Fig.~\ref{fig:hbn}) shows our best approximate to the sc$GW$. The obtained lattice spacing is very 
good agreement with respect to the experiment (6.18~bohr against 6.25~bohr).

This example shows that tuning the starting point in RPA does not destroy the
quantitative agreement with respect to the experiment.
This interesting conclusion calls for further studies in the future.

\section{Conclusion}
\label{sec:conclusion}

The linearized $GW$ density matrix $\gamma^{GW}$ has been introduced
in realistic solid-state systems.
We have carefully tested the norm-conserving pseudopotential approximation
and have concluded that QMC pseudopotentials, such as ccECP \cite{bennett_jchemphys2017},
are compulsory in this context for accurately determining RPA and $\gamma^{GW}$ lattice constants. 
On a benchmark of 7 covalent crystals (Si, C, SiC, zb-BN, AlP, AlAs, and Ge),
we have proven numerically that $\gamma^{GW}$ actually fulfils the exact constraints:
its natural occupation numbers range from 0 to 2 (when spin is summed)
and they sum up to the correct number of electrons.
In addition, the correlated nature of $\gamma^{GW}$ allows the electron occupancy 
to reorganize across the Brillouin zone, in strong contrast with all the mean-field
approaches, where the electron count remains constant in the Brillouin zone 
for crystals with a band gap.

The one-shot total energy expression $E_\mathrm{total}^{\gamma^{GW}}$ has been found to be superior to the usual RPA total energy expression in terms of sensitivity to the gKS starting point.

We provide strong evidence to support the assumption that $\gamma^{GW}$-based total energy
is a reliable substitute for sc$GW$, which remains unachievable at present.
As a cheaper alternative, RPA can also be used, but
we advocate applying it on top of gKS functionals with a large content of the exact exchange,
at least 75~\%, such as PBEh(0.75) or HF \fabien{to best approximate sc$GW$}.
Our last statement disagrees with the current wisdom that recommends RPA@PBE
\fabien{ based on comparison to experiment.}

Our implementation is available in the public version
the open source code ABINIT \cite{abinit_website}.

\begin{acknowledgments}
The authors acknowledge the financial support provided by the Cross-Disciplinary Program on Numerical Simulation of the French Alternative Energies and Atomic Energy Commission (CEA) (\textsc{ABIDM} project).
This work was performed using HPC resources from GENCI-CCRT-TGCC (Grants No. 2022-096018).
MRM thanks S.~Sharma for sharing her one-body-reduced-density-matrix expertise 
and FB thanks L.~Mitas for insights about the quantum Monte Carlo references.
\end{acknowledgments}

\appendix

\section{The unit cell density matrix in the natural orbital representation.}
\label{appexA}

\paragraph{The density matrix $\gamma(\br,\br')$ is diagonal in $\bk$}

Imposing the Born-van Karman periodic conditions,
we can write the double-Fourier expansion of $\gamma(\br,\br')$ in reciprocal space as
\begin{equation}
  \label{eq:gammakk}
    \gamma ({\bf r},{\bf r}') = \frac{1}{N_k \Omega} 
           \sum_{\bk \bk' \bG \bG'}  e^{\icomp (\bk + \bG) \cdot \br } 
              \gamma_{\bk\bk'}(\bG, \bG')
                e^{-\icomp (\bk' + \bG') \cdot \br' } ,
\end{equation}
where $N_k$ is the number of k-points and $\Omega$ is the volume of the unit cell
and $\bG$ and $\bG'$ are reciprocal lattice vectors.

By virtue of the translation invariance in the unit cell,
the shift of the two space indices with 
${\bf R}= n_1 {\bf a}_1 +n_2 {\bf a}_2 +n_3 {\bf a}_3 $ (with $n_1, n_2, n_3 \in \mathbb{Z}$ and ${\bf a}_1,{\bf a}_2,{\bf a}_3$ being the primitive lattice vectors)
does not change the density matrix \cite{sharma2008reduced}:
\begin{equation}
 \gamma ({\bf r},{\bf r}') = \gamma ({\bf r}+{\bf R},{\bf r}'+{\bf R}).
\end{equation}
When inserting the double Fourier transform, this implies
\begin{equation}
  e^{\icomp (\bk-\bk') \cdot \bR } = 1 .
\end{equation}
This condition is fulfilled \fabien{if and only if} $(\bk-\bk')$ belongs to the reciprocal lattice.
And as $\bk$ and $\bk'$ both belong to the first Brillouin zone, the only possible
reciprocal lattice vector the difference can match is $\mathbf{0}$.

As a consequence, one can insert the Kronecker sign $\delta_{\bk \bk'}$
in Eq.~(\ref{eq:gammakk}):
\begin{equation}
  \label{eq:gammagg}
    \gamma ({\bf r},{\bf r}') = \frac{1}{N_k \Omega} 
           \sum_{\bk \bG \bG'}  e^{\icomp (\bk + \bG) \cdot \br } 
              \gamma_{\bk}(\bG, \bG')
                e^{-\icomp (\bk + \bG') \cdot \br' } 
\end{equation}
and obtain the desired expression that shows
that $\gamma$ is block-diagonal with respect to $\bk$.

\paragraph{Natural orbitals and occupations}

Now for each discrete value of $\bk$, one can diagonalize $\gamma_\bk$:
\begin{equation}
  \sum_{\bG'} \gamma_{\bk}(\bG, \bG') \tilde u_{\bk \lambda} (\bG')
     = n_{\bk \lambda} \tilde u_{\bk \lambda} (\bG) ,
\end{equation}
where $\tilde u_{\bk \lambda} (\bG)$ are the eigenvectors and $n_{\bk \lambda}$
the eigenvalues, indexed with $\lambda$.

By construction:
\begin{equation}
  \gamma (\br, \br') = \gamma (\br', \br)^*
\end{equation}
implies
\begin{equation}
  \gamma_{\bk}(\bG, \bG') = \gamma_{\bk}(\bG', \bG)^*
\end{equation}
$\gamma_{\bk}(\bG, \bG')$ is an Hermitian matrix and thus 
 the $n_{\bk \lambda}$ are real-valued and
 the $\tilde u_{\bk \lambda}$ form a unitary matrix

Using this decomposition,
\begin{equation}
  \gamma_{\bk}(\bG, \bG')
     = \sum_\lambda  n_{\bk \lambda} \tilde u_{\bk \lambda}(\bG) \tilde u_{\bk \lambda}^*(\bG') ,
\end{equation}

Inserting this in Eq.~(\ref{eq:gammagg}), we obtain
\begin{multline}
  \gamma(\br, \br') = \frac{1}{N_k \Omega} \sum_{\bk \lambda} 
      \sum_\bG   e^{\icomp (\bk + \bG) \cdot \br } \tilde u_{\bk \lambda}(\bG) \\
        \times      n_{\bk \lambda}
              \sum_{\bG'}  \tilde u_{\bk \lambda}^*(\bG')
                e^{-\icomp (\bk + \bG') \cdot \br' } 
\end{multline}

Let us introduce the natural orbital in real-space $\phi_{\bk\lambda}(\br)$
that have a Bloch's wave form \cite{bloch1929quantenmechanik}:
\begin{align}
 \phi_{\bk\lambda}(\br)
    &= \frac{1}{\sqrt{N_k \Omega}} 
       \sum_\bG e^{\icomp (\bk + \bG) \cdot \br } \tilde u_{\bk \lambda}(\bG) \\
    &= \frac{1}{\sqrt{N_k \Omega}} 
       e^{\icomp \bk \cdot \br } \tilde u_{\bk \lambda}(\br)
\end{align}

The final expression in real space reads
\begin{equation}
  \gamma(\br, \br') = \sum_{\bk \lambda} 
       n_{\bk \lambda}  \phi_{\bk\lambda}(\br) \phi_{\bk\lambda}^*(\br') .
\end{equation}
that looks extremely similar to the mean-field (gKS) expression
\begin{equation}
  \gamma^\mathrm{gKS} (\br, \br') = \sum_{\bk i} 
       f_{\bk i}  \varphi_{\bk i}(\br) \varphi_{\bk i}^*(\br') ,
\end{equation}
where $f_{\bk i}$ are the Fermi-Dirac occupations and $\varphi_{\bk i}(\br)$
the mean-field wavefunctions.

However, there are subtle differences that are much more meaningful:
\begin{itemize}
    \item In gKS DFT, $\varphi_{\bk i}(\br)$ sum up to the exact electronic density,
    whereas the natural orbital sum up to the exact density \textit{and} density matrix.
    
    \item For the ground-state, the spin-summed $f_{\bk i}$ are constrained to be 0 or 2,
    whereas the spin-summed $n_{\bk \lambda}$ continuously span the range from 0 to 2
    (not proven here).

    \item Since $\int d\br \gamma(\br,\br)$ is normalized to the number of electrons $N_e$,
     both $f_{\bk i}$ and $n_{\bk \lambda}$ sum up to $N_e$.
     However, for insulators, while $\sum_i f_{\bk i} = N_e$, for each $\bk$ individually,
     no equivalent exists for the natural occupations $n_{\bk \lambda}$.
     as we show in the appendix~\ref{appexB}.
\end{itemize}

\section{Non-integer $N(\bk)$ values for individual $\bk$}
\label{appexB}
The non-integer values reported for the electron count, $N({\bf k})$, are a consequence of the electron correlation effects. In this appendix, we gain some insights into this result.

\subsection{The many-electron wavefunction in crystals} 
The basis of Bloch waves is complete; therefore, the real-space $N$-electron wavefunction can be written as a linear combination of Slater determinants $\left[ (N!)^{-1/2}|\varphi_{\bk _\mu i } (\br_1 )\ldots \varphi_{\bk _\nu j} (\br _N )| \right]$ built using Bloch waves, 
\begin{equation}
\varphi _{\bk i} (\br )= \frac{1}{\sqrt{N_k \Omega}} \sum _G e^{\icomp \bk \cdot \br} u _{\bk i} (\br),  
\end{equation}
which are usually the ones obtained from a mean-field method (like the ones obtained from a gKS DFT calculation). However, other basis sets can be used to build the Slater determinants, such as the Bloch waves corresponding to the natural orbitals. In this representation, the real-space (spin-less) $N$-electron wavefunction can be written as a configuration interaction (CI) expansion
\begin{widetext}
\begin{equation}
\Psi (\br_1,\br_2,...,\br_N)= 
\frac{1}{\sqrt{N!}(\sqrt{N_k \Omega})^N} \sum _{\bk _\mu 
 \ldots \bk _\nu } \sum _{
i \in \Omega _{\bk _\mu}} 
\ldots  \sum _{j \in \Omega _{\bk _\nu}}  C_{\bk _\mu i \ldots \bk _\nu j}
e^{\icomp \left( \bk _\mu  \cdot \br _1 + \ldots + \bk _\nu \cdot \br _N \right) }  u_{\bk _\mu i}(\br _1) \dots  u_{\bk _\nu j }(\br _N),
\label{eq:mb_wfn}
\end{equation}
\end{widetext}
where the sum $\sum _{\bk _\mu \ldots \bk _\nu }$ runs over all k-points in the first Brillouin zone, and the $C_{\bk _\mu i \ldots \bk _\nu j}$ are the expansion coefficients (that are adequately adjusted to ensure that $\Psi$ preserves the correct symmetries). Let us highlight that in Eq.~\eqref{eq:mb_wfn}, the Hartree product of Bloch's waves contains waves that belong to different k-points (i.e. $\Psi$ is the many-body wavefunction of the supercell).

The Born-van Karman periodic conditions imposed to the many-electron wavefunction~\cite{stoyanova2006delocalized} state that $\Psi (\br_1,\br_2,...,\br_N)= \Psi (\br_1+{\bf T},\br_2+{\bf T},...,\br_N+{\bf T})$ for ${\bf T}=n_1 N_1 {\bf a}_1 +n_2 N_2 {\bf a}_2 +n_3 N_3 {\bf a}_3 $ (with $N_k= N_1N_2N_3$). As a consequence, $e^{\icomp \left( \bk _\mu  + \ldots + \bk _\nu  \right) \cdot {\bf T}} = e^{\icomp {\bf K }\cdot {\bf T}}  = 1$ and 
\begin{align}
 {\bf K}&= \bk _\mu  + \ldots + \bk _\nu      \\
&= \frac{ \chi _1 }{N_1}  {\bf b}_1
  + \frac{ \chi _2 }{N_2} {\bf b}_2
  + \frac{ \chi _3 }{N_3} {\bf b}_3,
\end{align}
where $\chi_n \in \mathbb{Z}$. The many-electron wavefunctions are eigenfunctions of the many-body translation operator $\widehat{T}_{\bf R}$, i.e.  
\begin{align}
\widehat{T}_{\bf R} \Psi _{\bf K} (\br_1,\br_2,...,\br_N) &= \Psi _{\bf K} (\br_1 +{\bf R},\br_2 +{\bf R},...,\br_N +{\bf R})  \\
&=e^{\icomp {\bf K}\cdot {\bf R}}\Psi _{\bf K} (\br_1,\br_2 ,...,\br_N ), 
\end{align}
with $e^{\icomp {\bf K}\cdot {\bf R}}$ eigenvalues. Since the many-body Hamiltonian $\widehat{H}$ (within the Born-Oppenheimer approximation) commutes with the $\widehat{T}_{\bf R}$ operator~\cite{stoyanova2006delocalized}, the solutions to the many-body Hamiltonian can be taken associated with a given ${\bf K}$ value (i.e. $\widehat{H} \Psi _{\bf K}= E \Psi _{\bf K}$). Actually, the many-electron wavefunctions associated with ${\bf K} + \bG$  (with $\bG=\chi _1 {\bf b}_1+\chi _2 {\bf b}_2+\chi _3 {\bf b}_3$ being a reciprocal lattice vector) also lead to the same eigenvalue $e^{\icomp {\bf K}\cdot {\bf R}}$ upon application of the $\widehat{T}_{\bf R}$ operator and may contribute to the CI expansion.

\subsection{The density matrix and the second-order reduced density matrix in crystals} 

Let us define the density matrix elements (from the many-body wavefunction $\Psi$) as
\begin{equation}
\gamma _{\bk ij} = \langle \Psi | \widehat{b} ^\dagger _{\bk i} \widehat{b}  _{\bk j}   |   \Psi \rangle 
\end{equation}
and the second-order reduced density matrix (2-RDM) matrix elements as
\begin{equation}
\Gamma _{\bk_\mu i,\bk_\nu l} ^{\bk_\tau j,\bk_\vartheta m}= \frac{1}{2} \langle \Psi | \widehat{b} ^\dagger _{\bk_\mu i} \widehat{b} ^\dagger _{\bk_\nu l} \widehat{b} _{\bk_\vartheta m} \widehat{b}  _{\bk_\tau j}   |   \Psi \rangle 
\end{equation}
whose $\bk _n$ values fulfill the condition 
\begin{equation}
  \bk_\mu + \bk_\nu - \bk_\tau - \bk_\vartheta + {\bf G} = 0, 
 \label{eq:k4_2rdm}
\end{equation}
which ensures the correct translational symmetry of the 2-RDM, i.e. 
\begin{equation}
 \Gamma ({\bf r}_1,{\bf r}_2,{\bf r}_1',{\bf r}_2') = \Gamma ({\bf r}_1 + {\bf R},{\bf r}_2 + {\bf R},{\bf r}_1' + {\bf R},{\bf r}_2' + {\bf R}),   
\end{equation}
where $\Gamma ({\bf r}_1,{\bf r}_2,{\bf r}_1',{\bf r}_2')$ is the 2-RDM 
in space representation
(see for example Refs.~\cite{mcclain2017gaussian,schafer2017quartic} for more details). The 2-RDM contains more information about the system than the density matrix.
Indeed, the matrix elements of the density matrix can be obtained from the partial trace of the 2-RDM:
\begin{equation}
\gamma _{\bk ij} = \frac{2}{N-1} \sum _{\bk '} \sum _{l \in \Omega _{\bk'}} \Gamma_{\bk i,\bk 'l} ^{{\bk j,\bk 'l}},
\label{eq:2rdm_1rdm}
\end{equation}
where $\Omega _{\bk ' _\nu}$ is the subspace formed by all the one-electron wavefunctions sharing the same ${\bk ' _\nu}$ value. For completeness, let us mention that the matrix formed using the $\gamma _{\bk ij}$ elements is Hermitian and upon diagonalization produces the occupation numbers $n_{\bk \lambda }$ discussed in the appendix~\ref{appexA}. 

\subsection{The APSG \textit{ansatz} for crystals}

Here we prove in the specific case of an anti-symmetrized product of strongly-orthogonal geminals (APSG) \textit{ansatz} \cite{surjan1999introduction} for the many-electron wavefunction
that electron count transfer can occur across k-points due to electronic correlation effects. If it is true for this subclass of wavefunctions, then the statement also holds for the exact wavefunction.

The APSG \textit{ansatz} for the many-electron wavefunction with spin, where an even number of electrons present in the system is assumed (as we employed throughout this work) reads
\begin{equation}
\Psi ^{\textrm{APSG}} _{\bf K} (\bx_1,\bx_2,...,\bx_N)= \widehat{A} \prod ^{N/2} _{P=1} \psi_P (\bx_{2P-1},\bx_{2P}),
\label{eq:apsg}
\end{equation}
where $\bx = (\br,\sigma)$ is spatial and spin coordinate with $\sigma = \alpha, \beta$ referring to the spin index, $\widehat{A}$ stands for the antisymmetrizer responsible for inter-geminal permutations of electron coordinates and the geminal wavefunctions $ \psi_P (\bx_{2P-1},\bx_{2P})$ are wavefunctions containing one $\alpha$ and one $\beta$ electron. Because of this, the geminal wavefunction is a two-electron wavefunction; the sum of the occupation numbers for each spin channel must be 
\begin{equation}
\sum _{\left[\bk , i \in \Omega _\bk\right] \in P} n_{\bk i} =1   
\label{eq:occ_geminal}
\end{equation}
with $\left[\bk , i \in \Omega _\bk\right] \in P$ indicating that the $i$-th Bloch's wave belonging to the $\bk$-th k-point (i.e. the Bloch's wave natural orbital $\phi _{\bk i}$) is one of the Bloch's waves used in the construction of the $P$-th geminal. The $P$-th geminal wavefunction written in terms of the natural orbital Bloch's waves reads as
\begin{widetext}
\begin{equation}
\psi_P (\bx_{2P-1},\bx_{2P}) = 2^{-1/2} \sum _{\left[\bk  , i \in \Omega _\bk\right] \in P} c_{\bk i} \left[ \phi  _{\bk i} ({\bf r}_{2P-1})  \alpha _{2P-1} \phi ^* _{\bk i} ({\bf r}_{2P}) 
 \beta _{2P}  -  \phi _{\bk i} ({\bf r}_{2P}) \alpha _{2P}
 \phi ^*  _{\bk i} ({\bf r}_{2P-1}) \beta_{2P-1}  \right] ,
\label{eq:geminalwfn}
\end{equation}    
\end{widetext}
where the time-reversal symmetry is being employed to relate the degenerated Bloch's waves containing electrons with opposite spin~\cite{kramers1930theorie,aucar1995operator} forming a Kramers' pairs (i.e. $\phi  _{\bk i}$ and $\phi ^*  _{\bk i}$ form a Kramers' pair). Let us remark that the presence of complex-conjugated Bloch's waves (natural orbitals) is related to states filled by $\beta$ spin electrons; this choice is completely arbitrary. Also, the time-reversal symmetry imposed on the many-electron wavefunction leads to $\Psi _{{\bf K}={\bf 0}}$ for spin-compensated systems. 

Since the $\{\psi _P\}$ geminal wavefunctions are built with the strong orthonormality requirement, i.e. the condition that $\forall_{P\neq Q} \; \int d\bx _2 \psi_ P (\bx _1, \bx _2 ) \psi _Q (\bx _1',\bx _2) =\delta _{PQ} $, then the Bloch's waves natural orbitals are present in only one geminal wavefunction. When all the $\psi _P$ are built containing only one Kramers' pair as  
\begin{widetext}
\begin{align}
\psi_P (\bx_{2P-1},\bx_{2P}) &= 2^{-1/2}  \left[ \phi  _{\bk i} ({\bf r}_{2P-1})  \alpha _{2P-1} \phi ^* _{\bk i} ({\bf r}_{2P}) 
 \beta _{2P} - \phi _{\bk i} ({\bf r}_{2P}) \alpha _{2P}
 \phi ^*  _{\bk i} ({\bf r}_{2P-1}) \beta_{2P-1}  \right] ,
\end{align}    
\end{widetext}
the many-body wavefunction ($\Psi _{\bf K}$) defined in~\refeq{eq:apsg} corresponds to a single Slater determinant; thus,  the Hartree-Fock approximation is recovered, where the natural orbital basis and the so-called canonical orbitals (the mean-field ones) coincide.

The structure of the APSG wavefunction allows us to express the total energy in terms of the natural orbitals, the occupation numbers, and some undetermined phases. Then, the total APSG energy takes the following form
\begin{widetext}
\begin{align}
 \label{eq:etotal_apsg}
 E^{\textrm{APSG}} \left[ \{ f_{\bk _\mu i}\}, \{n_{\bk _\mu i}\},\{\phi _{\bk_\mu i} \} \right] &= 2 \sum _{\bk _\mu} \sum _{i \in \Omega _{\bk_\mu} } n_{\bk_\mu i} \langle \bk_\mu i |\widehat{h}| \bk_\mu i \rangle \nonumber \\ 
 &+ \sum ^{N/2} _P \sum \limits _{ \substack{\left[\bk_\mu  , i \in \Omega _{\bk_\mu} \right]  \\ \left[\bk_\nu  , j \in \Omega _{\bk_\nu} \right] } \in P} \zeta_{\bk _\mu i} \zeta_{\bk _\nu j} \sqrt{ n_{\bk _\mu i} n_{\bk _\nu j} } \langle  {\bk _\mu i} {\bk _\nu j} | {\bk _\nu j} {\bk _\mu i}  \rangle  \\ \nonumber
 &+  \sum ^{N/2} _{P\neq Q} \sum \limits _{ \substack{\left[\bk_\mu   , i \in \Omega _{\bk_\mu} \right] \in P \\ \left[\bk_\nu   , j \in \Omega _{\bk_\nu} \right] \in Q } }  n_{\bk _\mu i} n_{\bk _\nu j} (2 \langle  {\bk _\mu i \bk _\nu j}| {\bk _\mu i \bk _\nu j} \rangle - \langle  {\bk _\mu i \bk _\nu j}| {\bk _\nu j} {\bk _\mu i}  \rangle),
\end{align}    
\end{widetext}
where we have employed the known condition for APSG wavefunctions that allows us to express the CI coefficients in terms of occupation numbers ($c^2 _{\bk i} = n_{\bk i}$ or $c _{\bk i} = \zeta_{\bk i} \sqrt {n_{\bk i}} $), the phases $\zeta_{\bk _\mu i}^2 = 1$,  $\widehat{h}$ refers to all one-body operators of the electronic Hamiltonian (i.e. the kinetic energy and the interaction with the external potential),
$ \langle  {\bk _\mu i \bk _\nu j}| {\bk _\tau k \bk _\vartheta l} \rangle $ are the usual two-electron integrals. Notice that the energy minimization procedure implies optimization of the occupation numbers, natural orbitals, and phases. Since this wavefunction can be entirely written in terms of the natural orbitals and occupation numbers, it has been widely used in the context of reduced density matrix functional theory to propose energy functionals~\cite{lowdin1956natural,lowdin1955quantum,pernal2013equivalence,piris2013intrapair,piris2013natural,piris2017global,mitxelena2018phase,piris2021global,rodriguez2022relativistic}. 

The energy contribution in the second line of Eq.~\eqref{eq:etotal_apsg} is coming from the geminal wavefunctions and describes intra-geminal interactions (i.e. the ones among the two electrons belonging to the geminal wavefunction). On the other hand, energy contribution in the third line of Eq.~\eqref{eq:etotal_apsg} describes inter-geminal interactions, which are taken at the mean-field level (i.e. as Hartree--Fock interactions). Also, let us highlight that imposing the correct translation symmetry to $\Psi ^{\textrm{APSG}} _{\bf K}$ automatically enforces the correct symmetry in the 2-RDM elements. Making the 2-RDM elements fulfill the condition presented in Eq.~\eqref{eq:k4_2rdm}. 
 
Since the natural orbitals belonging to different $\Omega _\bk$ subspaces (e.g. $\bk$ and $\bk '$) can be employed in the construction of the geminal wavefunctions (see Eq.~\eqref{eq:geminalwfn}) the constraint given by Eq.~\eqref{eq:k4_2rdm} takes the following form for the geminals wavefunction 
\begin{equation}
 \bk ' - \bk ' - \bk +  \bk = {\bf 0} = \bG, 
 \label{eq:bkp_bk_gem}
\end{equation}
which ensures the correct translation symmetry in $\Psi ^{\textrm{APSG}} _{\bf K}$, the 2-RDM matrix elements, and the density matrix. Notice that the complex conjugation associated with the time-reversal symmetry of the Kramers' pairs was employed. 

The occupation numbers of the natural orbitals that belong to the geminal are optimized under the constraint given in Eq.~\eqref{eq:occ_geminal} during the energy minimization procedure (recalling that the CI coefficients can be written as $c _{\bk i} = \zeta_{\bk i} \sqrt {n_{\bk i}} $ for the $\Psi ^{\textrm{APSG}} _{\bf K} $ \textit{ansatz}). Hence, the coupling of natural orbitals belonging to different k-points is allowed; thus, a reorganization of electrons among k-points can take place during the energy minimization procedure.  Moreover, it is known~\cite{giesbertz2013natural} that the occupation numbers are not likely to become zero; then, the reorganization of electrons among k-points is not forbidden. 

In summary, the $\Psi ^{\textrm{APSG}} _{\bf K}$  \textit{ansatz} is a valid approximation to the many-electron wavefunction that permits us to illustrate the reasons leading to the reorganization of electrons among k-points. Obviously, a more general valid CI expansion \textit{ansatz} (or the exact full-CI expansion) could also lead to a reorganization of the electrons among k-points since the $\Psi ^{\textrm{APSG}} _{\bf K}$  \textit{ansatz} is a particular case of the exact $\Psi_ {\bf K}$, where the electron pairs do not interact. Then, let us conclude that the reorganization of electrons among k-points that lead to non-integer $N_e$ values for each $\bk$ value is purely a consequence of the electronic correlation effects. In this work, the electronic correlation effects are captured with the $\gamma ^{GW}$ approximation, which produces the reorganization of electrons among k-points. And, this leads to the non-integer $N_e$ values obtained for each k-point that were used to compute the $\Delta N({\bk})$ values presented in Fig.~\ref{fig:si_nk}. 

In the next section, we present an example based on the Si crystal where the reorganization of electrons among k-points  is allowed using a $\Psi ^{\textrm{APSG}} _{\bf K}$ \textit{ansatz}.

\subsection{Example of an allowed electronic density reorganization among the k-points in the Si crystal}
For a working example, let us take the Si crystal computed excluding all the core states (i.e. using a pseudo-potential and retaining only 8 electrons per unit cell). At the Hartree--Fock (or gKS DFT) level, eight states forming four Kramers' pairs are occupied for each k-point value. From the band structure, it is easy to recognize that the highest (in terms of energy) occupied state with an $\alpha$ electron localized at the $\Gamma$ point ($\bk = (0,0,0)=\bk _{\Gamma}$). On the other hand, the lowest (in terms of energy) unoccupied state for the electrons with $\alpha$ spin belongs to the $X$ point ($\bk = (0.5,0.5,0)=\bk _X$). Let us label these states as $\varphi _{\bk_{\Gamma} 4}$ and $\varphi _{\bk _X 5}$, respectively. The energy difference between the $\varphi _{\bk 4}$ and $\varphi _{\bk 5}$ states is small (the experimental value is approximately 1~eV), which leads to the small indirect band gap obtained for this system.  

Following, let us organize in ascending order in terms of energy all the mean-field Bloch's waves for the whole system (i.e. of the supercell). And, as it is usually done in the search for the optimal $\Psi ^{\textrm{APSG}} _{{\bf K}={\bf 0}}$, let us write the initial guess for the APSG \textit{ansatz} in terms of the mean-field Bloch's waves. But, let us search for a particular $\Psi ^\textrm{APSG} _ {\bf K}$ \textit{ansatz} where all the geminal wavefunctions contain only one Kramers' pair (i.e. are treated at the Hartree--Fock level) except for the last geminal (the $P=N/2$ in~\eqref{eq:geminalwfn}) that is built coupling the Bloch's waves $\varphi _{\bk_{\Gamma} 4}$ and $\varphi _{\bk _X 5}$, i.e.
\begin{widetext}
\begin{align}
 \psi _{N/2} ^{}  (\bx_{N-1},\bx_N)& = 2^{-1/2} \left[ c_{\bk _{\Gamma} 4} ( \varphi _{\bk _{\Gamma} 4}(\br _{N-1}) \alpha _{N-1} \varphi ^*  _{\bk _{\Gamma} 4}(\br _{N}) \beta _N + \varphi _{\bk _{\Gamma} 4}(\br _{N}) \alpha _{N} \varphi ^*  _{\bk _{\Gamma} 4}(\br _{N-1}) \beta _{N-1}  ) \right. \\ 
 &+ \left. c_{\bk _{X} 5} ( \varphi _{\bk _{X} 5}(\br _{N-1}) \alpha _{N-1} \varphi ^*  _{\bk _{X} 5}(\br _{N}) \beta _N + \varphi _{\bk _{X} 5}(\br _{N}) \alpha _{N} \varphi ^*  _{\bk _{X} 5}(\br _{N-1}) \beta _{N-1}  ) \right] ,    \nonumber   
\end{align}
\end{widetext}
with $c_{\bk _{\Gamma} 4} =  \zeta_{\bk _{\Gamma} 4} \sqrt{n_{\bk _{\Gamma} 4}} $ and $c_{\bk _{X} 5} = \zeta_{\bk _{X} 5} \sqrt{n_{\bk _{X} 5}} $ being variational parameters subject to the condition $n_{\bk _{\Gamma} 4} + n_{\bk _X 5}= 1 $ (to fulfill the  requirement presented in Eq.~\eqref{eq:occ_geminal}). This type of geminal approach, where only two states (four considering spin) are present in the geminal wavefunction is known as a perfect-pairing approach. Next, let us assume that the mean-field Bloch's waves $\varphi _{\bk_{\Gamma} 4}$ and $\varphi _{\bk _X 5}$ coincide with the optimal natural orbitals in order to skip the orbital optimization procedure. 

Next, let us focus on the energy contribution arising from the ${N/2}$ geminal to the second term in the r.h.s. of the APSG energy (see Eq.~\eqref{eq:etotal_apsg})  
\begin{widetext}
\begin{equation}
n_{\bk _{\Gamma} 4} \langle  {\bk _{\Gamma} 4}  {\bk _{\Gamma} 4}|  {\bk _{\Gamma} 4}  {\bk _{\Gamma} 4} \rangle +  n_{\bk _X 5}  \langle  {\bk _X 5} {\bk _X 5} | {\bk _X 5} {\bk _X 5} \rangle + 2 \zeta_{\bk _{\Gamma} 4} \zeta_{\bk _{X} 5} \sqrt{ n_{\bk _{\Gamma} 4} n_{\bk _X 5} } \langle  {\bk _{\Gamma} 4} {\bk _X 5} | {\bk _X 5} {\bk _{\Gamma} 4}  \rangle  .    
\end{equation}    
\end{widetext}
Setting the usual approximation for (fixing) the phases (i.e. $\zeta_{\bk _{\Gamma} 4} \zeta_{\bk _{X} 5}  =-1 $ \cite{lowdin1956natural,lowdin1955quantum,piris2013natural,piris2013intrapair}) for the interaction among the states above and below the Fermi level and letting all two-electron integrals to be equal (which can occur in the extreme case when degenerated states are involved). The occupation numbers that would minimize the energy contribution are $n_{\bk _{\Gamma} 4} =  n_{\bk _X 5}  = 1/2 $. Illustrating that a reorganization of electrons occurs among the $\Gamma$ and $X$ k-points. In the Si crystal, the mean-field Bloch's waves are not completely degenerate in terms of energy and do not correspond to the optimal natural orbitals; then, the actual optimal occupation numbers differ from $1/2$. But, they also differ from the initial values at the mean-field level, where $n_{\bk _{\Gamma} 4} =1 $ and $n_{\bk _X 5} =0$. Moreover, beyond the perfect pairing approach, the coupling of states to form a geminal wavefunction can include states belonging to other k-points. Since the geminal wavefunctions are built with states for the $\alpha$ and 
the $\beta$ electron; the state for the $\alpha$ electrons is associated with a $\bk ''$ k-point while the state for the $\beta$ is related to a $-\bk ''$ k-point. The Hartree product in Eq.~\eqref{eq:geminalwfn} conserves the ${\bf K}=\bk '' - \bk''={\bf 0}$ value, which could lead to further reorganization of electrons among different k-points beyond the perfect pairing approach. Thus, for example, the coupling of states belonging to $\Gamma$, $X$, and $\Delta$, etc. is allowed in the Si crystal. Actually, the coupling of all k-point values in the first Brillouin zone is valid to build geminal wavefunctions.  

Finally, let us remark that this example is based on a valid approximation to the many-electron wavefunction (i.e. a $\Psi ^\textrm{APSG} _ {\bf K}$ \textit{ansatz}), where we illustrate that the reorganization of electrons among k-points is purely a consequence of the electronic correlation effects. 

\bibliography{my_biblio}

\end{document}


\title{Supplemental Information: $GW$ density matrix in solids as
an estimate of self-consistent $GW$ total energies}

\author{Adam Hassan Denawi}
\affiliation{Universit\'e Paris-Saclay, CEA, Service de recherche en Corrosion et Comportement des Matériaux, SRMP, 91191 Gif-sur-Yvette, France}

\author{Fabien Bruneval}
\affiliation{Universit\'e Paris-Saclay, CEA, Service de recherche en Corrosion et Comportement des Matériaux, SRMP, 91191 Gif-sur-Yvette, France}

\author{Marc Torrent}
\affiliation{CEA, DAM, DIF, F-91297 Arpajon, France}
\affiliation{Universit\'e Paris-Saclay, CEA, Laboratoire Mati\`ere en Conditions Extr\^emes, 91680 Bruy\`eres-le-Ch\^atel, France}

\author{Mauricio Rodr{\'i}guez-Mayorga}
\affiliation{Universit\'e Paris-Saclay, CEA, Service de recherche en Corrosion et Comportement des Matériaux, SRMP, 91191 Gif-sur-Yvette, France\\ Department of Physical Chemistry,
University of Alicante, E-03080 Alicante, Spain}

\maketitle

\phantom{aaa}

\newpage

\section{Pseudopotential validation using Hartree-Fock}

\begin{figure*}[h!]
  \includegraphics[width=0.60\columnwidth]{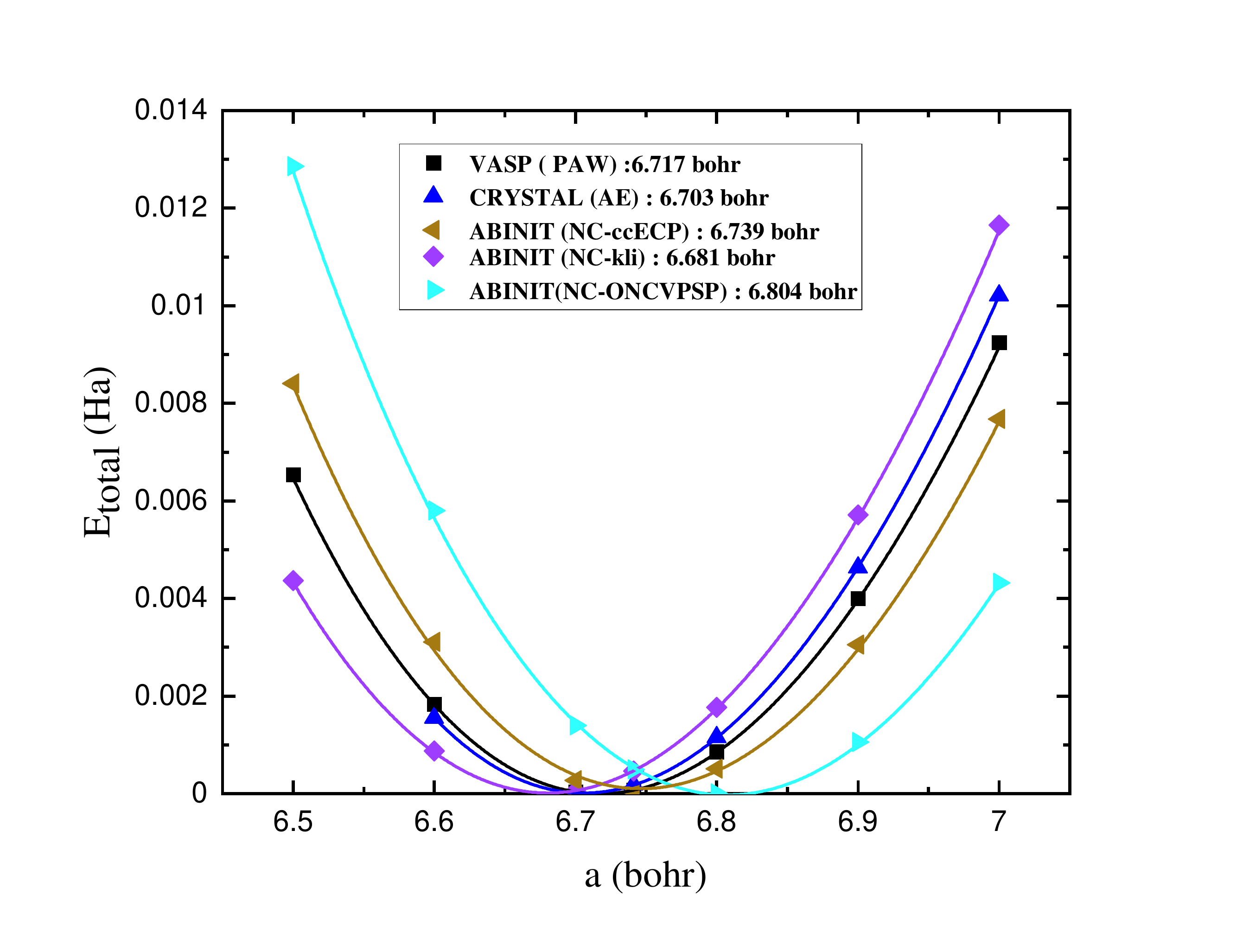}
  \includegraphics[width=0.60\columnwidth]{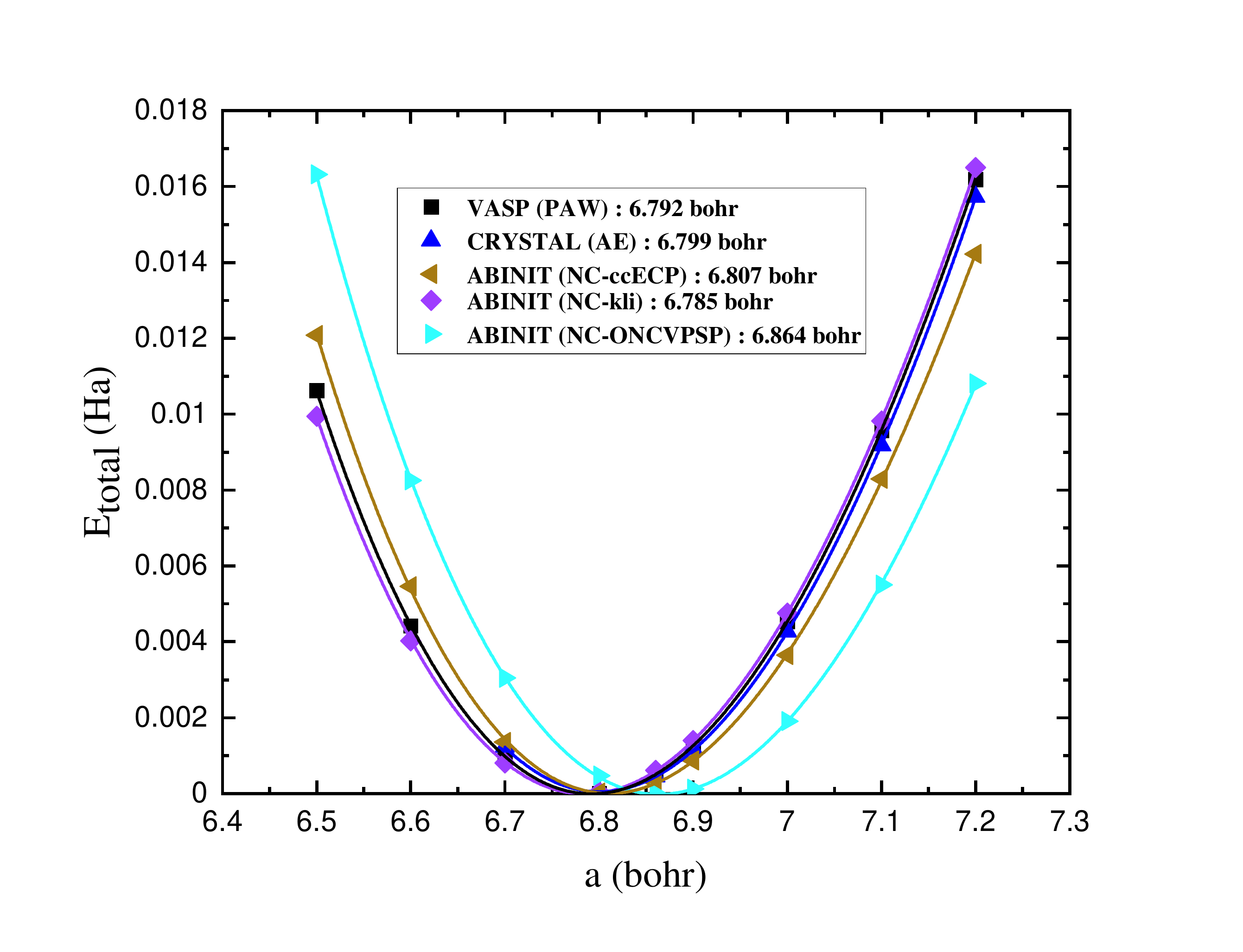}

    \caption{Hartree-Fock (HF) total energies as a function of lattice constants of diamond and zinc-blende boron nitride (zb-BN) with different codes and pseudopotentials. We used all-electron (AE) of CRYSTAL\cite{dovesi_jchemphys2020} with the accurate basis set designed by Heyd and coworkers \cite{heyd_jchemphys2005}, the projector-augmented wave (PAW) of VASP \cite{kresse_prb1996}, the correlation consistent effective core potentials (ccECP) \cite{bennett_jchemphys2017}, the pseudopotentials generated for KLI \cite{krieger_pla1990} and NC-ONCVPSP \cite{hamann_prb2013} of ABINIT code \cite{gonze_cpc2020}.
    \label{fig:S1}
  }
\end{figure*}

\clearpage
\newpage

\section{$\gamma^{GW}$ and RPA total energies for diamond, SiC, zb-BN, AlP, AlAs, and Ge}

\subsection{Diamond}

\begin{figure*}[h!]
  \includegraphics[width=0.40\columnwidth]{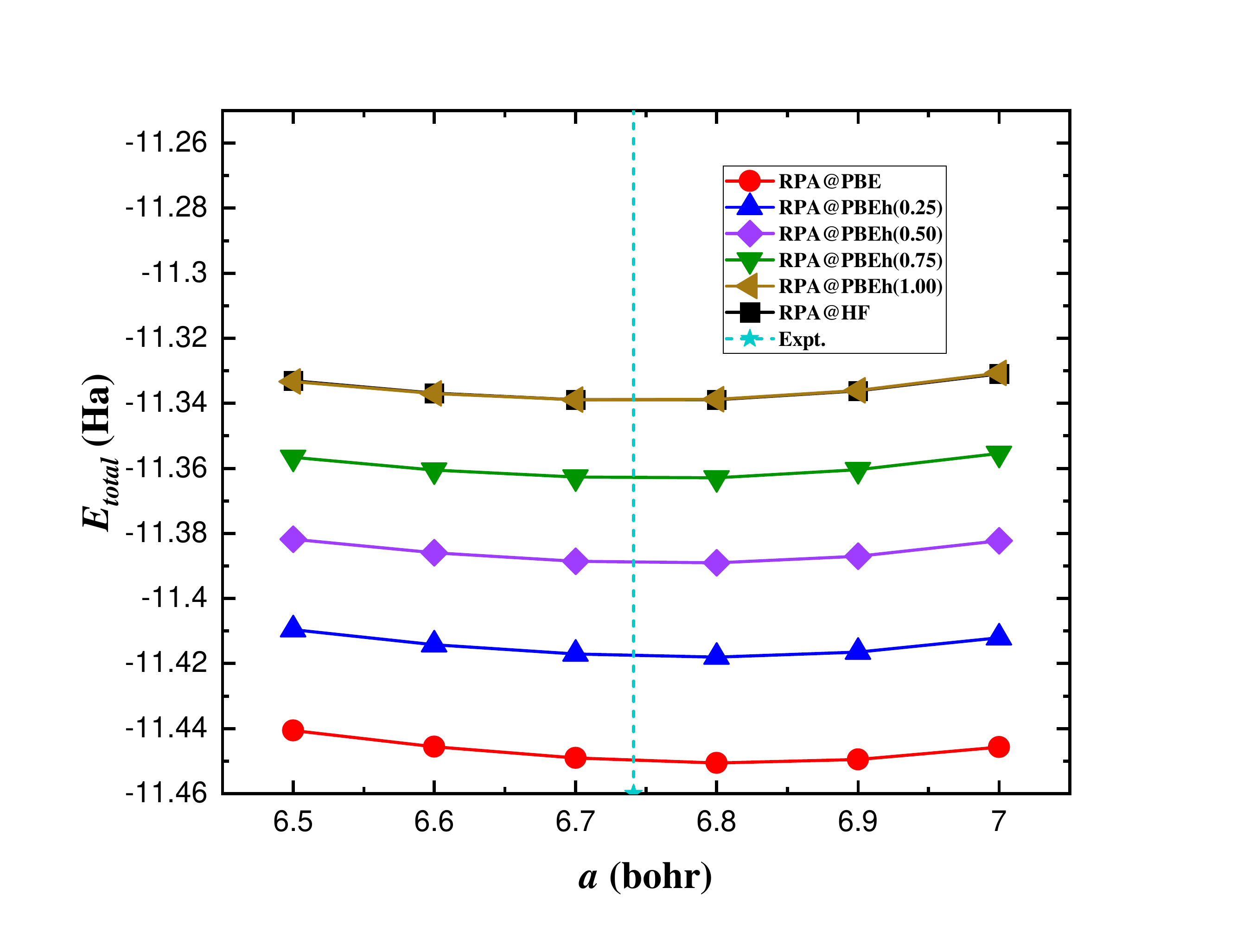}
  \includegraphics[width=0.40\columnwidth]{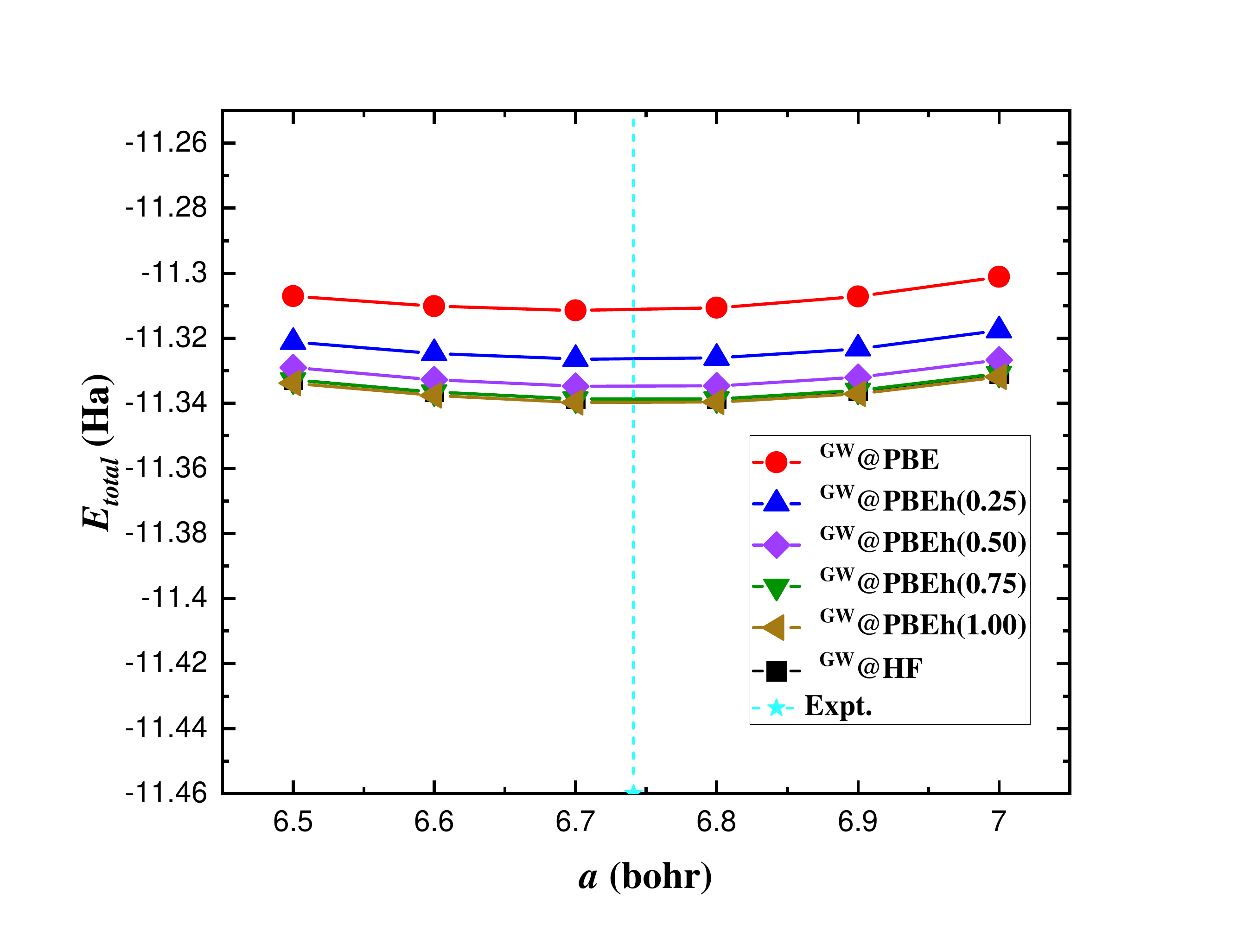}
  \includegraphics[width=0.40\columnwidth]{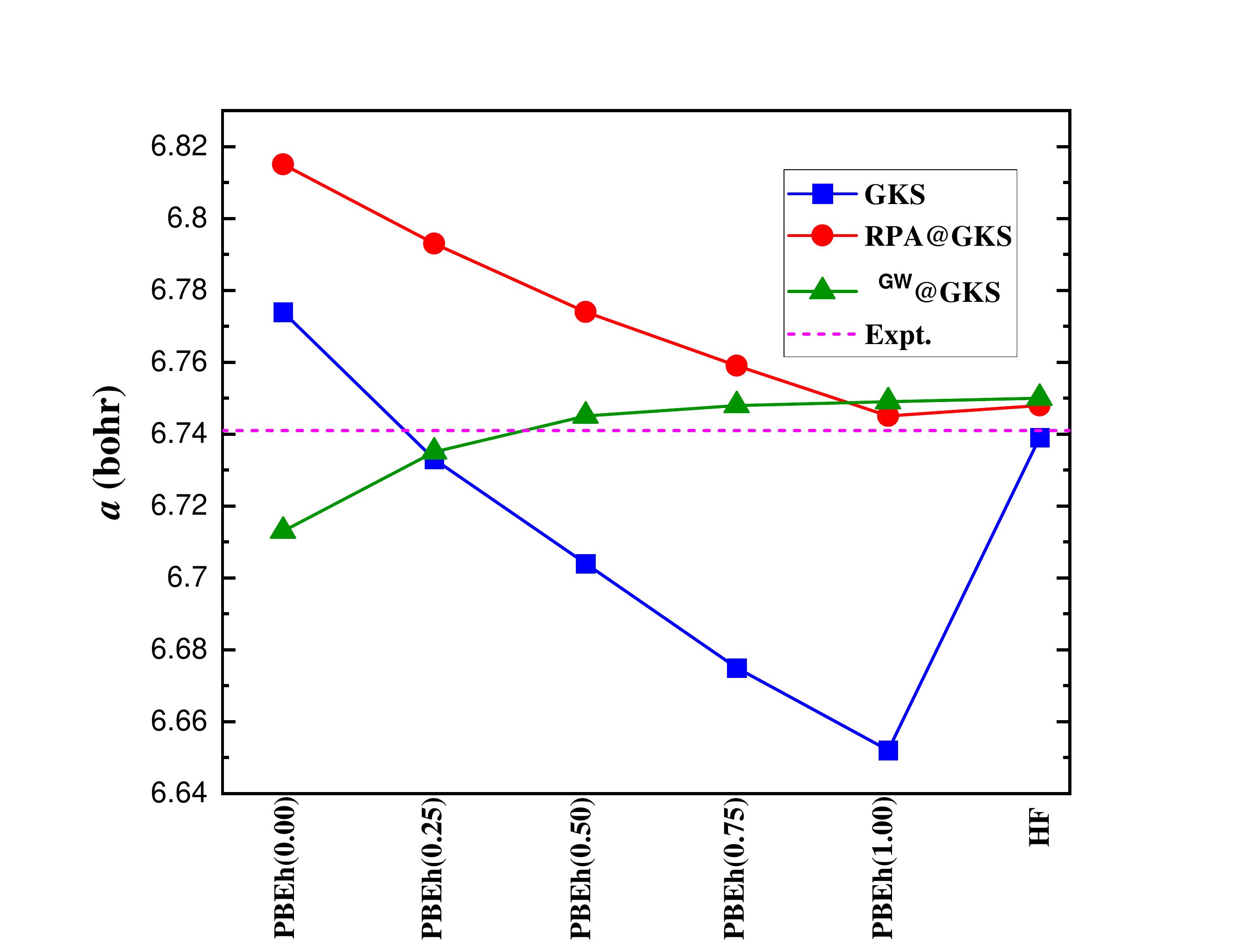}
  \includegraphics[width=0.40\columnwidth]{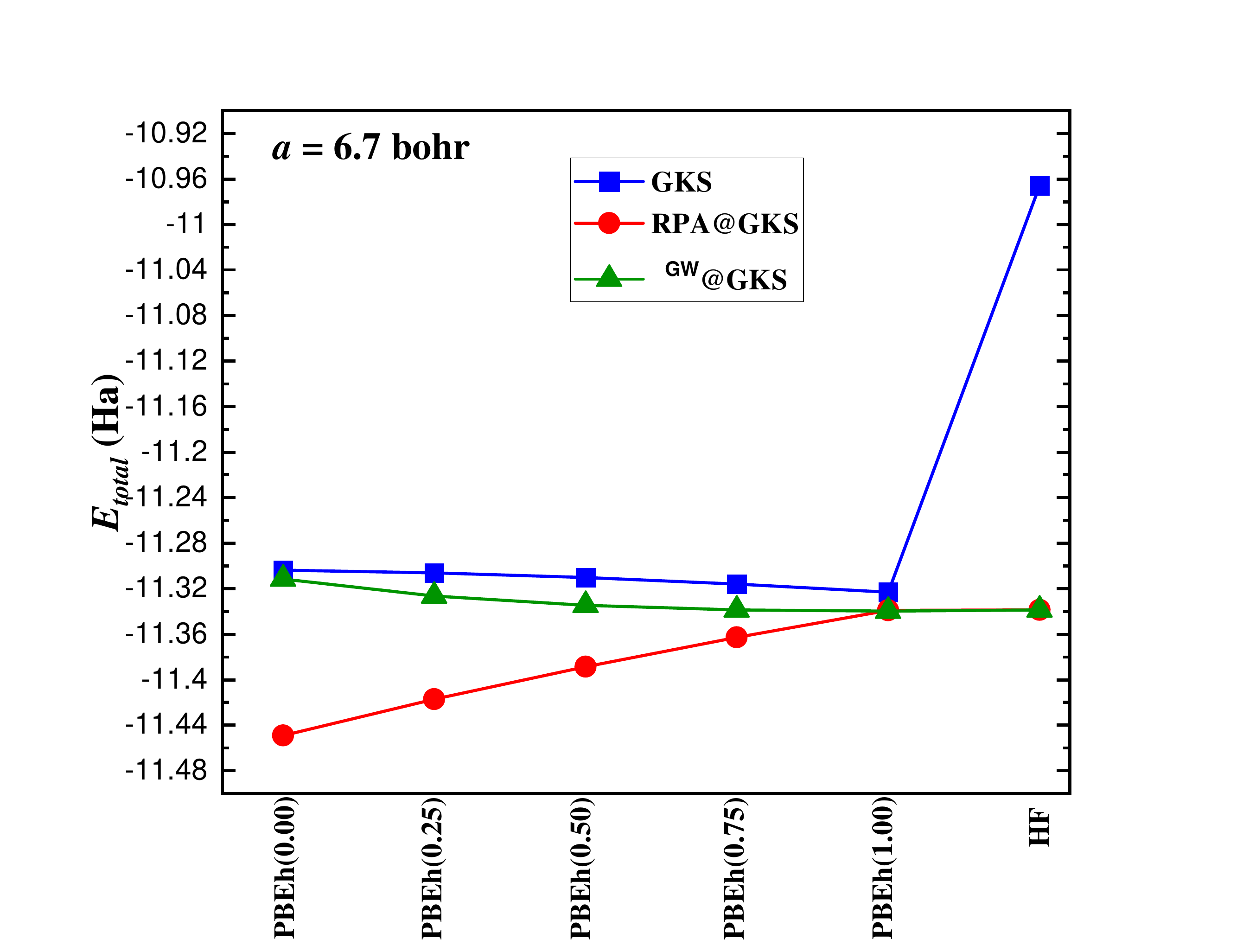}

    \caption{Calculated the equilibrium lattice constants $a$ (bohr) for Diamond, comparing with experiments and comparison of the total energy for the SCF, RPA and $\gamma^{GW}$ as a function of the starting SCF functional.
    \label{fig:S2}
  }
\end{figure*}

\newpage

\subsection{SiC}

\begin{figure*}[h!]
  \includegraphics[width=0.40\columnwidth]{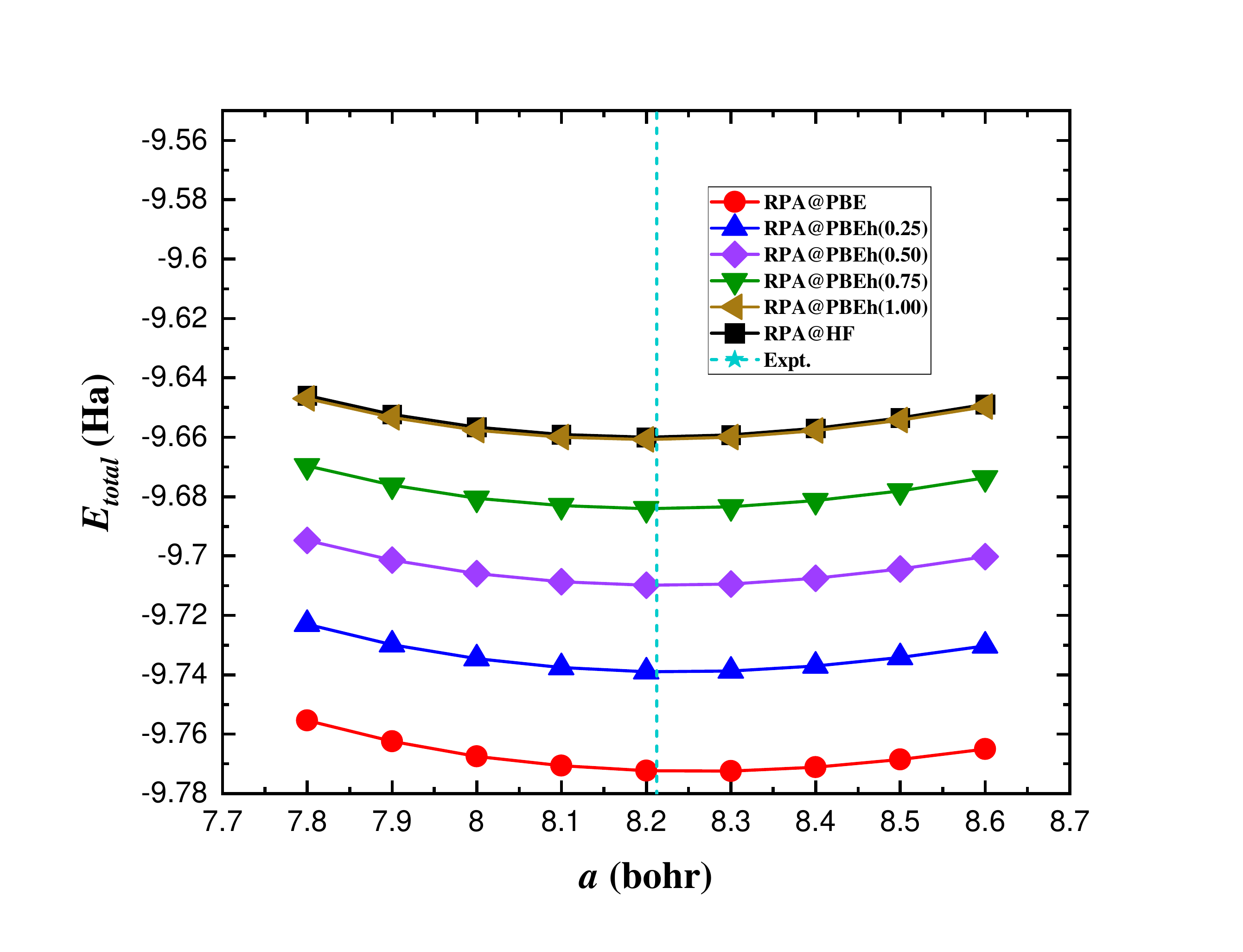}
  \includegraphics[width=0.40\columnwidth]{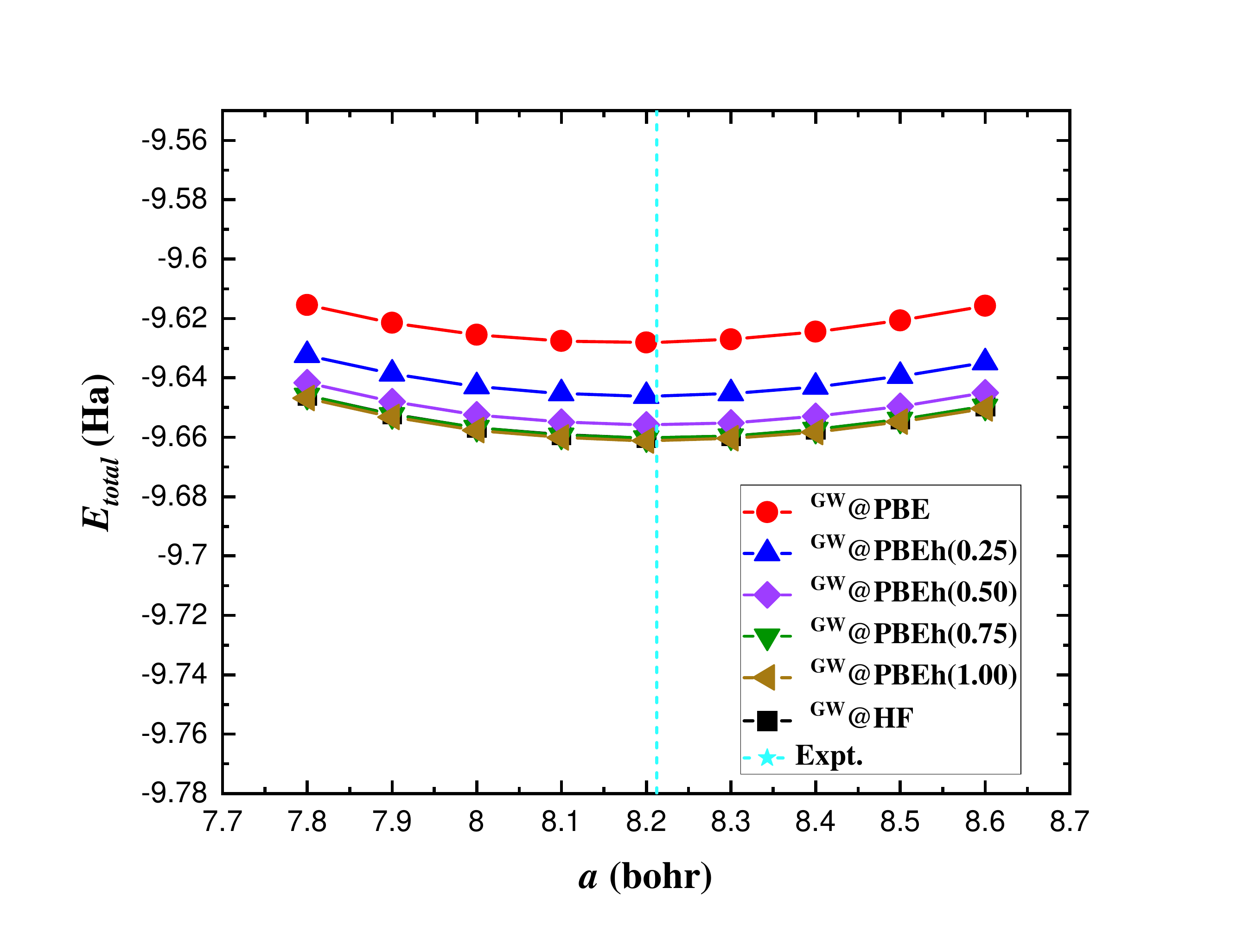}
  \includegraphics[width=0.40\columnwidth]{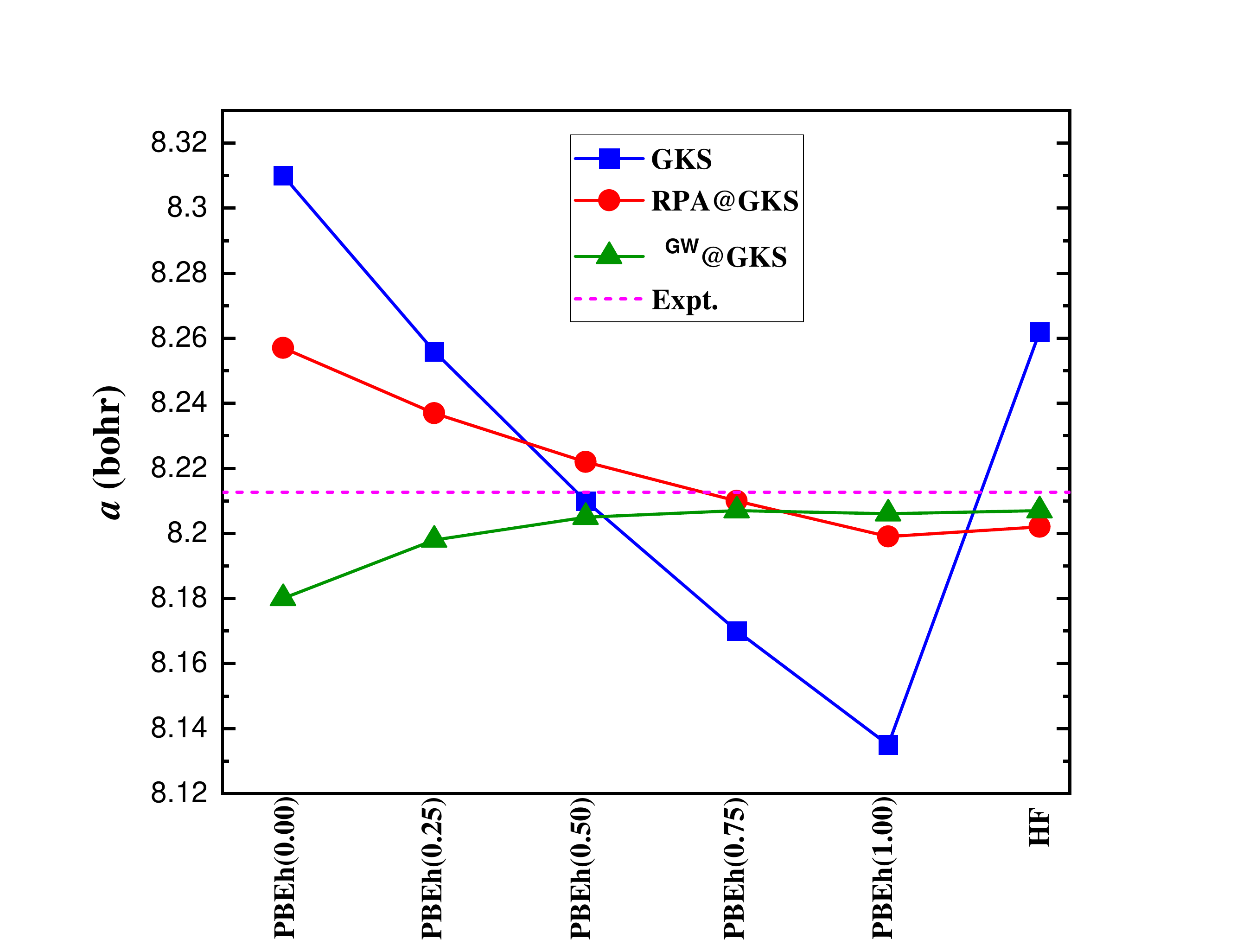}
  \includegraphics[width=0.40\columnwidth]{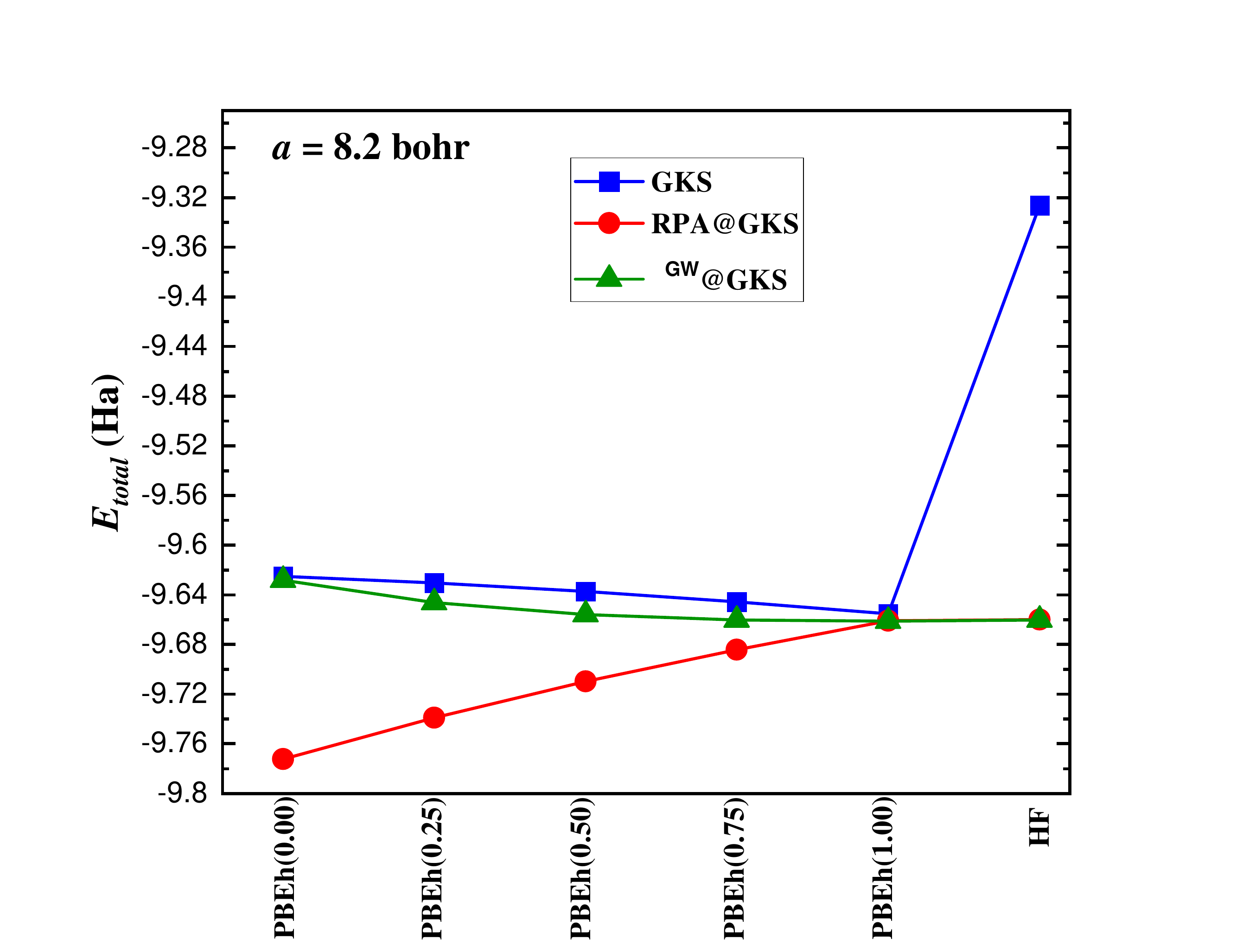}

    \caption{Calculated the equilibrium lattice constants $a$ (bohr) for Silicon Carbide, comparing with experiments and comparison of the total energy for the SCF, RPA and $\gamma^{GW}$ as a function of the starting SCF functional.
    \label{fig:S3}
  }
\end{figure*}

\newpage
\subsection{zb-BN}

\begin{figure*}[h!]
  \includegraphics[width=0.40\columnwidth]{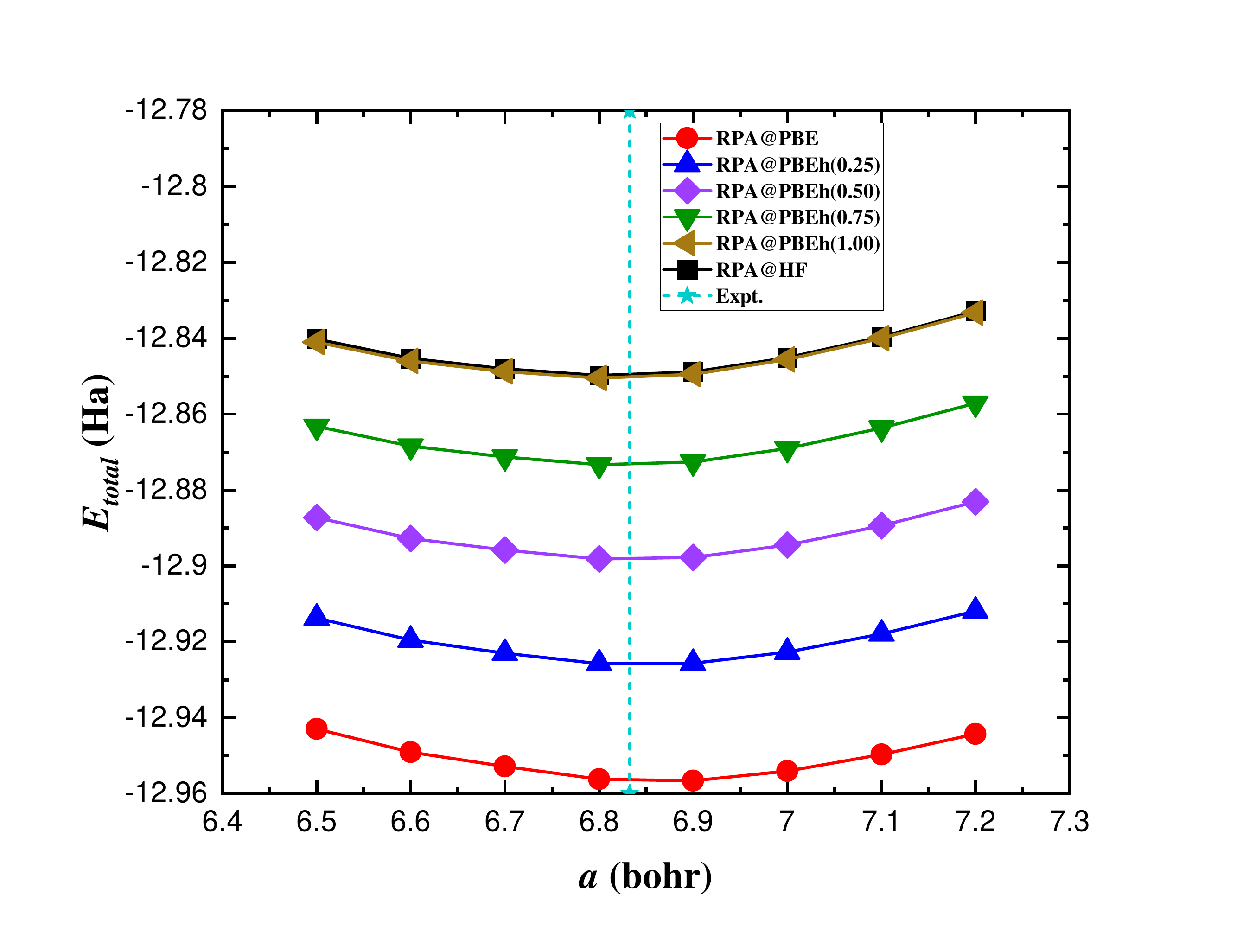}
  \includegraphics[width=0.40\columnwidth]{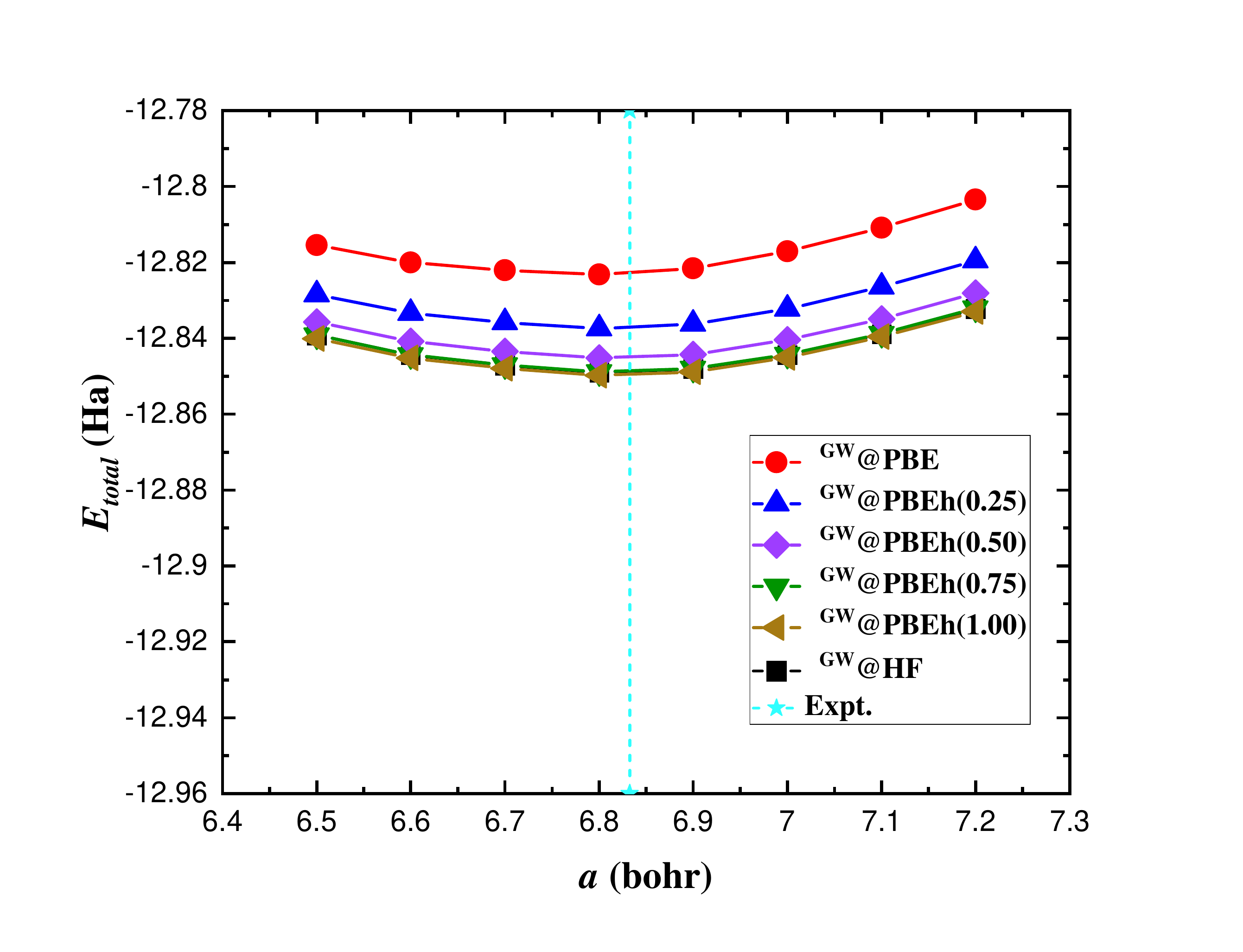}
  \includegraphics[width=0.40\columnwidth]{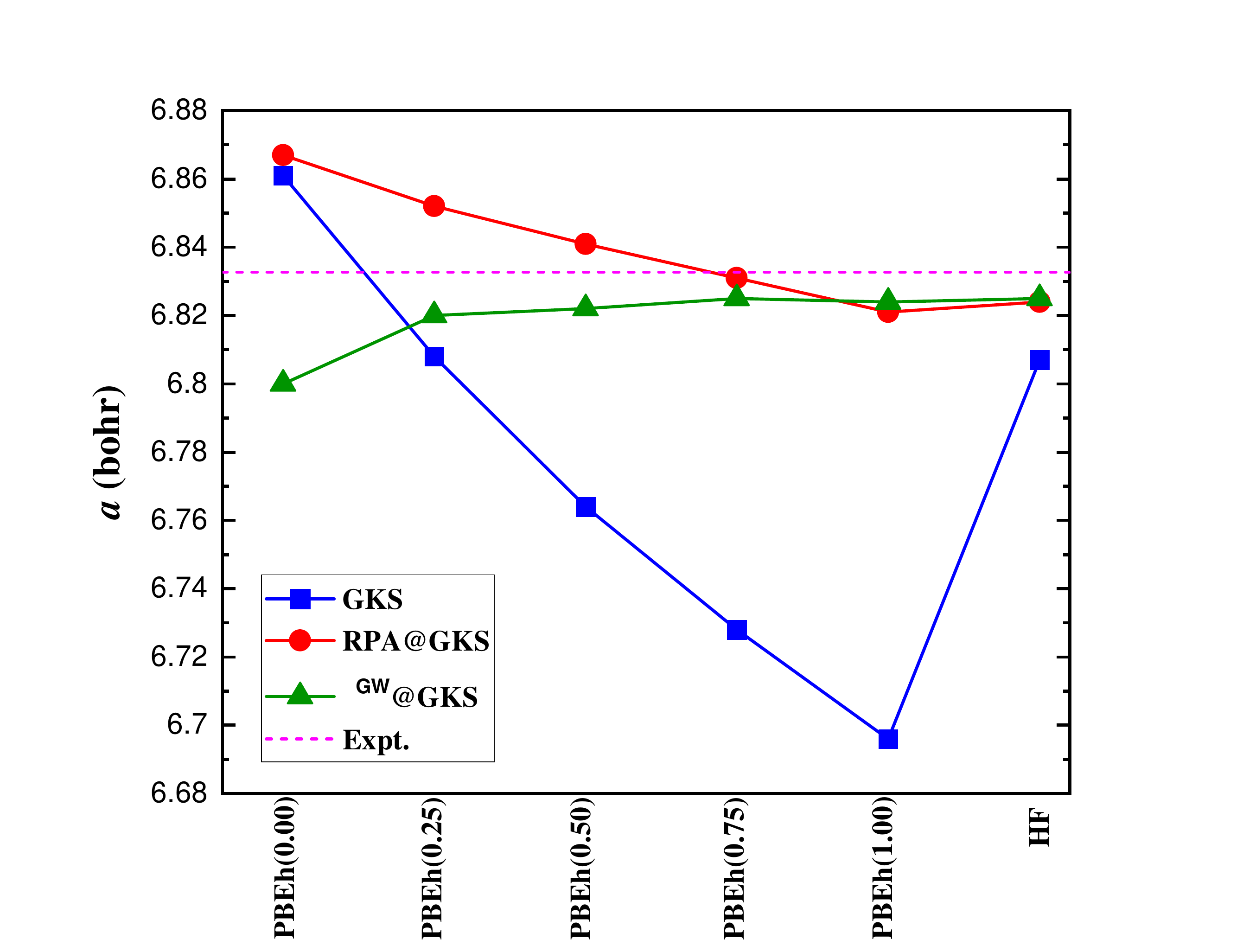}
  \includegraphics[width=0.40\columnwidth]{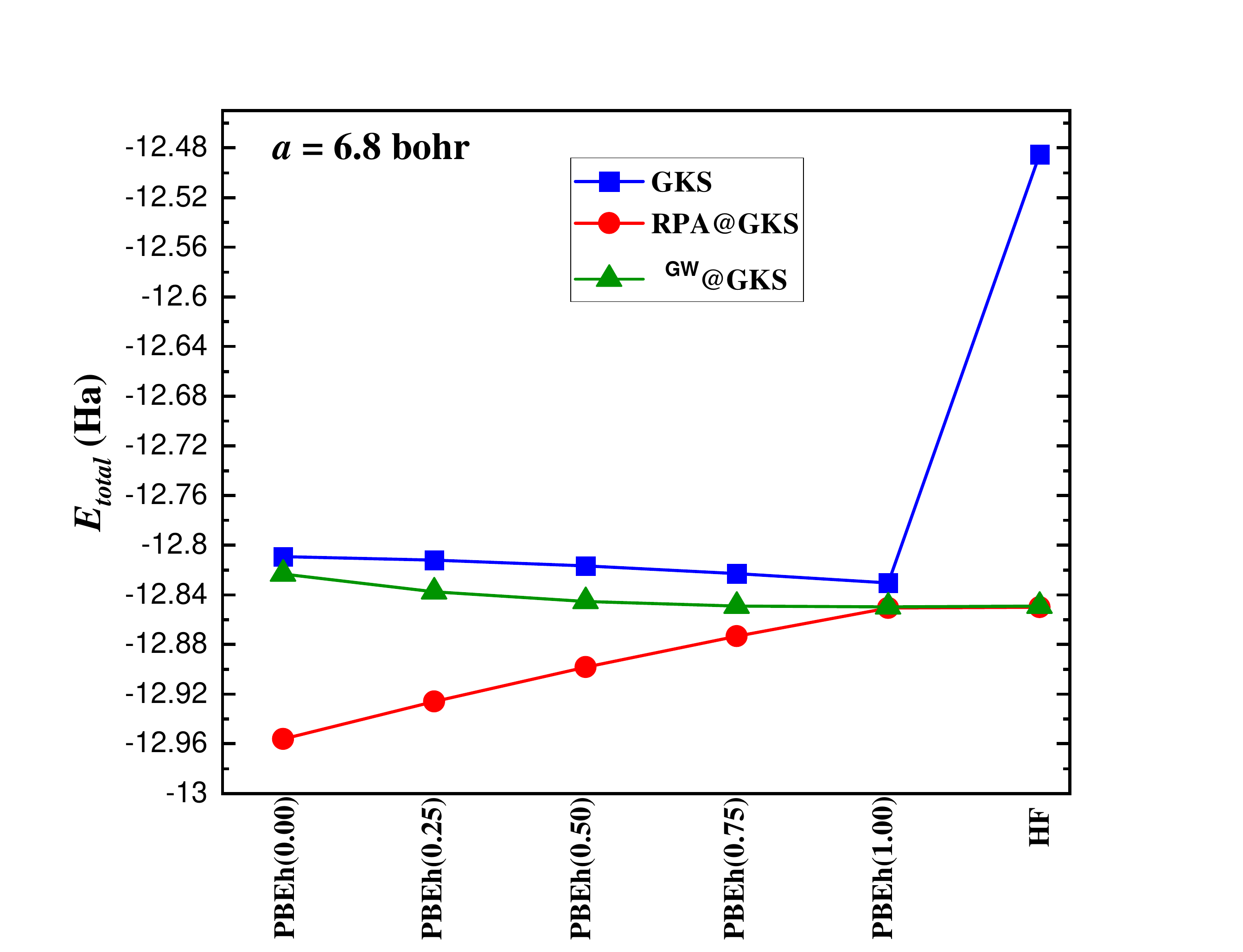}

    \caption{Calculated the equilibrium lattice constants $a$ (bohr) for Boron Nitride, comparing with experiments and comparison of the total energy for the SCF, RPA and $\gamma^{GW}$ as a function of the starting SCF functional.
    \label{fig:S4}
  }
\end{figure*}

\newpage 
\subsection{AlP}

\begin{figure*}[h!]
  \includegraphics[width=0.40\columnwidth]{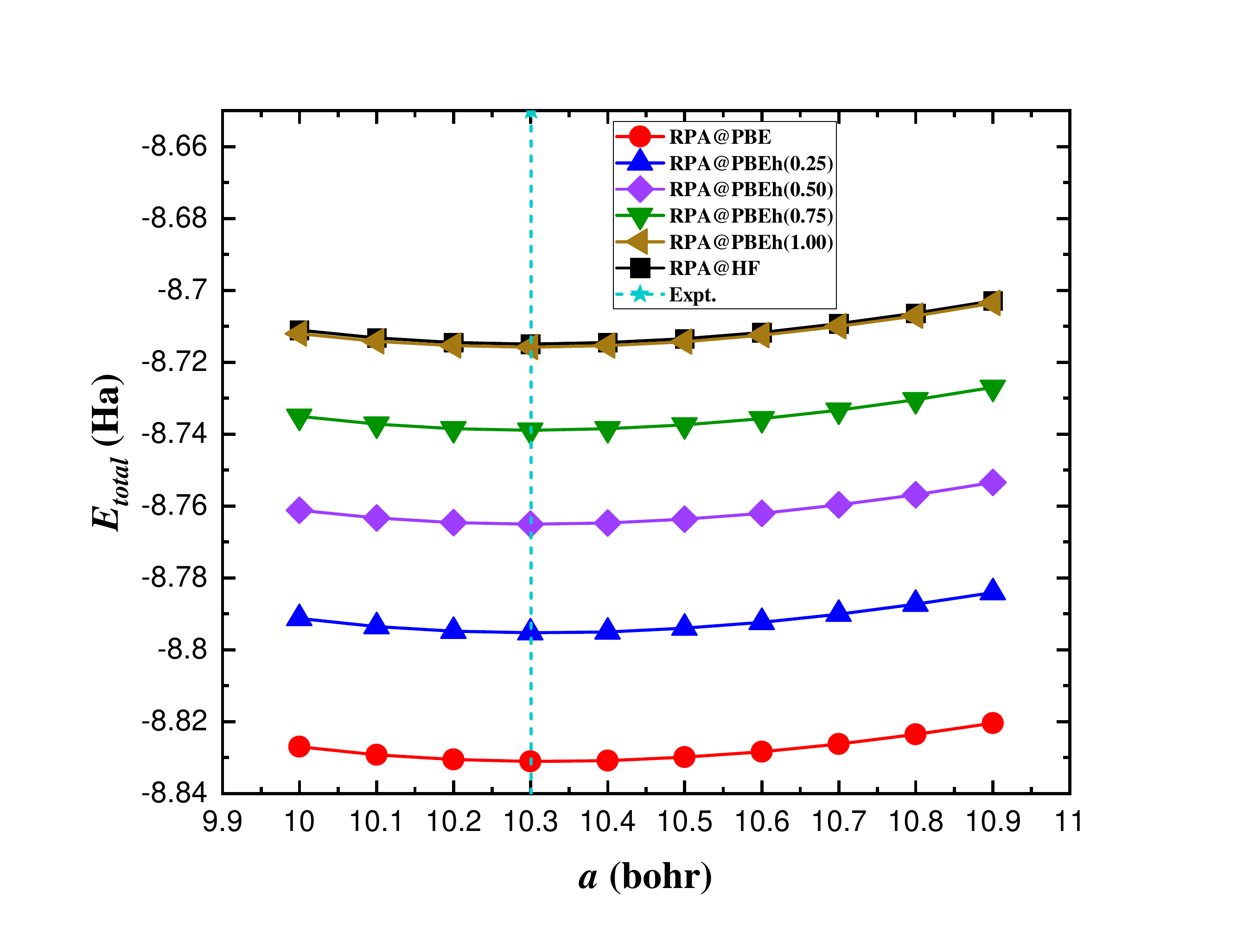}
  \includegraphics[width=0.40\columnwidth]{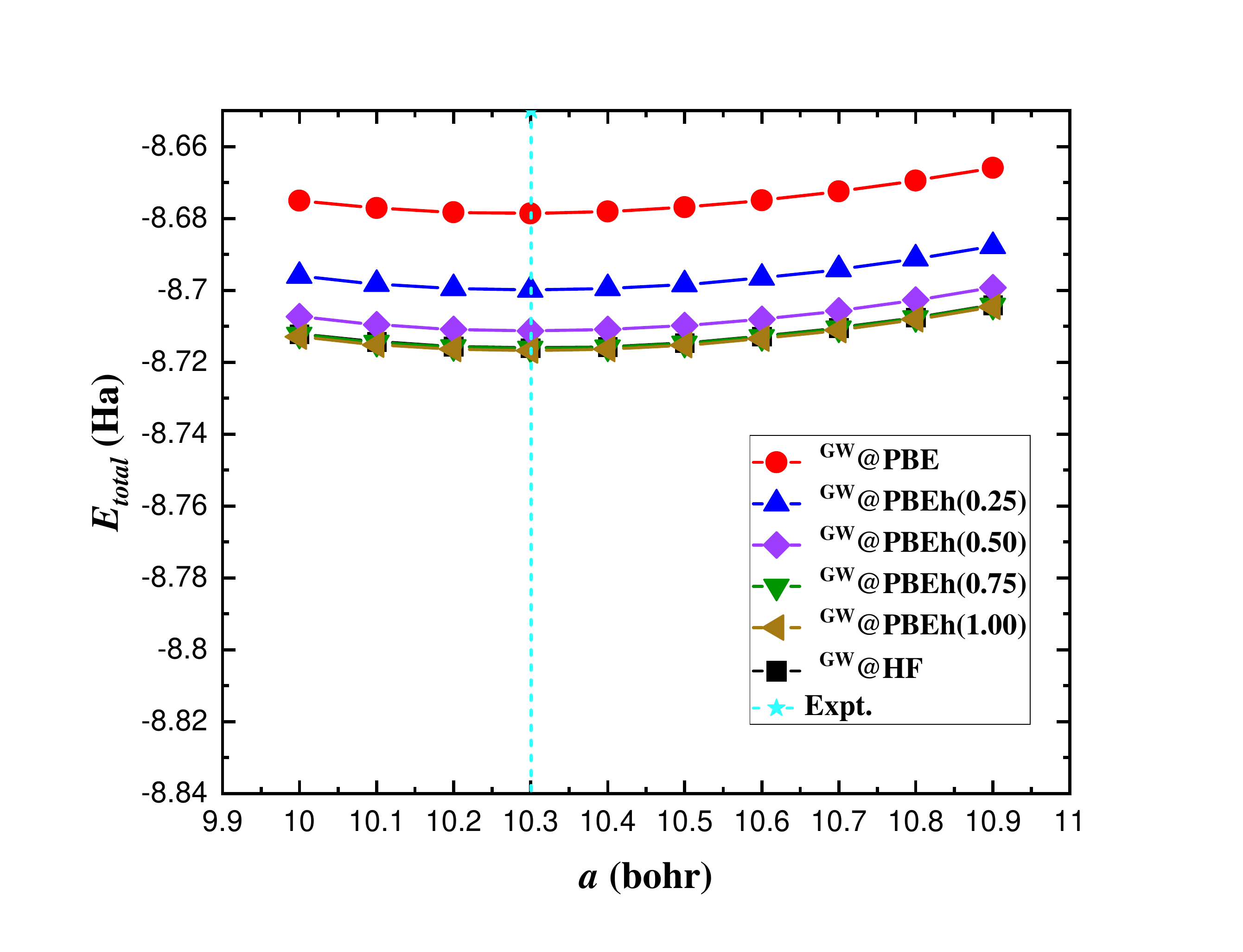}
  \includegraphics[width=0.40\columnwidth]{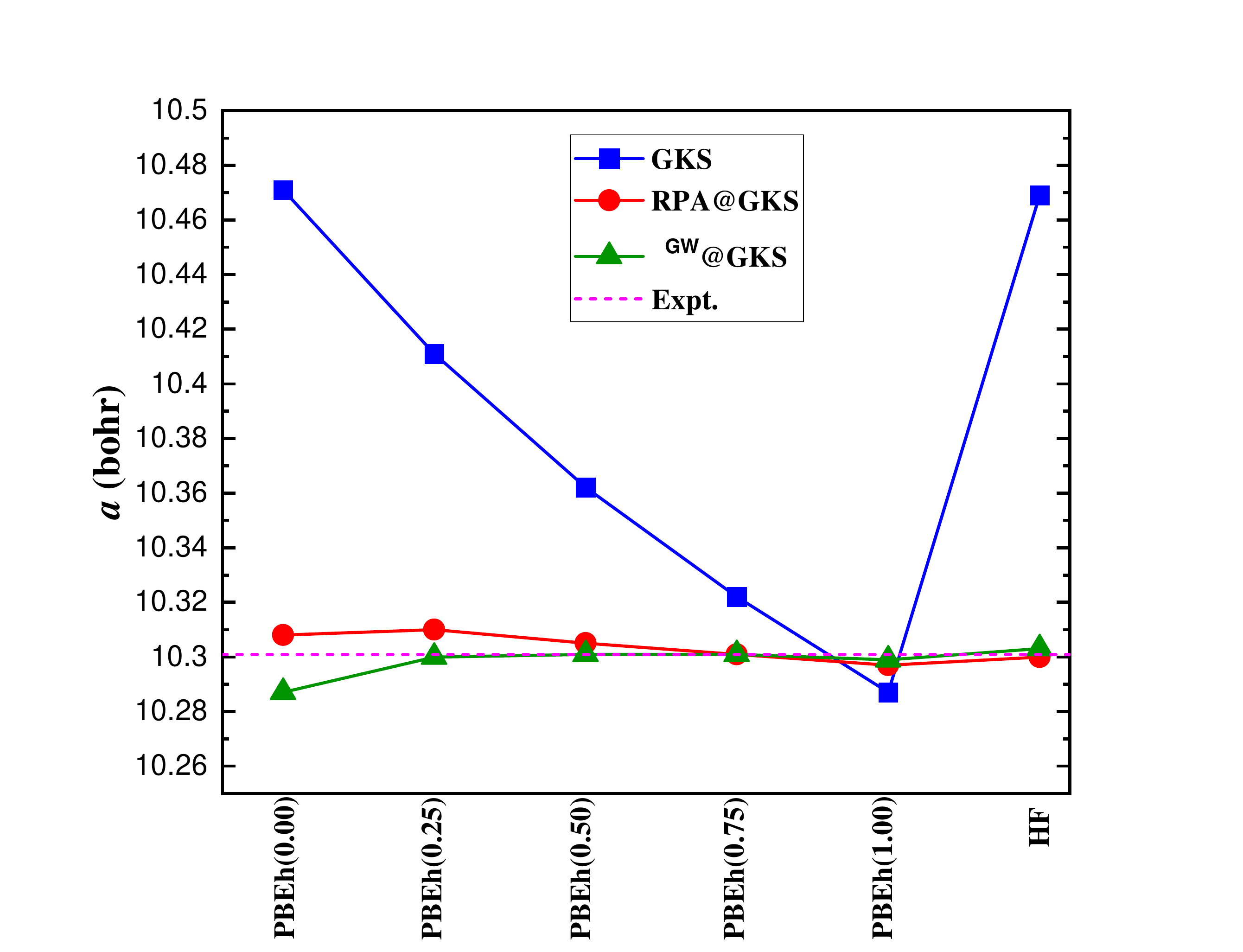}
  \includegraphics[width=0.40\columnwidth]{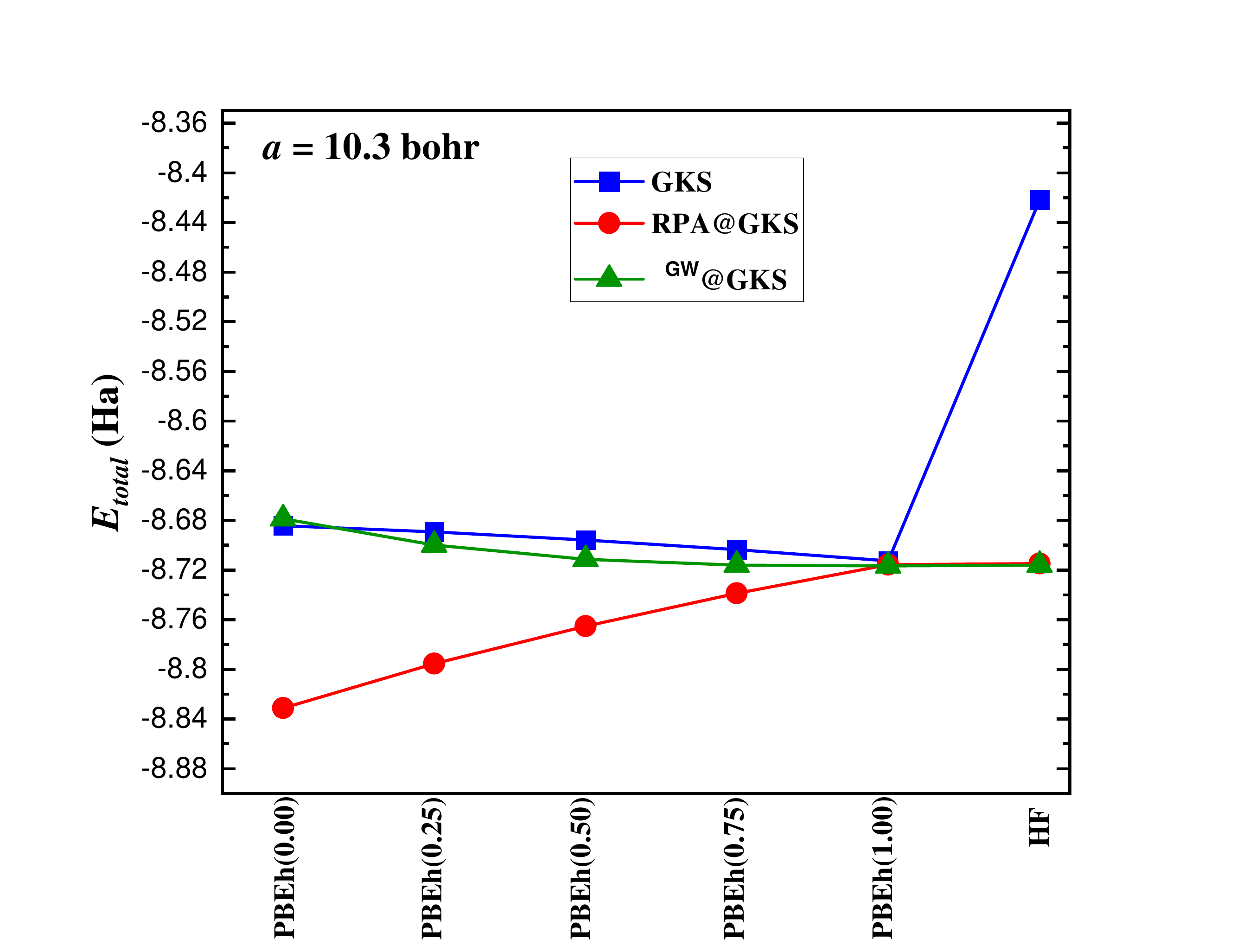}

    \caption{Calculated the equilibrium lattice constants $a$ (bohr) for Aluminium Phosphide, comparing with experiments and comparison of the total energy for the SCF, RPA and $\gamma^{GW}$ as a function of the starting SCF functional.
    \label{fig:S5}
  }
\end{figure*}

\newpage
\subsection{AlAs}

\begin{figure*}[h!]
  \includegraphics[width=0.40\columnwidth]{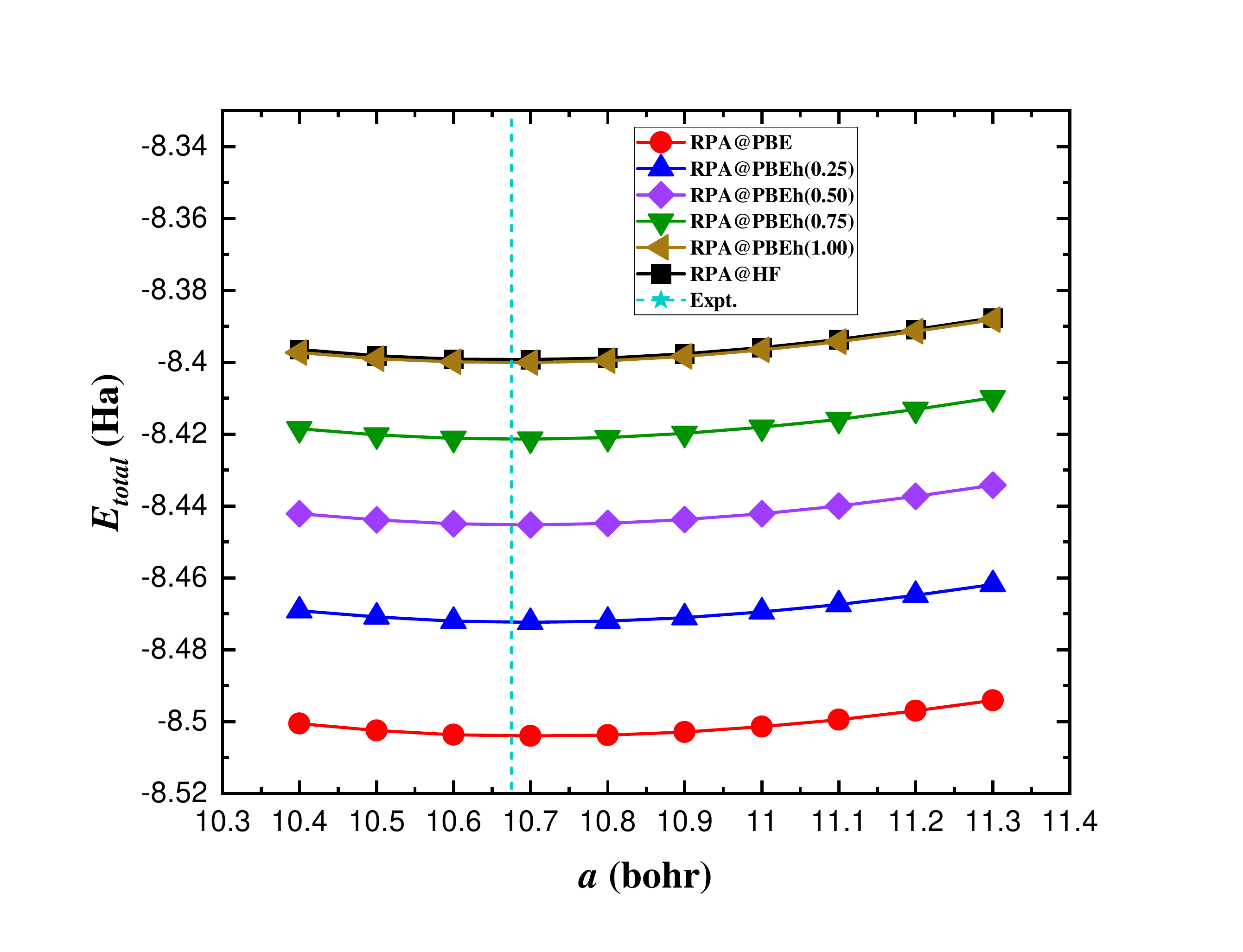}
  \includegraphics[width=0.40\columnwidth]{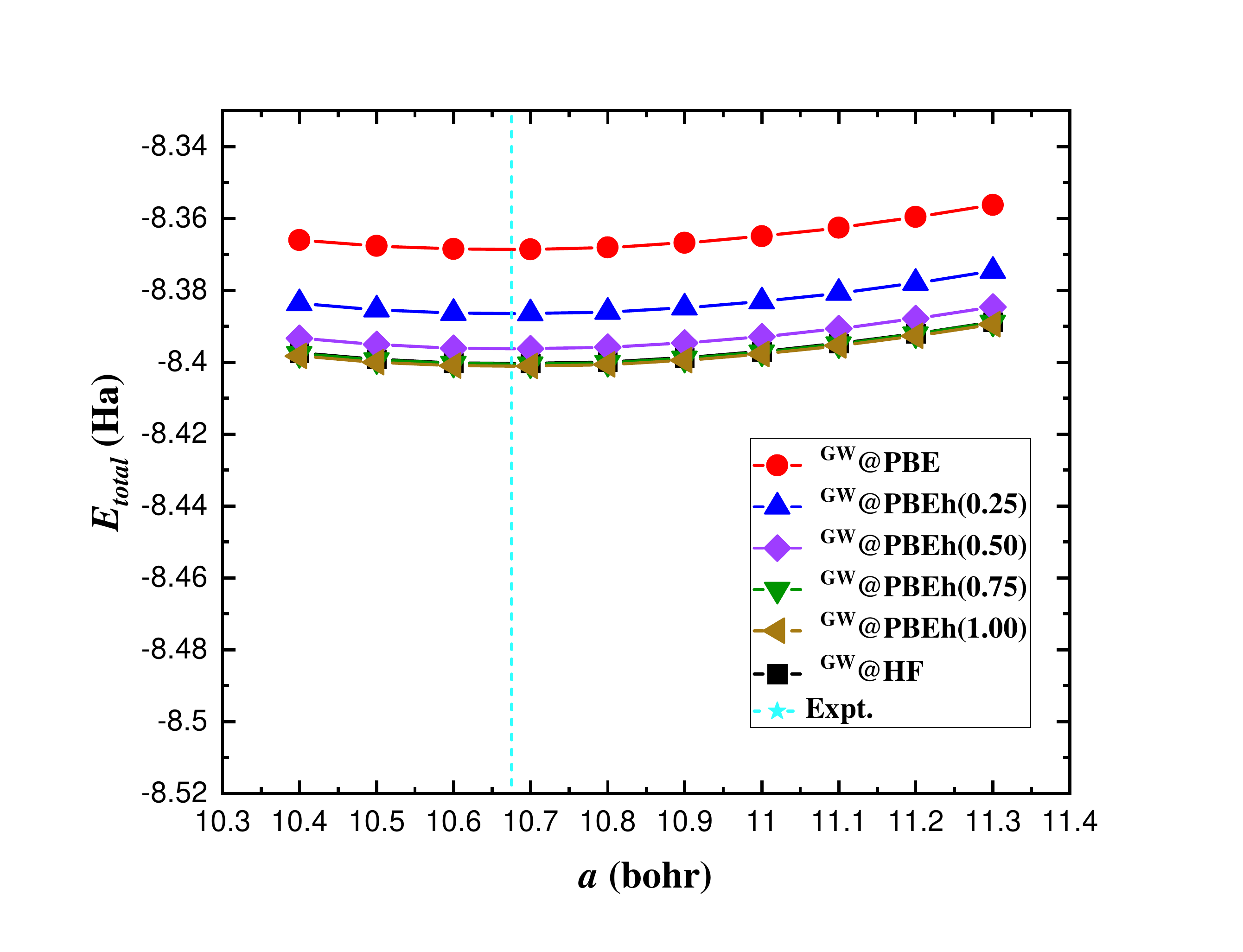}
  \includegraphics[width=0.40\columnwidth]{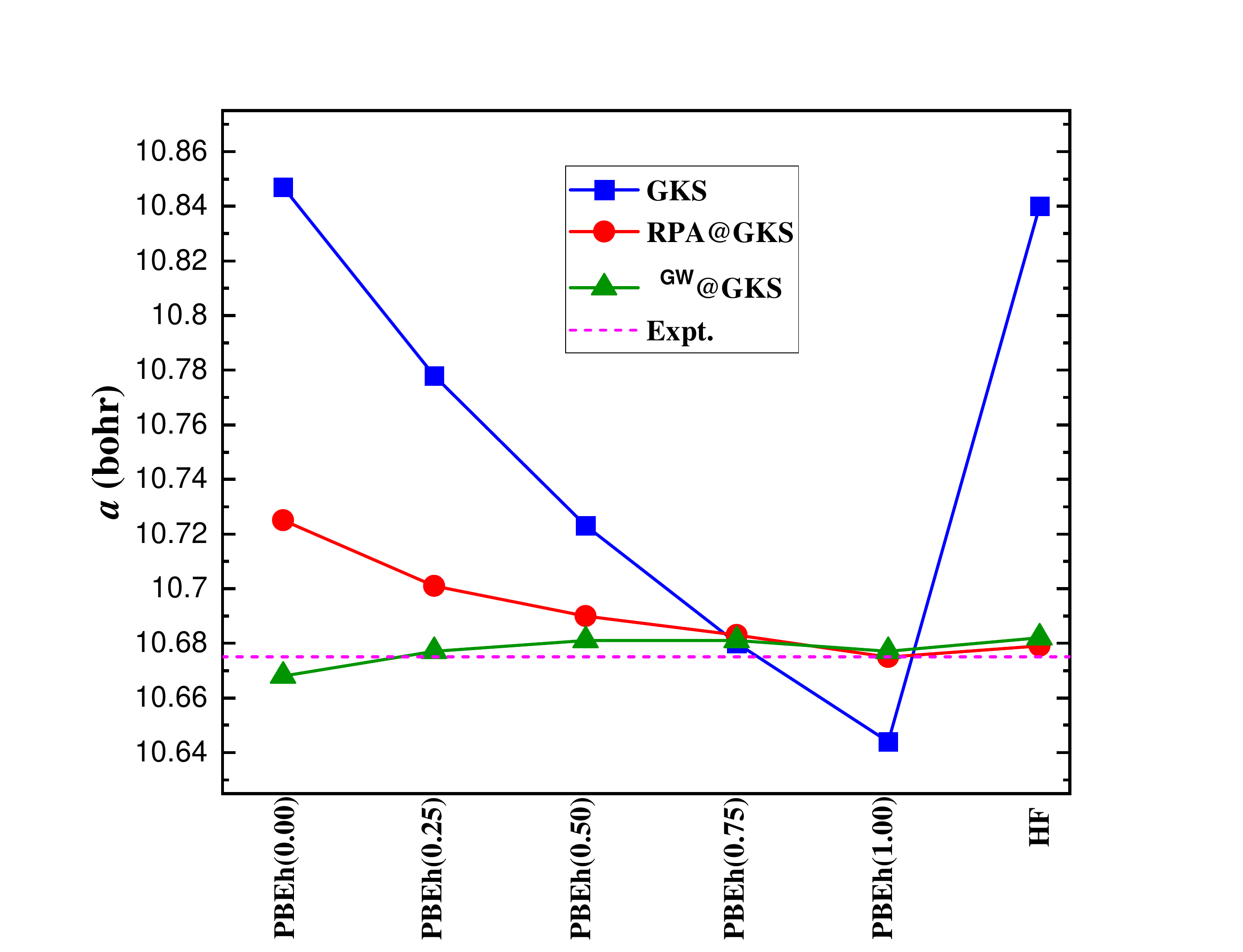}
  \includegraphics[width=0.40\columnwidth]{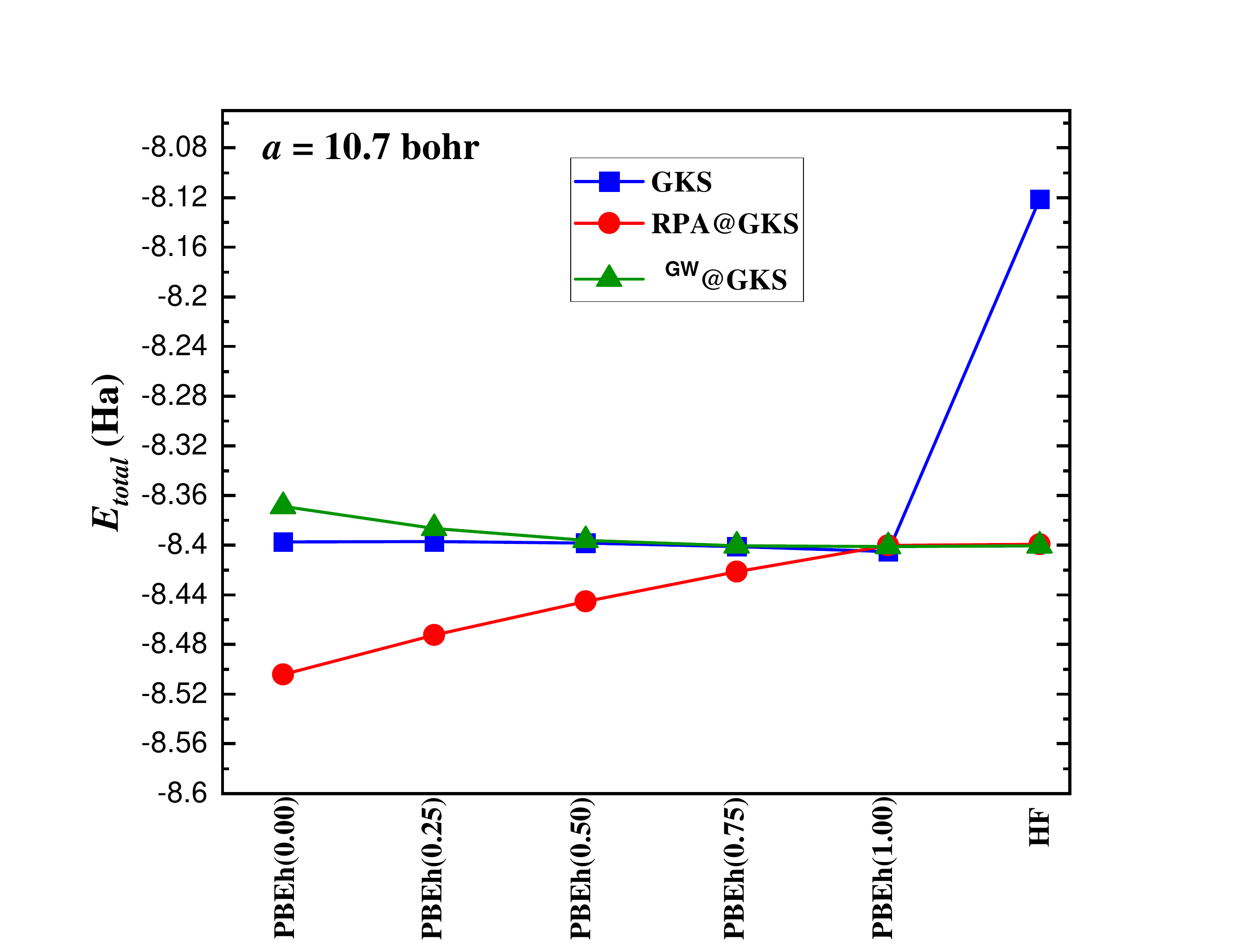}

    \caption{Calculated the equilibrium lattice constants $a$ (bohr) for Aluminum Arsenide, comparing with experiments and comparison of the total energy for the SCF, RPA and $\gamma^{GW}$ as a function of the starting SCF functional.
    \label{fig:S6}
  }
\end{figure*}

\newpage
\subsection{Ge}

\begin{figure*}[h!]
  \includegraphics[width=0.40\columnwidth]{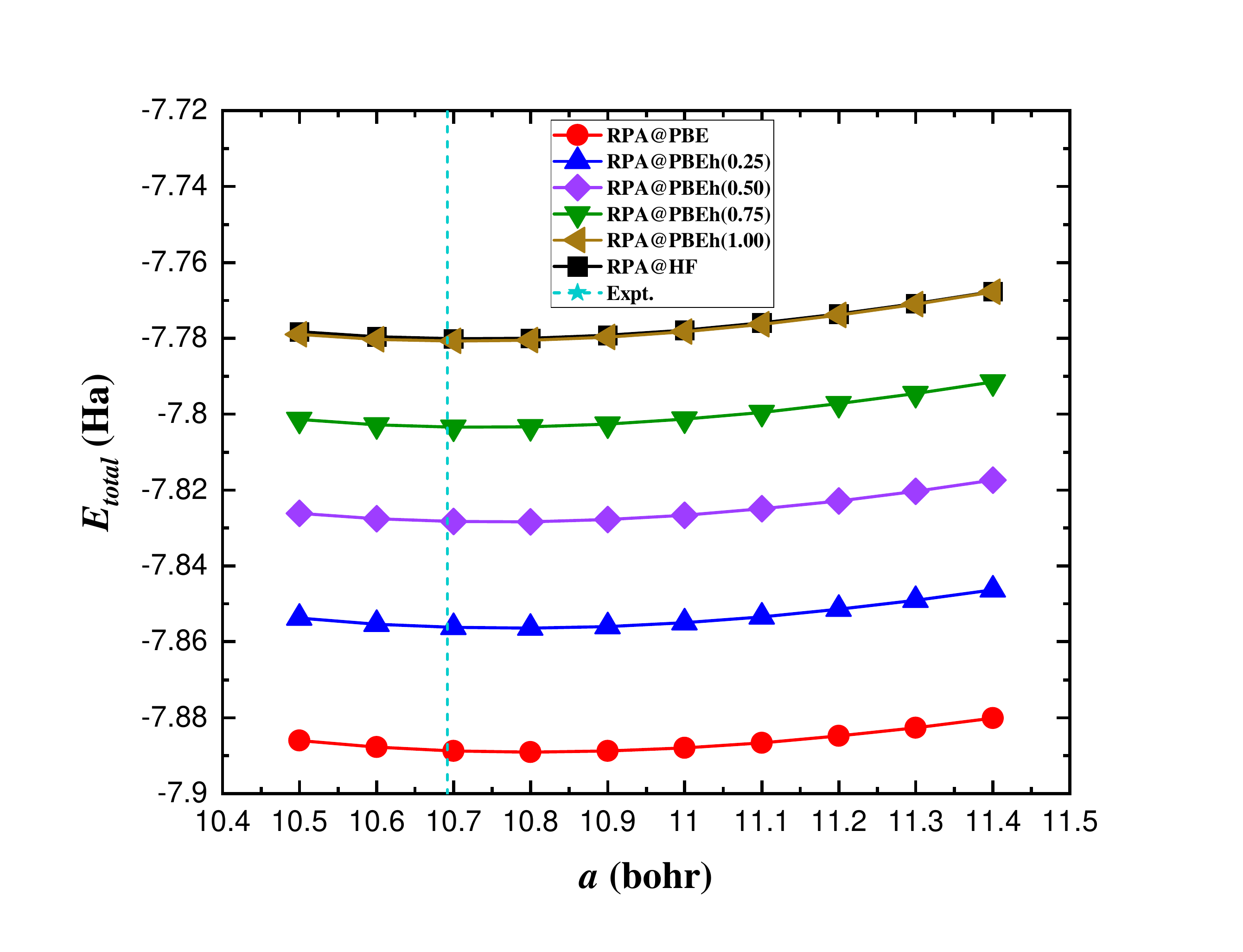}
  \includegraphics[width=0.40\columnwidth]{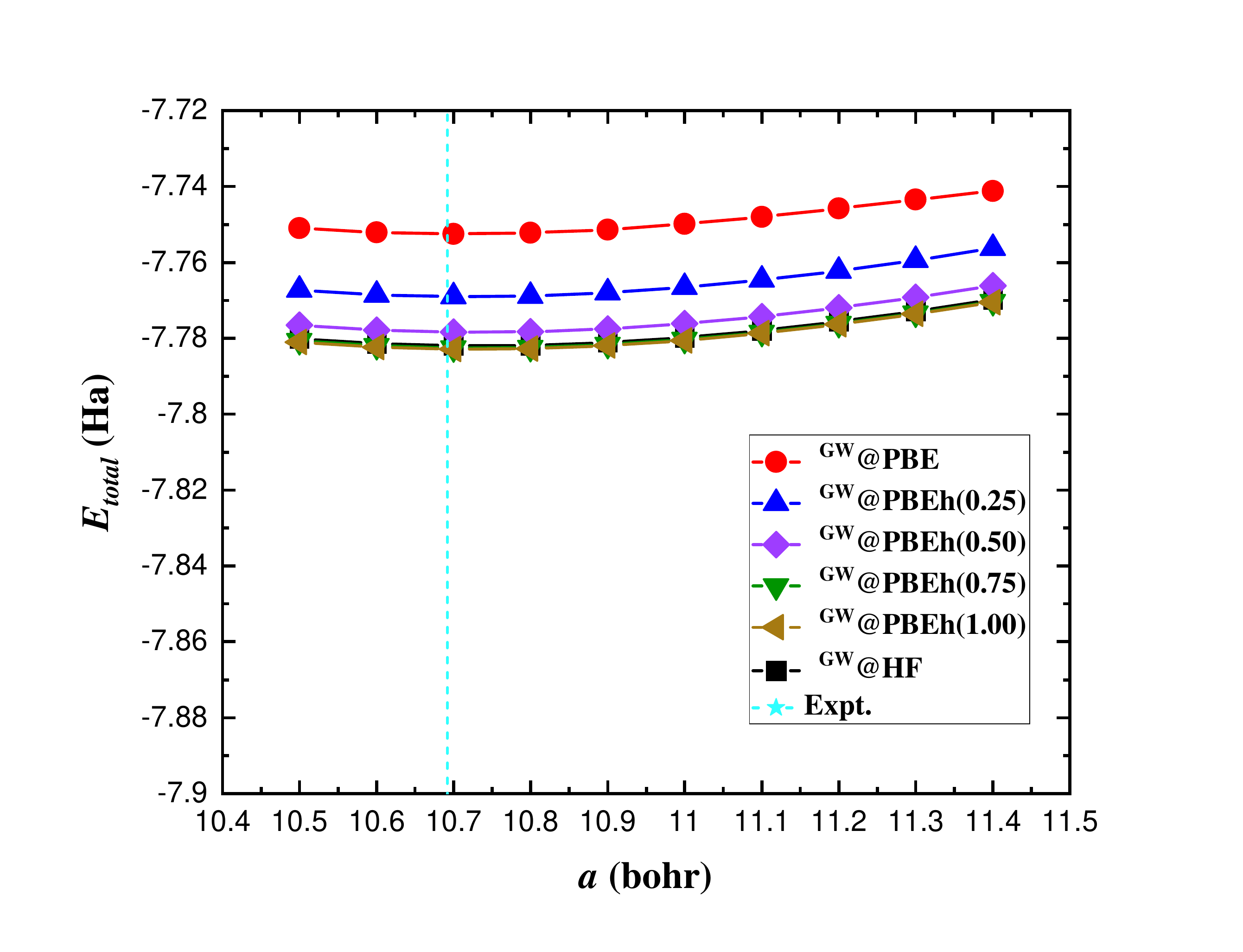}
  \includegraphics[width=0.40\columnwidth]{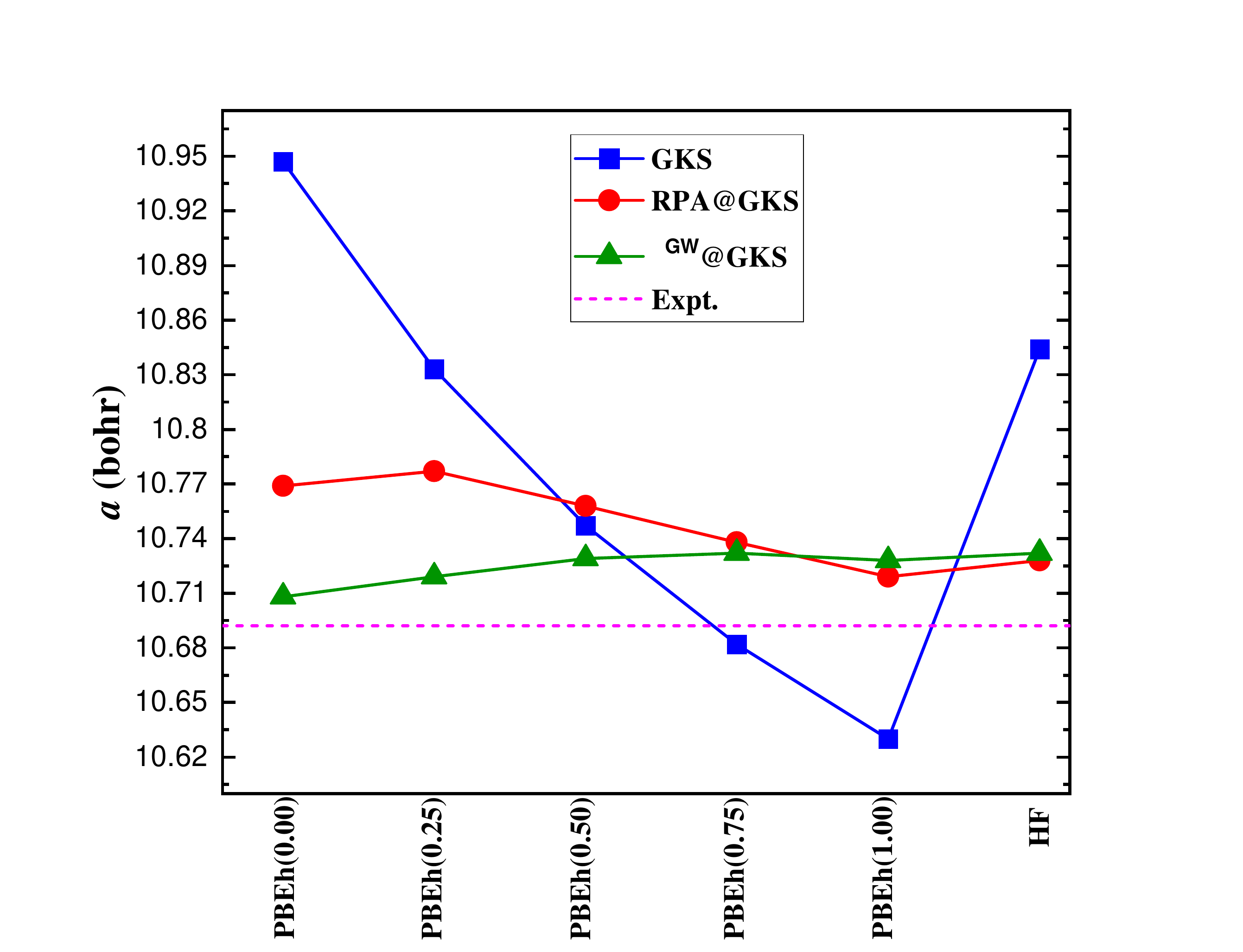}
  \includegraphics[width=0.40\columnwidth]{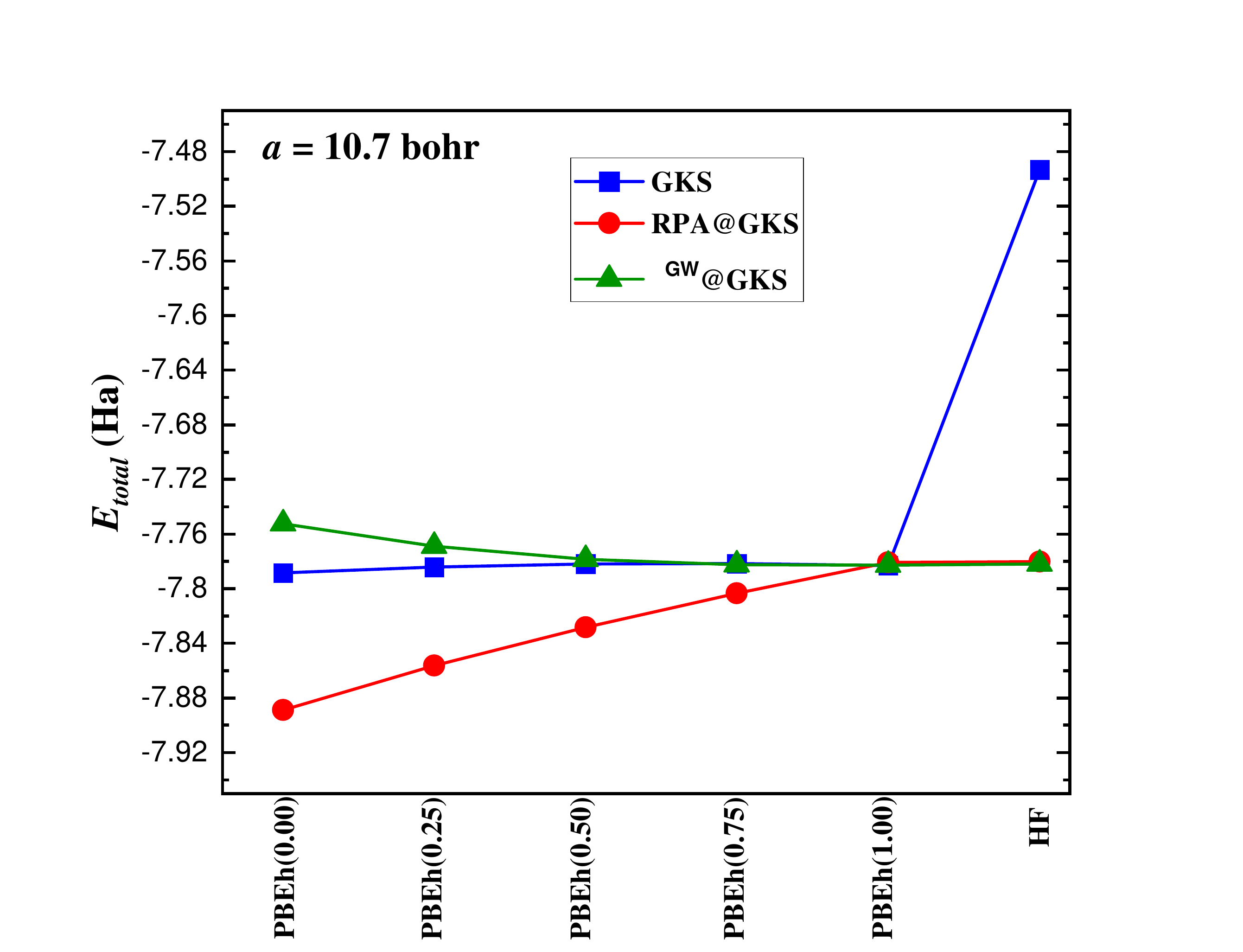}

    \caption{Calculated the equilibrium lattice constants $a$ (bohr) for Germanium, comparing with experiments and comparison of the total energy for the SCF, RPA and $\gamma^{GW}$ as a function of the starting SCF functional.
    \label{fig:S7}
  }
\end{figure*}

\newpage
\subsection{Summary of lattice constants and bulk moduli}

\begin{table*}[h!]
  \caption{
    \label{tab:codes}
    Lattice constant $a$ (bohr), and bulk modulus $B$ (GPa) of solids
  }
  \begin{ruledtabular}
  \begin{tabular}{L{2.7cm}|C{1cm}|C{1.5cm}C{1.5cm}C{1.5cm}C{1.5cm}C{1.5cm}C{1.5cm}C{1.5cm}}
    &    & Si   & C   & SiC  & BN  & AlP  & AlAs  & Ge      \\ \hline
\multirow{2}{2.7cm}{HF}    &  $a$  &  10.409  &  6.739  &  8.262 &  6.807 &  10.469 & 10.84  & 10.844      \\ 
                        &  $B$ & 106.82   &  495.38  & 253.43  & 440.06  & 97.30  & 83.73  & 83.01      \\ \hline
 \multirow{2}{2.7cm}{RPA@HF}   &  $a$  &  10.255  &  6.748  & 8.202  &  6.824 & 10.300  & 10.679  & 10.728      \\
                             &  $B$  &  104.10  &  435.18  & 255.40  & 411.36  & 100.40  &86.81   & 81.97      \\ \hline
 \multirow{2}{2.7cm}{$\gamma^{GW}$@HF}   &  $a$ &  10.258  & 6.750   & 8.207  & 6.825  &  10.303 & 10.682  &  10.732     \\
                               &   $B$ &  103.82  &  437.08  & 254.28  &  410.79 &  100.07 & 86.43  & 81.85      \\
   \hline\hline

\multirow{2}{2.7cm}{PBEh(0.00)}    &  $a$  & 10.407   & 6.774   &  8.310 & 6.861  &  10.471 &  10.847 & 10.947      \\ 
                         &  $B$ &  88.44  & 441.76   &  216.68 & 380.18  & 82.72  & 69.98  &   62.24    \\ \hline
 \multirow{2}{2.7cm}{RPA@PBEh(0.00)}   &  $a$  &  10.282  &  6.815  & 8.257  & 6.867  & 10.308  & 10.725  &   10.769    \\
                             &  $B$  & 97.08   &  477.44  &  235.25 & 401.35  & 96.78  & 83.57  &  86.10     \\ \hline
 \multirow{2}{2.7cm}{$\gamma^{GW}$@PBEh(0.00)}   &  $a$ &  10.232  &  6.713  &  8.180 &  6.800 & 10.287  & 10.668   &   10.708    \\
                               &   $B$ &  104.58  &  420.75  & 262.57  &  417.06 & 101.56  & 87.73  &   85.70    \\
   \hline\hline

\multirow{2}{2.7cm}{PBEh(0.25)}    &  $a$  &  10.345  &  6.733  &  8.256 & 6.808   &  10.411 &  10.778 &  10.833     \\ 
                         &  $B$ &  97.61  &  464.16  & 238.05  & 412.19  & 90.45  &  77.19 &  72.21     \\ \hline
 \multirow{2}{2.7cm}{RPA@PBEh(0.25)}   &  $a$  & 10.270   &  6.793  &  8.237 & 6.852  & 10.310  &  10.701 &  10.777     \\
                             &  $B$  &  99.57  &  457.03  & 242.46  & 406.42  & 97.12  & 83.81  &  77.74     \\ \hline
 \multirow{2}{2.7cm}{$\gamma^{GW}$@PBEh(0.25)}   &  $a$ &  10.250  &  6.735  &  8.198 &  6.820 & 10.300  &  10.677 &  10.719     \\
                               &   $B$ &   104.32 &   429.90 &  257.35 & 412.12  & 100.79  & 86.63  & 82.37      \\
   \hline\hline

   \multirow{2}{2.7cm}{PBEh(0.50)}    &  $a$  &  10.298  &  6.704  &  8.210 & 6.764  & 10.362  &   10.723 &  10.747     \\ 
                         &  $B$ &  105.31  & 484.26   &  255.53 & 438.60  & 96.91  & 83.08  & 80.41      \\ \hline
 \multirow{2}{2.7cm}{RPA@PBEh(0.50)}   &  $a$  &  10.262  &  6.774  & 8.222  & 6.841  & 10.305  & 10.69   &  10.758     \\
                             &  $B$  & 101.35   &  446.39  & 247.92  &  405.85 & 98.47  & 85.15   &  79.34     \\ \hline
 \multirow{2}{2.7cm}{$\gamma^{GW}$@PBEh(0.50)}   &  $a$ & 10.255  &  6.745  & 8.205  & 6.822  & 10.301  &  10.681 &  10.729     \\
                               &   $B$ &  103.90  &  434.05  &  255.04 & 410.84  & 100.15  & 86.38  &  81.98     \\
   \hline\hline

\multirow{2}{2.7cm}{PBEh(0.75)}    &  $a$  &  10.260  &  6.675  & 8.170  & 6.728  &  10.322 & 10.680  & 10.682      \\ 
                         &  $B$ &  111.93  & 502.28   & 270.10  & 461.11  & 102.44  & 88.02  &  87.30     \\ \hline
 \multirow{2}{2.7cm}{RPA@PBEh(0.75)}   &  $a$  & 10.254   &  6.759  & 8.210  & 6.831  & 10.301  & 10.683  &   10.738    \\
                             &  $B$  & 103.37   &  438.89  & 252.60  &  408.86 & 99.96  & 85.94  &    81.01   \\ \hline
 \multirow{2}{2.7cm}{$\gamma^{GW}$@PBEh(0.75)}   &  $a$ &   10.255 &  6.748  & 8.207  & 6.825  & 10.301  & 10.681  &  10.732     \\
                               &   $B$ &  103.91  &  436.08  & 254.39  & 410.23  & 100.22  & 86.59  &  82.07     \\
   \hline\hline

   \multirow{2}{2.7cm}{PBEh(1.00)}    &  $a$  &  10.229  &  6.652  & 8.135  & 6.696  &  10.287 & 10.644  & 10.630      \\ 
                         &  $B$ & 117.85   &  517.10  & 282.82  & 480.12  & 107.06  & 92.36  & 93.18      \\ \hline
 \multirow{2}{2.7cm}{RPA@PBEh(1.00)}   &  $a$  &   10.248 &  6.745  &  8.199 &  6.821 & 10.297  & 10.675  &   10.719    \\
                             &  $B$  &  105.12  &  433.99  & 256.73  & 412.00  & 100.89  & 87.25  &  82.95     \\ \hline
 \multirow{2}{2.7cm}{$\gamma^{GW}$@PBEh(1.00)}   &  $a$ &  10.252  & 6.749    & 8.206  &  6.824 & 10.299  & 10.677  &  10.728     \\
                               &   $B$ &  104.10  &  436.92  & 254.72  & 410.61  & 100.33  & 86.60  & 82.24      \\
   \hline\hline

   \multirow{2}{2.7cm}{Expt.}    &  $a$  &  10.261  & 6.741   &  8.213 & 6.833  &  10.301 &  10.675 & 10.692      \\ 
                         &  $B$ &  99  &  443  & 225  &  369-400 & 86  & 77  & 76      \\    
  \end{tabular}
  \end{ruledtabular}
\end{table*}


\clearpage

\bibliography{my_biblio}